\documentclass[aps,prb,twocolumn,superscriptaddress]{revtex4-2}
\usepackage{amsmath,amssymb,amsfonts}
\usepackage{bm, latexsym}
\usepackage{graphicx}
\usepackage{bbm,times}
\usepackage[normalem]{ulem}
\usepackage[pdftex]{xcolor}
\usepackage[pdftex]{hyperref}
\usepackage{braket}

\begin{document}

\newcommand{\cred}[1]{\textcolor{black}{#1}}

\title{Kondo breakdown induced by non-Hermitian complex hybridization}
\author{Kazuki Yamamoto}
\email{kazuki-yamamoto@omu.ac.jp}
\affiliation{Research Institute for Innovation and Co-Creation, Osaka Metropolitan University, Sakai, Osaka 599-8531, Japan}
\affiliation{Department of Physics, Osaka Metropolitan University, Sumiyoshi, Osaka 558-8585, Japan}
\affiliation{Nambu Yoichiro Institute of Theoretical and Experimental Physics (NITEP), Osaka Metropolitan University, Sumiyoshi, Osaka 558-8585, Japan}
\affiliation{Department of Physics, Institute of Science Tokyo, Meguro, Tokyo 152-8551, Japan}
\author{Masaya Nakagawa}
\affiliation{Department of Physics, University of Tokyo, Hongo, Tokyo 113-0033, Japan}
\author{Norio Kawakami}
\affiliation{Fundamental Quantum Science Program, TRIP Headquarters, RIKEN, Wako 351-0198, Japan}
\affiliation{Department of Physics, Ritsumeikan University, Kusatsu, Shiga 525-8577, Japan}
\affiliation{Department of Materials Engineering Science, University of Osaka, Toyonaka, Osaka 560-8531, Japan}

\date{\today}

\begin{abstract}
Recently, a non-Hermitian Anderson impurity model with one-body loss has been studied in [\href{https://journals.aps.org/prb/abstract/10.1103/PhysRevB.111.125157}{Phys. Rev. B \textbf{111}, 125157 (2025)}], and it has been demonstrated that the renormalization effect generated by strong correlations counterintuitively changes the nature of dissipation into an emergent many-body dissipation that causes a Kondo breakdown. In a closely related context, it is also known that two-body loss in a non-Hermitian Kondo model triggers the Kondo breakdown. To elucidate the essence of these phenomena, we study the Anderson impurity model with a non-Hermitian complex hybridization as an effective model that provides a simple understanding of the Kondo breakdown. Using the slave-boson mean-field theory, we show that this model can explain the Kondo breakdown with a single complex parameter. Furthermore, we provide the exact Bethe ansatz solutions that support the results obtained by the slave-boson mean-field theory.
\end{abstract}

\maketitle

\section{Introduction}
The Kondo effect has been of central interest in condensed matter physics as it reflects renormalization effects essential in strongly correlated phenomena \cite{Kondo64, Hewson97, Coleman15}. The Anderson impurity model (AIM) is known to capture the basic properties of an impurity spin interacting with itinerant fermions via the hybridization coupling \cite{Anderson61, Tsvelick83, Okiji84}. This model exhibits rich physical properties unique to strongly correlated systems, e.g., in the conductance of a quantum dot in and out of equilibrium \cite{Meir91, Beenakker91, Meir92, Meir93, Ralph94, Lopez03}. Importantly, the renormalization effect describes a crossover from the Kondo regime to the valence fluctuation regime, where many-body effects are gradually smeared and charge fluctuations govern the dynamics \cite{Coleman84, Coleman87, Newns87}. Such an interplay between the impurity fermion and the itinerant fermions offers an ideal platform to investigate many-body physics and has been experimentally studied not only in condensed matter \cite{Schroder00, Aynajian12, Gordon98, Cronenwett98, Borzenets20} but also in ultracold atoms \cite{Scazza14, Riegger18, Ono21}. 

On another front, non-Hermitian (NH) systems have been actively investigated in open quantum systems \cite{Daley14, Ashida20, Fazio24}. In particular, rapid progress in dissipation engineering with ultracold atoms \cite{Syassen08, Yan13, Patil15, Schneider17, Tomita17, Spon18, Gerbier20, Honda23, Ott13, Labouvie16, Benary22, Takasu20, Yamamoto21, Ren22, Liang22, Tsuno24, Jo25} has offered the possibility to explore NH many-body phenomena \cite{Yamamoto19, Hamazaki19, Ashida16, Ashida17, Xu20, Zhang20, Liu20, Yamamoto22, Yamamoto23, Kawabata22, TYoshida24, Kim24, Takemori25, Takemori25L, Hanai19, Gopa21, Yang21, Sun23, Yu24, Yamamoto26}. Recently in Ref.~\cite{Yamamoto24}, the NH-AIM with one-body loss has been studied to investigate the impact of dissipation on the Kondo effect by incorporating the effect of valence fluctuations. Intriguingly, in the Kondo regime, Ref.~\cite{Yamamoto24} has demonstrated that the single-body dissipation is renormalized to zero and generates an emergent many-body dissipation, which induces the Kondo breakdown signaled by a vanishing resonance width. Such a result is counterintuitive because the one-body loss typically shortens the lifetime of the impurity, while the Kondo breakdown rather means that the lifetime is infinitely enhanced.

The Kondo effect in ultracold alkaline-earth atoms has been widely explored in theory \cite{Gorshkov10, ARey10, ARey10A, Demler13, ARey15, Nakagawa15, Nishida16, Zhai16, Scazza25, Ashida18, Schiro24, Schiro24arXiv, Stefanini25, Vanh25, Werner25, Qu25, Hasegawa21, Han23, Kulkarni25, Kulkarni25light, Kattel25, Lourenco18, Kulkarni22} and the Kondo Hamiltonian has been realized in experiments though the direct observation of the Kondo resonance is not yet achieved mainly due to thermal fluctuations \cite{Riegger18, Ono21}. In this context, the NH Kondo model has been investigated \cite{Nakagawa18, Kattel25L, Chen25, Burke25}, where non-Hermiticity arises from two-body loss caused by inelastic scattering between a ground-state atom and a metastable excited-state atom \cite{Riegger18, Ono21}. Such two-body loss induces a breakdown of the Kondo effect associated with an anomalous reversion of the renormalization group flow \cite{Nakagawa18}. The Kondo breakdown induced by non-Hermiticity offers a prototypical phenomenon in the strong correlation effects on NH quantum impurity physics. However, since the specific form of dissipation differs between the NH-AIM with one-body loss \cite{Yamamoto24} and the NH Kondo model with two-body loss \cite{Nakagawa18}, the core mechanism behind the Kondo breakdown is still elusive.

In this paper, we study the AIM with the NH complex hybridization and demonstrate that the renormalization effect of the complex hybridization gives a simple mechanism of the Kondo breakdown. We first analyze the model by means of the NH generalization of the slave-boson (SB) mean-field theory and elucidate that the model can describe the Kondo breakdown with a single complex parameter, in contrast to the case of one-body loss. We also employ the second-order perturbation theory and show that the Kondo breakdown induced by the complex hybridization is relevant to that of the NH Kondo model with two-body loss. Moreover, we obtain the exact Bethe ansatz results that support the transition point obtained by the SB mean-field theory.


The rest of this paper is organized as follows. In Sec.~\ref{sec_NHSB}, we introduce the NH-AIM with a complex hybridization and apply the NH-SB theory. We also give numerical and analytical results for the effective ground state of the NH-AIM. Sec.~\ref{sec_nhkondo} presents the relation between the NH-AIM and the NH-Kondo model by means of the second-order perturbation theory. Sec.~\ref{sec_bethe} is devoted for the exact Bethe ansatz results for the Kondo breakdown.
Finally, we summarize the results in Sec.~\ref{sec_conclusion}.

\section{Non-Hermitian slave-boson mean-field theory}
\label{sec_NHSB}
In this section, we introduce the NH-AIM by generalizing the hybridization coupling to a complex value of NH type and apply the NH-SB mean-field theory. Then, we give both numerical and analytical results by solving the self-consistent equations for physical parameters.

\subsection{Model and implication from the noninteracting case}
\label{sec_model}

The AIM includes an on-site Coulomb interaction $U$ at an impurity site and a coupling to itinerant fermions. Here, we study the case of a single impurity. Recently, NH generalization of the Kondo problem has been studied with the AIM and the Kondo model, and it has been shown that non-Hermiticity causes the breakdown of the Kondo effect \cite{Yamamoto24, Lourenco18, Kulkarni22, Nakagawa18, Kattel25L, Chen25, Burke25}. In this paper, we propose that an essential aspect of the Kondo breakdown in NH systems can be understood by using the NH-AIM with a complex hybridization, which is defined by
\begin{align}
H_\mathrm{eff}=&\sum_{\bm k \sigma}\epsilon_{\bm k} c_{\bm k \sigma}^\dag c_{\bm k \sigma} +\sum_\sigma E_d  n_{d\sigma}\notag\\
&+ U n_{d\uparrow} n_{d\downarrow} + \sum_{\bm k \sigma}\left[\tilde V_{\bm k d} c_{\bm k \sigma}^\dag c_{d\sigma} + \tilde V_{d \bm k} c_{d\sigma}^\dag c_{\bm k \sigma}\right],
\label{eq_NHAnderson}
\end{align}
where $c_{d\sigma}$ and $c_{\bm k\sigma}$ denote annihilation operators for fermions at an impurity site and in a fermion reservoir, $n_{d\sigma}=c_{d\sigma}^\dag c_{d\sigma}$ is the particle number operator at the impurity site, $E_d$ is the impurity level, and $\tilde V_{\bm k d}$ and $\tilde V_{d \bm k}$ are the complex hopping amplitude \cite{PT}. \cred{In Eq.~\eqref{eq_NHAnderson}, we write $\tilde{V}_{d\bm{k}}=V_{d\bm{k}}e^{i\theta_{\bm{k}}}$ and $\tilde{V}_{\bm{k}d}=V_{\bm{k}d} e^{-i\theta_{\bm{k}}}$ by using a gauge transformation $c_{\bm{k}\sigma}\to c_{\bm{k}\sigma} e^{i\theta_{\bm{k}}}$ associated with the U(1) symmetry of the Hamiltonian. We then assume that $V_{d\bm{k}}=V_{\bm{k}d}$ takes a complex value, which makes the Hamiltonian~\eqref{eq_NHAnderson} non-Hermitian.}

The motivation to introduce the model in Eq.~\eqref{eq_NHAnderson} is twofold. First, it shows the Kondo breakdown qualitatively similar to that in the NH-AIM with one-body loss \cite{Yamamoto24} by a simpler mechanism, as shown in Sec.~\ref{sec_numerics}. Second, it reduces to the NH Kondo model with two-body loss studied in Ref.~\cite{Nakagawa18} in the Kondo limit (see Sec.~\ref{sec_nhkondo}). Thus, the NH-AIM \eqref{eq_NHAnderson} with a complex hybridization serves as an effective model that enables a simple understanding of the Kondo breakdown induced by non-Hermiticity. We note that Eq.~\eqref{eq_NHAnderson} can be also derived from the Lindblad master equation \cite{Daley14} (see Appendix \ref{app_lindblad}).

Though the precise form of $\tilde V_{\bm k d}$ and $\tilde V_{d \bm k}$ would be model dependent, the physics studied below is determined only by $\tilde V^2= \tilde V_{\bm k d}\tilde V_{d\bm k}=V_{\bm{k}d}V_{d\bm{k}}$. \cred{We assume that $\tilde V$ does not depend on momentum, which holds if $V_{d\bm{k}}$ is $\bm k$-independent.} Thus, in the following calculations, we tune the parameter $\tilde V\equiv V_0-iV$ with real $V_0$ and $V$, whose effects are reflected in the complex hybridization $\tilde \Delta\equiv\pi\rho\tilde V^2$. Physically, the imaginary part $V$ reflects the phase mismatch between $\tilde{V}_{\bm{k}d}$ and $\tilde{V}_{d\bm{k}}$, and therefore controls the strength of non-Hermiticity of the model. Here, we have assumed that the $\bm k$-dependence of $\tilde V_{\bm k d}\tilde V_{d\bm k}$ can be ignored, and the density of states is given by a constant value $\rho=1/(2D)$ with a cutoff $D$.

To understand the phenomena induced by the complex hybridization, it is helpful to study the noninteracting case $U=0$ in Eq.~\eqref{eq_NHAnderson}. The impurity Green function is calculated as \cite{Yamamoto24}
\begin{align}
\tilde G_{d,U=0}^{R\sigma}(\epsilon)&=\frac{1}{\epsilon+i\eta- E_d-\sum_{\bm k}[\tilde V^2/(\epsilon+i\eta-\epsilon_{\bm k})]}\notag\\
&=\frac{1}{\epsilon-E_d+i\tilde \Delta-\tilde V^2 P\sum_{\bm k}[1/(\epsilon-\epsilon_{\bm k})]},\label{eq_green0i}
\end{align}
where $\eta\to+0$ and $P$ stands for the principal value. We find that the width of the impurity level is given by 
\begin{align}
\Delta^\prime\equiv\mathrm{Re}[\tilde{\Delta}]=\pi\rho(V_0^2-V^2).
\label{eq_RWNH}
\end{align}
Here, we have ignored the contribution from $2V_0VP\sum_{\bm k}[1/(\epsilon-\epsilon_{\bm k})]$ in Eq.~\eqref{eq_green0i} by assuming that the density of states $\rho$ is constant with the large band width $D$, and that $\epsilon_{\bm k}$ is close to the Fermi energy. In the Hermitian limit $V=0$, $\Delta^\prime$ is positive and the impurity fermion has a finite lifetime due to the tunneling into the reservoir. However, when the non-Hermiticity is introduced, $\Delta^\prime$ in Eq.~\eqref{eq_RWNH} decreases as $V$ is increased and vanishes at $V=V_0$, where the lifetime of impurity fermions diverges. From this result, we can see a unique role of the complex hybridization in NH systems: while the phase factors of $\tilde{V}_{\bm{k}d}$ and $\tilde{V}_{d\bm{k}}$ cancel out in $\tilde{\Delta}$ for the Hermitian case, the phase mismatch in the NH case leads to the suppression of the resonance width. Such an enhancement of the impurity lifetime induced by non-Hermiticity captures the key mechanism of the Kondo breakdown in the interacting case as shown below. \cred{We note that the divergence of the impurity lifetime occurs only when $V=V_0$, and we obtain the negative $\Delta^\prime$ for $V>V_0$ in Eq.~\eqref{eq_RWNH}, which may indicate the unstable Kondo state that was studied in Ref.~\cite{Yamamoto24}.}

\subsection{Method}
\label{sec_sb}
We employ the SB theory for the AIM in the strong correlation limit ($U\to\infty$), which was first developed by Coleman \cite{Coleman84, Coleman87, Newns87}. We briefly summarize the NH generalization of the SB theory \cite{Yamamoto24} to the NH-AIM with the complex hybridization (see Appendix \ref{App_SB} for details). The calculations are similar to those for the case of one-body loss studied in Ref.~\cite{Yamamoto24}. However, we note that the imaginary part of the impurity level arises only from the renormalization effect in the current model, while it is initially introduced in the model as one-body loss in Ref.~\cite{Yamamoto24}. As we will see later, this difference of the model enables us to simplify the understanding of the renormalization effects induced by strong correlations in the NH impurity phenomena. 

In the NH SB mean-field theory, we introduce the complex SB field $b$ and the complex Lagrange multiplier $\tilde \lambda$, the latter of which enforces a constraint on the total particle number at the impurity site. Then, the NH-AIM with the Lagrange multiplier is written as
\begin{align}
&H_\mathrm{eff}(\tilde \lambda)=\sum_{\bm k \sigma} \epsilon_{\bm k} c_{\bm k \sigma}^\dag c_{\bm k \sigma} + \sum_\sigma E_d d_\sigma^\dag d_\sigma +\sum_{\bm k \sigma}[\tilde V_{\bm k d} c_{\bm k \sigma}^\dag b^\dag d_\sigma \notag\\
&+ \tilde V_{d\bm k} d_\sigma^\dag b c_{\bm k \sigma}]+\tilde \lambda\big(\sum_\sigma d_\sigma^\dag d_\sigma + b^\dag b -1\big),
\label{eq_NHAnderson_vtilde}
\end{align}
from which we can read off that the impurity level is renormalized to a complex value $E_d+\tilde \lambda$. \cred{The SB field $b^\dag$ and $b$ are introduced as $c_{d\sigma}=b^\dag d_\sigma$ and $c_{d\sigma}^\dag = d_\sigma^\dag b$ with a constraint $\sum_\sigma d_\sigma^\dag d_\sigma +b^\dag b = 1$. Here, $d_\sigma$ is the new fermion operator in the restricted Hilbert space spanned by $\ket{\uparrow}=d_\uparrow^\dag|\Omega\rangle$, $\ket{\downarrow}=d_\downarrow^\dag|\Omega\rangle$, and $|0\rangle=b^\dag|\Omega\rangle$, where $|\Omega\rangle$ is a vacuum state.} For this model, the retarded (advanced) NH Green function of an impurity fermion is given by
\begin{align}
\tilde G_d^{R(A)\sigma}(\omega)=[\omega-E_d^\prime\mp\Delta_b^\mathrm{Im}\pm i(\Delta_b^\mathrm{Re}\mp\mathrm{Im}\tilde\lambda)]^{-1},
\label{eq_greenimpurity}
\end{align}
where the upper (lower) sign is for the retarded (advanced) Green function. Here, we have defined four important physical quantities. The first one is the real part of the renormalized impurity level $E_d^\prime\equiv E_d+\mathrm{Re}\tilde\lambda$, and the second one is the imaginary part of the renormalized impurity level $-\mathrm{Im}\tilde\lambda$ (whose sign is reversed for later convenience). The latter can be regarded as effective single-body dissipation, since the imaginary part of the impurity level arises in the effective Hamiltonian derived from the Lindblad equation with one-body loss \cite{Yamamoto24}. The rest two quantities are the renormalized resonance widths $\Delta_b^\mathrm{Re}\mp\mathrm{Im}\tilde\lambda$ and the renormalized peak position $E_d^\prime\pm\Delta_b^\mathrm{Im}$ of the NH impurity Green function with $\Delta_b^\mathrm{Re}\equiv\mathrm{Re}\Delta_b$ and $\Delta_b^\mathrm{Im}\equiv\mathrm{Im}\Delta_b$, where $\Delta_b\equiv b_0^2 \tilde\Delta$ is the renormalized complex hybridization. \cred{Here, we note that $b_0$ is defined via the U(1) transformation of the SB field $b$ as shown in Appendix \ref{App_SB}.} These parameters are determined from the following self-consistent equations (SCEs) for the effective ground state (see Appendix \ref{App_SB} for details):
\begin{align}
\tilde \lambda &+\frac{\tilde\Delta}{\pi}\log\left[\frac{(E_d^\prime\pm\Delta_b^\mathrm{Im})^2+(\Delta_b^\mathrm{Re}\mp\mathrm{Im}\tilde\lambda)^2}{(D+E_d^\prime\pm\Delta_b^\mathrm{Im})^2+(\Delta_b^\mathrm{Re}\mp\mathrm{Im}\tilde\lambda)^2}\right]\notag\displaybreak[2]\\
&\pm\frac{2i\tilde\Delta}{\pi}\Bigg[\tan^{-1}\Bigg(\frac{E_d^\prime\pm\Delta_b^\mathrm{Im}}{\Delta_b^\mathrm{Re}\mp\mathrm{Im}\tilde\lambda}\Bigg)\notag\displaybreak[2]\\
&-\tan^{-1}\Bigg(\frac{D+E_d^\prime\pm\Delta_b^\mathrm{Im}}{\Delta_b^\mathrm{Re}\mp\mathrm{Im}\tilde\lambda}\Bigg)\Bigg]\mp i\Delta_b=\mp i\tilde\Delta,
\label{eq_self}
\end{align}
\cred{where the cutoff $D$ is introduced to ensure the convergence.} Below, we analyze a crossover from the Kondo regime to the valence fluctuation regime by solving Eq.~\eqref{eq_self} with changing the impurity level $E_d$. \cred{Here, we note that all mean-field variables are completely determined from Eq.~\eqref{eq_self} .}

\subsection{Numerical results}
\label{sec_numerics}
We numerically solve the SCEs \eqref{eq_self} and investigate how the renormalization effect appears in the physical parameters. First, we summarize our main findings. From the numerical solution, we obtain the following results:
\begin{itemize}
\item \textit{Kondo breakdown}--The renormalized complex hybridization $\Delta_b$ induces a quantum phase transition from the Kondo phase to the unscreened phase. This is characterized by the vanishment of the renormalized resonance width $\Delta_b^\mathrm{Re}\mp\mathrm{Im}\tilde\lambda$. \\
\item \textit{Simplification of the renormalization mechanism}--In the Kondo regime for the deep impurity level with $|E_d|\gg\Delta_0\equiv\pi\rho V_0^2$, the real and the imaginary parts of the renormalized impurity level, $E_d^\prime$ and $\mathrm{Im}\tilde \lambda$, are almost pinned to zero, and $\Delta_b$ is the only complex renormalized parameter. This fact distills the key insights into the microscopic mechanism of the Kondo breakdown.
\item \textit{Crossover to the valence fluctuation regime}--As the impurity level $E_d$ is raised, the Kondo regime crossovers to the valence fluctuation regime, where charge fluctuations gradually become dominant.
\end{itemize}

\begin{figure}[t]
\includegraphics[width=8.5cm]{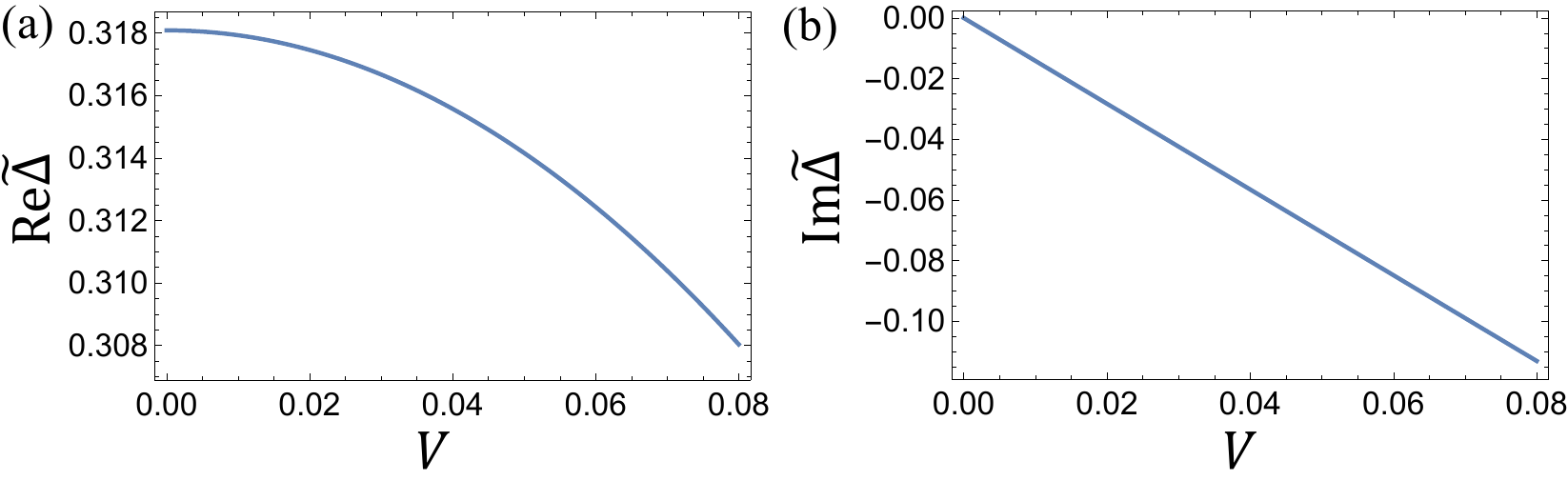}
\caption{(a) Real and (b) imaginary parts of $\tilde\Delta$ as a function of the imaginary coupling $V$. The parameters are set to $D=1$ and $V_0=0.45$.}
\label{fig_DeltaVtilde}
\end{figure}

\begin{figure*}[t]
\includegraphics[width=17cm]{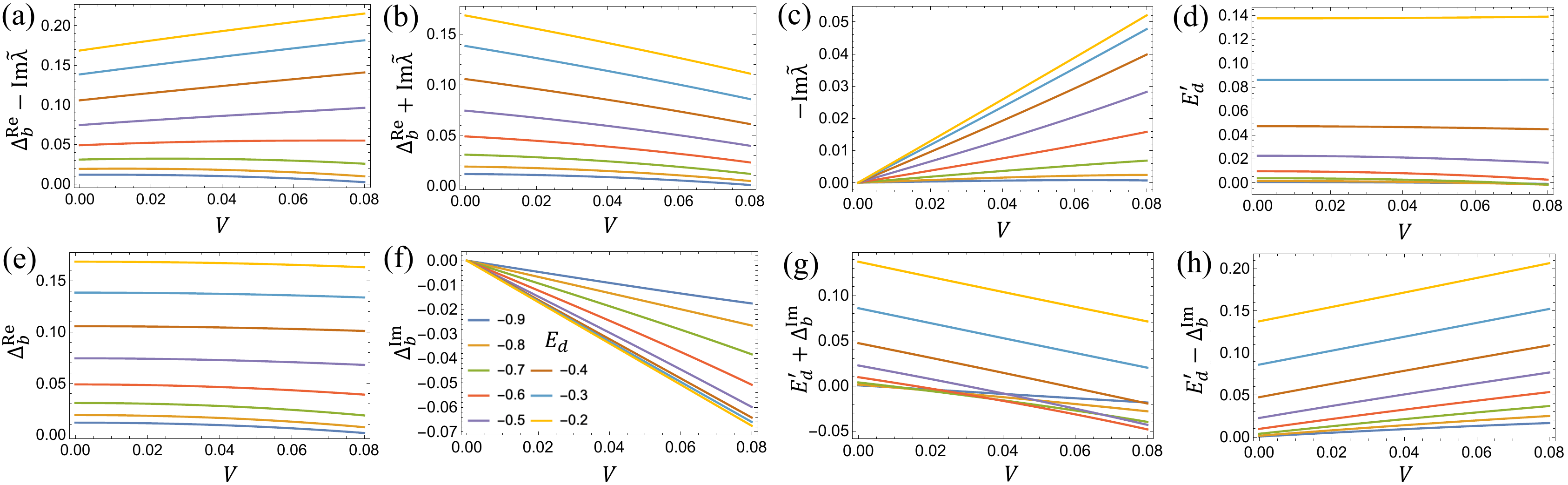}
\caption{Numerical solutions of the SCEs \eqref{eq_self}. (a), (b) Resonance width for the retarded and the advanced Green functions. (c), (d) Imaginary and real parts of the renormalized complex impurity level (the imaginary part is reversed for convenience). (e), (f) Real and imaginary parts of the renormalized complex hybridization. (g), (h) Peak position for the retarded and the advanced Green functions. For the deep impurity level $E_d$, we see in (a) and (b) the suppression of the Kondo effect characterized by the decrease of the renormalized resonance width. In (c), we find that $-\mathrm{Im}\tilde\lambda$ is almost pinned to zero for the deep impurity level $E_d$, which highlights that the renormalized hybridization $\Delta_b$ is the only renormalized complex parameter that describes the Kondo breakdown. \cred{The values of $E_d$ are shown in (f).} The other parameters are set to the same values as those in Fig.~\ref{fig_DeltaVtilde}.}
\label{fig_KondoResonance}
\end{figure*}

In Fig.~\ref{fig_DeltaVtilde}, to highlight the renormalization effect in the NH-AIM, we show the dependence of the complex hybridization $\tilde{\Delta}$ on the imaginary coupling $V$. For the parameter region shown in Fig.~\ref{fig_DeltaVtilde}, the real part of $\tilde \Delta$ is suppressed as $V$ is increased, but it remains nonzero as $V<V_0$ (see Sec.~\ref{sec_model}). As shown below, the correlation effect renormalizes $\mathrm{Re}\tilde{\Delta}$ and $\mathrm{Im}\tilde{\Delta}$ into $\Delta_b^{\mathrm{Re}}$ and $-\Delta_b^{\mathrm{Im}}$, respectively, leading to the Kondo breakdown. 

The numerical solutions of the SCEs \eqref{eq_self} are shown in Fig.~\ref{fig_KondoResonance}. By comparing Fig.~\ref{fig_DeltaVtilde}(a) and \ref{fig_DeltaVtilde}(b) with Fig.~\ref{fig_KondoResonance}(e) and \ref{fig_KondoResonance}(f), we find that the complex hybridization is strongly renormalized due to the correlation effect and that the renormalization effect significantly depends on the value of $E_d$. We first investigate the Kondo resonance that emerges for deep impurity level $E_d$ (e.g., see the blue curve for $E_d=-0.9$). In Figs.~\ref{fig_KondoResonance}(a) and \ref{fig_KondoResonance}(b), we find that the resonance widths $\Delta_b^\mathrm{Re}\mp\mathrm{Im}\tilde\lambda$ are suppressed by the NH hybridization $V$. This decrease of the resonance width demonstrates that the Kondo effect is suppressed by non-Hermiticity. If $V$ is further increased, the system exhibits a quantum phase transition from the Kondo phase to the unscreened phase near $V\simeq 0.08$, which is signaled by the vanishing resonance width. \cred{We remark that the critical value of $V$ is quite different from that in the noninteracting case due to the strong correlation effect (see Sec.~\ref{sec_model}). We also note that the critical value of $V$ depends on $V_0$.} Similar breakdown phenomena have been intensively studied in NH quantum impurity systems \cite{Lourenco18, Hasegawa21, Kulkarni22, Han23, Kattel25, Kulkarni25, Kulkarni25light, Nakagawa18, Kattel25L, Chen25, Burke25, Yamamoto24}.

Remarkably, as shown in Fig.~\ref{fig_KondoResonance}(c), the imaginary part of the renormalized complex impurity level almost stays zero for the deep impurity level, which reflects the Kondo effect. As such an imaginary impurity level can be regarded as a single-body dissipation, this fact demonstrates that the complex hybridization does not evoke emergent single-body dissipation. Accordingly, in Fig.~\ref{fig_KondoResonance}(d), the real part of the renormalized impurity level $E_d^\prime$ is pushed up to be just around the Fermi level \cred{$E_d^\prime=0$}, for the deep impurity level. Importantly, as shown in Figs.~\ref{fig_KondoResonance}(e) and \ref{fig_KondoResonance}(f), we find that the complex hybridization $\Delta_b$ is renormalized to a finite value satisfying 
\begin{align}
|\mathrm{Im}\tilde\lambda|, |E_d^\prime|\ll|\Delta_b|.
\end{align}
This demonstrates that, though we have originally introduced two complex parameters $\Delta_b$ and $\tilde \lambda$ in the SB mean-field theory, the hybridization $\Delta_b$ is the only complex parameter for describing the NH Kondo effect.

We emphasize that, in Ref.~\cite{Yamamoto24}, the resonance width is suppressed by one-body loss which induces three nonzero complex quantities: the Lagrange multiplier $\tilde\lambda$, the renormalized complex hybridization $\Delta_b$, and the one-body loss rate of an impurity fermion. However, in the current complex hybridization model, $-\mathrm{Im}\tilde\lambda$ is pinned to zero, and the one-body loss term is absent by definition. This indicates that the suppression of the Kondo effect is characterized solely by the renormalized complex hybridization $\Delta_b$. Thus, the current complex-hybridization model simplifies the mechanism of the Kondo breakdown and serves as a prototypical model for describing the NH Kondo effect.

In addition, for the deep impurity level $E_d$, we find in Fig.~\ref{fig_KondoResonance}(d) that $E_d^\prime$ is decreased by the imaginary hybridization $V$ and crosses the Fermi level to become a negative value for large non-Hermiticity, while it is always located just above the Fermi level in the Hermitian limit. Also, we see in Figs.~\ref{fig_KondoResonance}(g) and \ref{fig_KondoResonance}(h) that the behavior of the renormalized peak positions for the retarded and the advanced Green functions are roughly reversed with respect to the horizontal axis. Reflecting that $-\mathrm{Im}\tilde \lambda$ and $E_d^\prime$ are almost pinned to zero in the Kondo regime as shown in Figs.\ref{fig_KondoResonance}(c) and \ref{fig_KondoResonance}(d), $\Delta_b^\mathrm{Re}$ in Fig.~\ref{fig_KondoResonance}(e) and $\Delta_b^\mathrm{Im}$ in Fig.~\ref{fig_KondoResonance}(f) exhibit the behavior qualitatively similar to $\Delta_b^\mathrm{Re}-\mathrm{Im}\tilde\lambda$ in Fig.~\ref{fig_KondoResonance}(a) and $E_d^\prime+\Delta_b^\mathrm{Im}$ in Fig.~\ref{fig_KondoResonance}(g), respectively. We note that $\Delta_b^\mathrm{Im}$ does not vanish at the phase transition point in the Kondo regime though $\Delta_b^\mathrm{Re}$ vanishes \cite{BCS}.

When the impurity level $E_d$ is raised, the valence fluctuation gradually governs the dynamics, and the Kondo effect is smeared. As shown in Fig.~\ref{fig_KondoResonance}(c), the renormalized single-body dissipation is enhanced as we increase the non-Hermiticity $V$ for shallow impurity level $E_d$ (e.g., see the yellow curve for $E_d=-0.2$). Then, we see in Fig.~\ref{fig_KondoResonance}(d) that the real part of the renormalized impurity level $E_d^\prime$ is ramped up as $E_d$ is raised, and the value seems not much sensitive to $V$. Also, in the valence fluctuation regime in Figs.~\ref{fig_KondoResonance}(a) and \ref{fig_KondoResonance}(g), the resonance width $\Delta_b^\mathrm{Re}-\mathrm{Im}\tilde\lambda$ is enhanced as $V$ is increased, and the peak position $E_d^\prime+\Delta_b^\mathrm{Im}$ is far above the Fermi level. These results demonstrate that the emergent single-body dissipation, which is generated by the renormalization effect of the complex hybridization, gradually dominates the physics as the impurity level $E_d$ is raised. Accordingly, the resonance width $\Delta_b^\mathrm{Re}+\mathrm{Im}\tilde\lambda$ for the advanced Green function shown in Fig.~\ref{fig_KondoResonance}(b) gradually decreases with increasing $V$ in the valence fluctuation regime, and the peak position $E_d^\prime-\Delta_b^\mathrm{Im}$ in Fig.~\ref{fig_KondoResonance}(h) is increased far above the Fermi level with increasing $V$.


\subsection{Analytical results}
\label{sec_analytics}
Next, we analytically obtain the complex energy scale in the Kondo regime by evaluating the resonance width for the deep impurity level $E_d$. We call this energy scale the NH Kondo scale, which can be interpreted as a generalization of the Kondo temperature to the NH realm. The calculation proceeds with evaluating the SCEs \eqref{eq_self} under the assumption that the cutoff $D$ is much larger than the energy scale of the other parameters as
\begin{align}
\mathrm{Im}\tilde \lambda &+\frac{\mathrm{Im}\tilde\Delta}{\pi}\log\left(\frac{(E_d^\prime\pm\Delta_b^\mathrm{Im})^2+(\Delta_b^\mathrm{Re}\mp \mathrm{Im}\tilde \lambda)^2}{D^2}\right)\notag\\
&\pm\frac{2\mathrm{Re}\tilde\Delta}{\pi}\tan^{-1}\left(\frac{E_d^\prime\pm\Delta_b^\mathrm{Im}}{\Delta_b^\mathrm{Re}\mp \mathrm{Im}\tilde \lambda}\right)
\mp\Delta_b^\mathrm{Re}=0,
\label{eq_self_imag}\displaybreak[2]\\
\mathrm{Re}\tilde \lambda &+\frac{\mathrm{Re}\tilde\Delta}{\pi}\log\left(\frac{(E_d^\prime\pm\Delta_b^\mathrm{Im})^2+(\Delta_b^\mathrm{Re}\mp \mathrm{Im}\tilde \lambda)^2}{D^2}\right)\notag\\
&\mp\frac{2\mathrm{Im}\tilde\Delta}{\pi}\tan^{-1}\left(\frac{E_d^\prime\pm\Delta_b^\mathrm{Im}}{\Delta_b^\mathrm{Re}\mp \mathrm{Im}\tilde \lambda}\right)\pm\Delta_b^\mathrm{Im}=0.
\label{eq_self_real}\displaybreak[2]
\end{align}
\cred{Then, assuming that the resonance widths $\Delta_b^{\mathrm{Re}}\mp\mathrm{Im}\tilde \lambda$ are much smaller than $\Delta_0=\pi\rho V_0^2$, which is the width of the impurity level in the limit of $U\to0$ and $V\to0$ in Eq.~\eqref{eq_NHAnderson}, we arrive at the following equations from $\mathrm{Eq.~\eqref{eq_self_imag}}+\mathrm{Eq.~\eqref{eq_self_real}}\times\mathrm{Re}\tilde\Delta/\mathrm{Im}\tilde\Delta$ and $\mathrm{Eq.~\eqref{eq_self_imag}}-\mathrm{Eq.~\eqref{eq_self_real}}\times\mathrm{Im}\tilde\Delta/\mathrm{Re}\tilde\Delta$ as}
\begin{align}
&\Delta_b^\mathrm{Re}\mp\mathrm{Im}\tilde\lambda\sim D\cos\left(\frac{\pi\mathrm{Im}\tilde\Delta E_d}{2|\tilde\Delta|^2}\right)\exp\left(\frac{\pi\mathrm{Re}\tilde\Delta E_d}{2|\tilde\Delta|^2}\right),
\label{eq_self_an1}\\
&E_d^\prime\pm\Delta_b^\mathrm{Im}\sim\mp D\sin\left(\frac{\pi\mathrm{Im}\tilde\Delta E_d}{2|\tilde\Delta|^2}\right)\exp\left(\frac{\pi\mathrm{Re}\tilde\Delta E_d}{2|\tilde\Delta|^2}\right),
\label{eq_self_an2}
\end{align}
where
we have used the fact that the impurity level $E_d$ should be deep enough in the Kondo regime.
Equations \eqref{eq_self_an1} and \eqref{eq_self_an2} suggest that $|\mathrm{Im}\tilde\lambda|, |E_d^\prime|\ll |\Delta_b|,$
which supports the numerical results in the Kondo regime obtained in Sec.~\ref{sec_numerics}. Finally, we obtain the NH Kondo scale as
\begin{align}
\tilde T_K^\mathrm{NH}\sim\Delta_b^\mathrm{Re}+i\Delta_b^\mathrm{Im}=D\exp\left(\frac{\pi E_d}{2\tilde \Delta}\right),
\label{eq_NHTK}
\end{align}
which is the form of the analytic continuation of the Kondo temperature \cite{Coleman15} to complex parameters, and therefore Eq.~\eqref{eq_NHTK} can be regarded as a generalized complex energy scale that characterizes the NH Kondo effect. 

\begin{figure}[t]
\includegraphics[width=7cm]{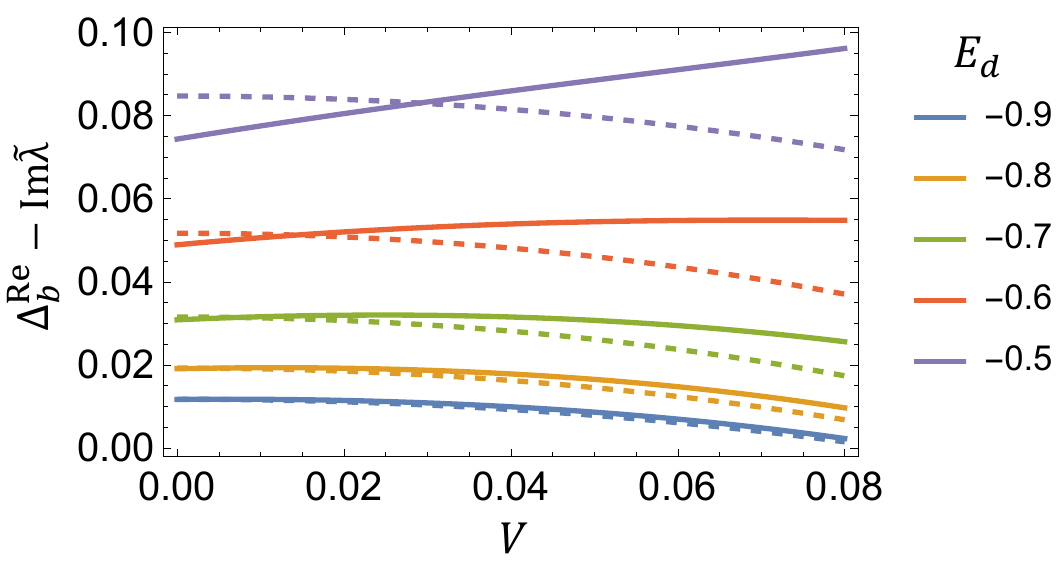}
\caption{Comparison between $T_K^\mathrm{NH}$ analytically obtained from Eq.~\eqref{eq_NHTK} (dashed curves) and the resonance width $\Delta_b^\mathrm{Re}-\mathrm{Im}\tilde\lambda$ (solid curves) obtained from the numerical calculation of the SCEs \eqref{eq_self}. For the deep impurity level $E_d$, both results agree well. The parameters are set to the same values as those in Fig.~\ref{fig_DeltaVtilde}.}
\label{fig_KondoTemperatureVComplex}
\end{figure}

In Fig.~\ref{fig_KondoTemperatureVComplex}, we compare the numerical results for the resonance width $\Delta_b^\mathrm{Re}-\mathrm{Im}\tilde\lambda$ and the analytical results for $T_K^\mathrm{NH}\equiv\mathrm{Re}\tilde T_K^\mathrm{NH}$. We find that the analytical results for $T_K^\mathrm{NH}$ agree with the numerical results for $\Delta_b^\mathrm{Re}-\mathrm{Im}\tilde\lambda$ quite well in the Kondo regime for the deep impurity level $E_d$. \cred{We also find that a phase transition occurs when the renormalized resonance width $\Delta_b^\mathrm{Re}-\mathrm{Im}\tilde\lambda$ vanishes at}
\begin{align}
T_K^\mathrm{NH}=0,
\label{eq_kondotemp}
\end{align}
or 
\begin{align}
\mathrm{Im}\left(\frac{1}{\tilde\Delta}\right)=-\frac{1}{E_d},
\label{eq_breakdown}
\end{align}
which is consistent with the numerical results shown in Fig.~\ref{fig_KondoResonance}. \cred{Though Eq.~\eqref{eq_kondotemp} indicates an infinite set of solutions as $\mathrm{Im}(1/\tilde\Delta)=(1+2m)/E_d\:(m\in\mathbb Z)$, physical solutions correspond to negative $m$ [see Fig.~\ref{fig_DeltaVtilde}(b)], and $\mathrm{Im}(1/\tilde\Delta)>-1/E_d$ does not induce the stable Kondo states \cite{Yamamoto24}. Thus, the Kondo breakdown only occurs when Eq.~\eqref{eq_breakdown} is satisfied.} We note that the resonance width becomes negative in the strong dissipation regime beyond the phase transition, and this raises an issue concerning the analyticity of NH Green functions as shown in Appendix~\ref{sec_Lehmann}.

\section{Relation to the non-Hermitian Kondo model with two-body loss}
\label{sec_nhkondo}
\begin{figure}[t]
\includegraphics[width=8.5cm]{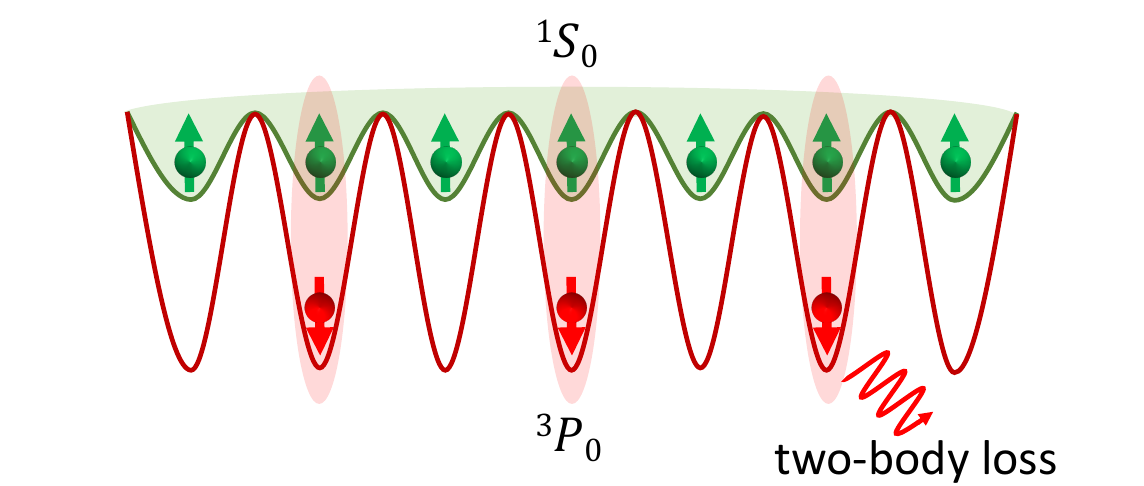}
\caption{Schematic illustration of the Kondo model implemented with ultracold atoms. In the case of alkaline-earth atoms, the ground state ${}^1S_0$ and the metastable state ${}^3P_0$ are trapped in a state-dependent optical lattice and serve as itinerant fermions and localized impurities, respectively. Interorbital spin-exchange interactions have been experimentally observed, where two-body loss is caused by inelastic collisions between the stable ${}^1S_0$ state and the metastable ${}^3P_0$ state \cite{Riegger18, Ono21}.}
\label{fig_Experiment}
\end{figure}

The NH-AIM with the complex hybridization reduces to the NH Kondo model with two-body loss proposed in Ref.~\cite{Nakagawa18}. \cred{The Kondo model with two-body loss due to inelastic scattering is schematically shown in Fig.~\ref{fig_Experiment}. In ultracold alkaline-earth atoms, itinerant fermions in the ground state and localized fermions in the excited state undergo inelastic scattering and are lost into the surrounding environment (vacuum), and such two-body loss has been experimentally observed \cite{Riegger18, Ono21}.} The NH Hamiltonian is obtained when we focus on a surviving impurity atom under the assumption that impurity atoms are independent with each other \cite{Nakagawa18}. In the following, we show that the NH Kondo model with the complex Kondo coupling is obtained by applying a second-order perturbation theory with respect to $\tilde V_{{\bm k}d}$ and $\tilde V_{d \bm k}$ in the NH-AIM \eqref{eq_NHAnderson} \cite{SW}.

Because the doubly-degenerate ground-state wave function at $\tilde V_{\bm k d}=\tilde V_{d \bm k}=0$ is given by $\psi_\sigma=c_{d \sigma}^\dag \psi_V$, where $\psi_V$ represents the Fermi sea of reservoir fermions, we start from $\psi_\sigma$ and consider the second-order processes with respect to $\tilde V_{{\bm k}d}$ and $\tilde V_{d \bm k}$, where a state with a single impurity fermion is retrieved. We consider the $U\to\infty$ limit and omit the doubly occupied impurity site in the virtual processes. By assuming that the impurity level $E_d$ is sufficiently deep $|E_d|\gg\Delta_0$, the effective Hamiltonian is written in the $n_d=\sum_\sigma n_{d\sigma}=1$ sector as 
\begin{align}
H_\mathrm{eff}^\prime=H_\mathrm{imp}+H_\mathrm{ex},
\label{eq_KondoHamiltonian}
\end{align}
where $H_\mathrm{imp}$ denotes the potential scattering term and $H_\mathrm{ex}$ stands for the spin exchange interaction, given by
\begin{align}
H_\mathrm{imp}=&-\sum_{\bm k \bm k^\prime}\tilde V_{\bm k^\prime d}\tilde V_{d\bm k}\frac{n_d}{2(E_d-\epsilon_{\bm k^\prime})}(c_{\bm k^\prime\uparrow}^\dag c_{\bm k \uparrow}+c_{\bm k^\prime\downarrow}^\dag c_{\bm k \downarrow}),\displaybreak[2]\label{eq_pot}\\
H_\mathrm{ex}=&-\sum_{\bm k \bm k^\prime}\tilde V_{\bm k^\prime d}\tilde V_{d\bm k}\frac{1}{E_d-\epsilon_{\bm k^\prime}}[(c_{\bm k^\prime\uparrow}^\dag c_{\bm k \uparrow}-c_{\bm k^\prime\downarrow}^\dag c_{\bm k \downarrow})S_z\notag\displaybreak[2]\\
&+c_{\bm k^\prime\uparrow}^\dag c_{\bm k \downarrow}S_- + c_{\bm k^\prime\downarrow}^\dag c_{\bm k \uparrow}S_+],
\label{eq_exchange}
\end{align}
respectively. Here, we have introduced the impurity spin operators as
\begin{align}
&S_z=\frac{1}{2}(c_{d\uparrow}^\dag c_{d\uparrow}-c_{d\downarrow}^\dag c_{d\downarrow}),\displaybreak[2]\\
&S_+=c_{d\uparrow}^\dag c_{d\downarrow},\displaybreak[2]\\
&S_-=c_{d\downarrow}^\dag c_{d\uparrow}.
\end{align}
Equation \eqref{eq_KondoHamiltonian} is rewritten as
\begin{align}
H_\mathrm{eff}^\prime=\frac{\tilde v}{2N_s}\sum_{\bm k \bm k^\prime \sigma}c_{\bm k^\prime\sigma}^\dag c_{\bm k\sigma}-\frac{\tilde J}{2N_s}\sum_{\bm k \bm k^\prime \sigma \sigma^\prime}c_{\bm k^\prime \sigma^\prime}^\dag \bm \sigma_{\sigma^\prime\sigma}c_{\bm k \sigma}\cdot \bm S
\end{align}
which is nothing but the spin-exchange interaction term with the potential scattering term in the NH Kondo Hamiltonian with two-body loss \cite{Nakagawa18}. Here, $N_s$ is the number of sites, $\bm \sigma$ is the three-component Pauli vector, and $\bm S$ is the impurity spin vector. From Eqs.~\eqref{eq_pot} and \eqref{eq_exchange}, we find that the spin-independent complex scattering rate $\tilde v$ and the complex Kondo coupling $\tilde J$ are given by 
\begin{gather}
\frac{\tilde v}{2N_s}\simeq - \frac{\tilde V^2}{2E_d},\displaybreak[2]\\
\frac{\tilde J}{2N_s}\simeq\frac{\tilde V^2}{E_d},
\label{eq_KondoAnderson}
\end{gather}
where we have assumed that $\tilde V^2\simeq\tilde V_{\bm k^\prime d}\tilde V_{d\bm k}$, and $\epsilon_{\bm k(\bm k^\prime)}$ is ignored by assuming that it is close to the Fermi energy. We find that the range of real and imaginary parts of $\tilde V$ used in Fig.~\ref{fig_DeltaVtilde} gives $\mathrm{Re}\tilde J<0$ and $\mathrm{Im}\tilde J>0$, which actually corresponds to the parameter region for the Kondo breakdown with two-body loss studied in Ref.~\cite{Nakagawa18}.

Both the NH-AIM with complex $E_d$ induced by one-body loss \cite{Yamamoto24} and that with the complex hybridization obtained in the current study reduce to the same NH Kondo Hamiltonian with the complex Kondo coupling. This means that different effective Anderson models can capture the same Kondo breakdown. Therefore, though both NH-AIMs with the complex $E_d$ and the complex hybridization can describe the Kondo breakdown, the latter is much simpler to explain the renormalization mechanism as detailed in Sec.~\ref{sec_numerics}. Finally, recent numerical renormalization group analysis has shown that the phase diagrams of the NH-AIM with the complex hybridization and the NH Kondo model with two-body loss have the same structure, which supports the robustness of our study \cite{Burke25}.


\section{Non-Hermitian Kondo Breakdown: Exact Bethe ansatz results}
\label{sec_bethe}
In this section, we give exact Bethe ansatz results for the Kondo breakdown that support the NH-SB mean-field results obtained in Sec.~\ref{sec_NHSB}. The AIM was shown to be exactly solvable both for symmetric and asymmetric cases \cite{Tsvelick83, Okiji84, Wiegmann80, Kawakami81, Kawakami82, Kawakami82lett, Okiji82, Okiji82solid, Wiegmann83C, Tsvelick83C}. Here, we generalize the exact solution of the infinite-$U$ AIM for the ground state \cite{Schlottmann83, Kawakami86, Schlottmann89} to NH systems, where several studies have shown that the Bethe ansatz method is still applicable to NH quantum many-body problems \cite{Fukui98, Nakagawa18, Shibata19, Nakagawa21, Buca20, Yamamoto22, Yamamoto23, Mao23, Marche24}.

Let us assume that the scattering by the impurity is of $s$-wave type, which allows us to map the three-dimensional problem in Eq.~\eqref{eq_NHAnderson} to a chiral one-dimensional problem. Then, by linearizing the energy dispersion of itinerant fermions around the Fermi energy as $\epsilon_k=k$ \cite{Schlottmann83}, \cred{we obtain the $S$-matrices of the NH-AIM in the infinite-$U$ limit, from the analytic continuation of those in the Hermitian limit \cite{Schlottmann83}, as}
\begin{gather}
S_{ij}=\frac{k_i-k_j+2i\tilde \Delta P_{ij}}{k_i-k_j-2i\tilde \Delta},\label{eq_Sij}
\end{gather}
where $P_{ij}$ is an operator that exchanges $i$th and $j$th particles. \cred{When linearizing the energy dispersion, we take the energy unit to be $v_F=1$, where $v_F$ is the Fermi velocity. Then, the complex hybridization is given by $\tilde \Delta=\tilde V^2/2$, and the parameters in the Hamiltonian should be translated to correctly reflect their dimensions.} In the Bethe ansatz formalism, all the interaction effects are incorporated via the quasimomentum $k_j$, which takes a complex value due to the complex hybridization $\tilde \Delta$ in general. \cred{Here, the energy of the system is given by $E=\sum_{j=1}^N k_j$, where $N$ is the total particle number of the system.} The $S$-matrices in Eq.~\eqref{eq_Sij} take the same form as the NH generalization of those for a one-dimensional Fermi gas with an attractive $\delta$-function interaction \cite{Yang67, Schlottmann83} and satisfy the Yang-Baxter equation as $S_{jk}S_{ik}S_{ij}=S_{ij}S_{ik}S_{jk}$. Hence, we can construct the nested Bethe equations for the NH-AIM as
\begin{gather}
e^{ik_j L}\:t(k_j-E_d)=\prod_{\alpha=1}^{M} t(k_j-\lambda_\alpha),\quad(1\le j \le N),\label{eq_bethe1}\displaybreak[2]\\
\prod_{\beta(\neq \alpha)}^{M}t\left(\frac{\lambda_\alpha-\lambda_\beta}{2}\right)=\prod_{j=1}^{N}t(\lambda_\alpha-k_j),\quad(1\le \alpha \le M).\label{eq_bethe2}
\end{gather}
Here, we have introduced $t(x)\equiv(x-i\tilde \Delta)/(x+i\tilde \Delta)$, $L$ is the length of the system, and $M$ is the number of the charge degrees of freedom corresponding to the bound states with the quasimomentum $k_\alpha^\pm=\lambda_\alpha \pm i\tilde \Delta$, where we redefine $k_j$ as $k_\alpha^\pm$ for charge excitations. \cred{We note that $k_\alpha^\pm$ becomes complex even in the Hermitian limit.} Similarly, $N-2M$ stands for the number of unpaired spins with the quasimomentum $k_j$. Here, $\lambda_\alpha$ and $k_j$ take complex values due to non-Hermiticity. We remark that the non-Hermiticity should be sufficiently weak so that it does not change the sign of $\mathrm{Im}k_\alpha^\pm$. A similar condition has been employed in the previous studies on NH many-body systems \cite{Ashida16, Yamamoto22}, and physically, this corresponds to the assumption that the bound states are not broken due to dissipation.

To proceed, we first rewrite Eqs.~\eqref{eq_bethe1} and \eqref{eq_bethe2} to obtain Bethe equations for $N-2M$ unpaired spin states and $M$ bound states. For $N-2M$ unpaired spins, we can employ Eq.~\eqref{eq_bethe1} directly. For $M$ bound states, we first multiply Eq.~\eqref{eq_bethe1} for $k_\alpha^\pm$ with each other, obtaining
\begin{align}
e^{2i\lambda_\alpha L}t\left(\frac{\lambda_\alpha -E_d}{2}\right)=&\prod_{\beta(\neq \alpha)}^{M}t\left(\frac{\lambda_\alpha-\lambda_\beta}{2}\right)\notag\\
&\times t(k_\alpha^+-\lambda_\alpha) t(k_\alpha^- -\lambda_\alpha).
\label{eq_bethe1prime}
\end{align}
To calculate the second line in the right-hand side of Eq.~\eqref{eq_bethe1prime}, we use Eq.~\eqref{eq_bethe2}, which is rewritten as
\begin{align}
1=\prod_{j=2M+1}^{N}t(\lambda_\alpha-k_j)\:t(\lambda_\alpha-k_\alpha^+) t(\lambda_\alpha-k_\alpha^-).\label{eq_bethe2prime}
\end{align}
Finally, we arrive at the nested Bethe equations respectively for $N-2M$ spin modes and $M$ charge modes as
\begin{gather}
e^{ik_j L}\:t(k_j-E_d)=\prod_{\alpha=1}^{M} t(k_j-\lambda_\alpha),\label{eq_bethespin}\displaybreak[2]\\
e^{2i\lambda_\alpha L}\:t\left(\frac{\lambda_\alpha -E_d}{2}\right)=\prod_{\beta(\neq \alpha)}^{M}t\left(\frac{\lambda_\alpha-\lambda_\beta}{2}\right)\notag\\
\hspace{4.5cm}\times \prod_{j=2M+1}^{N}t(\lambda_\alpha-k_j),\label{eq_bethecharge}
\end{gather}
where $2M+1\leq j \leq N$ in Eq.~\eqref{eq_bethespin} and $1\leq \alpha \leq M$ in Eq.~\eqref{eq_bethecharge}. Equations \eqref{eq_bethespin} and \eqref{eq_bethecharge} are rewritten by taking the logarithm as
\begin{align}
k_j L + \theta (k_j-E_d)=&2\pi I_j + \sum_{\alpha=1}^M \theta(k_j-\lambda_\alpha),\label{eq_bethespinlog}\\
2\lambda_\alpha L + \theta \left(\frac{\lambda_\alpha-E_d}{2}\right)=&2\pi J_\alpha + \sum_{\beta(\neq\alpha)}^M \theta\left(\frac{\lambda_\alpha-\lambda_\beta}{2}\right)\notag\\
&+\sum_{j=2M+1}^{N}t(\lambda_\alpha-k_j),\label{eq_bethechargelog}
\end{align}
where we have introduced the phase shift as $\theta(x)=2\tan^{-1}(x/\tilde \Delta)$. Here, the quantum number $I_j$ that characterizes $N-2M$ spin excitations takes half-odd integers (integers) for even (odd) $M$, and $J_\alpha$ for $M$ charge excitations becomes integers (half-odd integers) for even (odd) $N-M$.

Next, we introduce distribution functions of charge and spin excitations to characterize physical quantities. We first consider the case where quasimomenta $k$ for spin modes are distributed over the region $(-\infty-iQ_-, \: Q)_c$ and rapidities $\lambda$ for charge modes span the range $(-\infty-iB_-, B)_c$. Here, $(X, Y)_c$ denotes a complex path that connects $X$ and $Y$ in the complex plane and is determined so that the real part of the energy $E$ is minimized. In the Hermitian limit, the zero magnetization is realized in the $Q\to-\infty$ limit with $Q_-\to0$, and the Kondo limit is achieved for $B\to\infty$ with $B_-\to0$; however, introduction of non-Hermiticity can make the complex paths $(-\infty-iQ_-, \: Q)_c$ and $(-\infty-iB_-, B)_c$ deviate from the real axis. Accordingly, we can rewrite Eqs.~\eqref{eq_bethespinlog} and \eqref{eq_bethechargelog} in the thermodynamic limit as
\begin{widetext}
\begin{gather}
\frac{1}{2\pi} + \frac{\tilde \Delta}{\pi L[(k-E_d)^2+\tilde \Delta^2]}=\rho(k) + \int_{-\infty-iB_-}^B \frac{\tilde \Delta}{\pi[(k-\lambda)^2+\tilde \Delta^2]}\sigma(\lambda)d\lambda,\label{eq_bethespinthermo}\displaybreak[2]\\
\frac{1}{\pi} + \frac{2\tilde \Delta}{\pi L[(\lambda-E_d)^2+(2\tilde \Delta)^2]}=\sigma(\lambda) + \int_{-\infty-iB_-}^B \frac{2\tilde \Delta}{\pi[(\lambda-\lambda^\prime)^2+(2\tilde \Delta)^2]}\sigma(\lambda^\prime)d\lambda^\prime+\int_{-\infty-iQ_-}^Q \frac{\tilde \Delta}{\pi[(\lambda-k)^2+\tilde \Delta^2]}\rho(k)dk,\label{eq_bethechargethermo}
\end{gather}
\end{widetext}
where $\rho(k)$ [$\sigma(\lambda)$] represents the distribution function for spin (charge) modes, and the contour integration along the complex path $\int_{(X,Y)_c}$ is denoted as $\int_X^Y$ for readability. We find that the distribution functions are separated into the host part and the impurity part that involve O($1$) and O($1/L$) contributions, respectively. In the following, we use superscripts $(\mathrm i)$ and $(\mathrm c)$ to represent the impurity and host parts. Then, the energy density of the system is expressed as
\begin{align}
\frac{E}{L}= \int_{-\infty-iQ_-}^Q k\rho(k) dk + 2\int_{-\infty-iB_-}^B \lambda\sigma(\lambda) d\lambda, 
\label{eq_energy}
\end{align}
and we obtain the particle number density as
\begin{align}
\frac{N}{L} = \int_{-\infty-iQ_-}^Q \rho(k) dk + 2\int_{-\infty-iB_-}^B\sigma(\lambda) d\lambda. \label{eq_ndensity}
\end{align}
Finally, the magnetization density of the system is represented as
\begin{align}
\frac{M_z}{L} = \frac{1}{2}\int_{-\infty-iQ_-}^Q \rho(k) dk.
\label{eq_magnetizationdensity}
\end{align}
By using Eqs.~\eqref{eq_bethespinthermo}-\eqref{eq_magnetizationdensity}, we can in principle calculate physical quantities such as the energy and the spectrum of elementary excitations.

Though it is difficult to solve the Bethe equations \eqref{eq_bethespinthermo} and \eqref{eq_bethechargethermo} for general cases, we can evaluate several important quantities for the Kondo breakdown with the help of them. We focus on the excitation energy $\Delta E$ with respect to a small change $\Delta S_z$ of magnetization, where $\Delta E \propto D_s[\epsilon(Q)]^{-1}(\Delta S_z)^2$ and $D_s[\epsilon(Q)]$ is the density of states at the quasi-Fermi point. By decomposing the density of states in $\Delta E$ into the host and impurity parts, we determine the Kondo breakdown from the inverse of the impurity part of the density of states as $\mathrm{Re}[D_s^{(\mathrm i)}[\epsilon(Q)]^{-1}]=0$, where $D_s^{(\mathrm i)}[\epsilon(Q)]^{-1}$ is proportional to the NH Kondo scale in the Kondo limit. We find that
\begin{align}
D_s[\epsilon(Q)]=\left.\frac{\partial z}{\partial \epsilon}\right|_{\epsilon=\epsilon(Q)}=\left.\frac{\partial z}{\partial k}\frac{\partial k}{\partial \epsilon}\right|_{k=Q}=\frac{\rho(Q)}{\epsilon^\prime(Q)},
\label{eq_DOS}
\end{align}
where $\epsilon(k)$ is the dressed energy that incorporates the interaction effect in the spin excitation energy, and $z(k_j)=I_j/L$. Since $\epsilon^\prime(k)=2\pi\rho^{(\mathrm{c})}(k)$ is derived from the Bethe equations in the current case (see Appendix \ref{App_bethe}), we obtain
\begin{align}
D_s^{(\mathrm i)}[\epsilon(Q)]=\lim_{Q\to-\infty-iQ_-}\frac{\rho^{(\mathrm i)}(Q)}{2\pi \rho^{(\mathrm c)}(Q)},
\label{eq_susceptibility}
\end{align}
where we have focused on the zero magnetization sector. Then, Eq.~\eqref{eq_susceptibility} is calculated in the Kondo limit as
\begin{align}
D_s^{(\mathrm i)}[\epsilon(Q)]\propto e^{-\frac{\pi \tilde E_d}{2\tilde\Delta}},
\label{eq_KondoScalE}
\end{align}
whose reciprocal is the NH Kondo scale given in Eq.~\eqref{eq_NHTK} (see Appendix \ref{App_bethe} for the detailed calculation).
Here, $\tilde E_d\equiv \tilde E_d^{\mathrm R}+i\tilde E_d^{\mathrm I}$ is the renormalized complex impurity level. As $|\tilde E_d^{\mathrm I}|$ is small compared to $|\tilde E_d^{\mathrm R}|$ in the Kondo limit, the condition for the Kondo breakdown that emerges at $\mathrm{Re}[D_s^{(\mathrm i)}[\epsilon(Q)]^{-1}]=0\Leftrightarrow\mathrm{Im}(\pi \tilde E_d/2\tilde\Delta)=-\pi/2$ is estimated as
\begin{gather}
\mathrm{Im}\left(\frac{1}{\tilde \Delta}\right)=-\frac{1}{\tilde E_d^{\mathrm R}},
\label{eq_breakdownexact}
\end{gather}
which reduces to the result in Eq.~\eqref{eq_breakdown} obtained from the SB mean-field theory, except for the renormalization of the impurity level. This supports the SB mean-field results of the Kondo breakdown induced by the complex hybridization.

\section{Conclusions}
\label{sec_conclusion}
\cred{In this paper, we have studied the NH-AIM with a complex hybridization as a unified framework for understanding the Kondo breakdown in NH systems. Based on the SB mean-field theory, we have shown that the simplified mechanism underlying the Kondo breakdown is captured by a single complex hybridization parameter, which governs the suppression and eventual vanishing of the Kondo resonance. We have further demonstrated that the transition point obtained from the mean-field analysis is supported by the exact Bethe ansatz solution, confirming the robustness of the Kondo breakdown. In addition, we have clarified that the present model reduces to the NH Kondo Hamiltonian with two-body loss by applying the second-order perturbation theory, which means that the present model is useful to obtain the simple understanding of the previously studied Kondo breakdown.} As the Kondo effect in open quantum systems is one of the actively studied research topics in recent years \cite{Schiro24, Schiro24arXiv, Stefanini25, Vanh25, Werner25}, our result deepens the understanding of dissipative impurity phenomena.

\begin{acknowledgments}
K.Y. was supported by JSPS Program for Forming Japan's Peak Research Universities (J-PEAKS) Grant No.\ JPJS00420230008,
KAKENHI Grant No.\ JP25K17327,
Hirose Foundation,
Precise Measurement Technology Promotion Foundation,
Fujikura Foundation,
Toyota Riken Scholar Program,
and Support Center for Advanced Telecommunications Technology Research. M.N. was supported by KAKENHI Grant No.\ JP24K16989.
N.K. was supported by the RIKEN TRIP initiative.
\end{acknowledgments}

\section*{Data Availability}
The data that support the findings of this article are openly available \cite{Zenodo}.


\appendix

\section{Microscopic derivation of Eq.~\eqref{eq_NHAnderson} from the Lindblad equation}
\label{app_lindblad}
\cred{In this appendix, we derive the NH-AIM in Eq.~\eqref{eq_NHAnderson} from the Lindblad equation. Our system is directly relevant to ultracold atoms in open quantum systems, whose Markovian dynamics under dissipation is described by the Lindblad equation \cite{Daley14, Yamamoto23L, Yamamoto25}.}
Let us consider the time evolution of the density matrix $\rho$ given by
\begin{align}
\frac{d\rho}{dt}
&=-i[H,\rho]-\frac{\gamma}{2}\sum_\sigma (\{L_\sigma^\dag {L}_\sigma, \rho\} -2L_\sigma \rho L_\sigma^\dagger),\notag\displaybreak[2]\\
&=-i(H_\mathrm{eff}^\prime\rho-\rho{H_\mathrm{eff}^{\prime\dagger}})+\gamma\sum_\sigma{L}_\sigma\rho{L}_\sigma^\dagger,\displaybreak[2]
\label{eq_Lindblad}
\end{align}
where $H$ is the AIM, \cred{$c_{0\sigma}$ is the fermionic annihilation operator at the origin, and the Lindblad operator $L_\sigma = c_{0\sigma} +c_{d\sigma}$ describes the collective loss of an itinerant fermion and an impurity fermion,} and an effective NH Hamiltonian is given by $H_\mathrm{eff}^\prime=H-\frac{i}{2}\gamma\sum_\sigma L_\sigma^\dagger{L}_\sigma$ \cite{Yamamoto19, Yamamoto25}. \cred{Such an NH Hamiltonian is typically realized when we focus on the short-time dynamics, where no quantum jump is detected \cite{Yamamoto24}.} We note that an experimental setup for similar collective loss was proposed for ultracold atoms in the context of asymmetric hopping in NH systems by employing postselections \cite{Gong18}. The NH contribution to $H_\mathrm{eff}^\prime$ is calculated as
\begin{align}
-\frac{i}{2}\gamma\sum_\sigma L_\sigma^\dagger{L}_\sigma
=&-\frac{i\gamma}{2\sqrt {N_s}} \sum_{\bm k \sigma}(c_{\bm k \sigma}^\dag c_{d\sigma} + c_{d\sigma}^\dag c_{\bm k \sigma})\notag\\
&-\frac{i}{2}\gamma\sum_\sigma(c_{0\sigma}^\dag c_{0\sigma} + c_{d\sigma}^\dag c_{d\sigma}),
\label{eq_NHtermAIM}
\end{align}
where \cred{$N_s$ is the number of sites}, and the first term on the right-hand side is nothing but the NH contribution to the hybridization term. Then, $H_\mathrm{eff}^\prime$ reduces to the NH-AIM \eqref{eq_NHAnderson} ignoring the imaginary potential in the second term on the right-hand side of Eq.~\eqref{eq_NHtermAIM}, \cred{by assuming that it does not affect the results \cite{Gong18}}.
\cred{This means that we can tune the complex hybridization that causes the Kondo breakdown by introducing the collective one-body loss in ultracold atoms.}

\section{Formulation of the non-Hermitian slave-boson mean-field theory}
\label{App_SB}
In this appendix, we summarize the formulation of the NH SB mean-field theory and obtain the self-consistent equations for the renormalized physical quantities. \cred{Since the calculations closely follow those for the one-body loss case studied in \cite{Yamamoto24}, we refer the readers to that work for more detailed calculations.} We focus on the infinite-$U$ limit in the NH-AIM \eqref{eq_NHAnderson}, where the double occupancy at the impurity site is prohibited. Importantly, in the NH-SB theory for the NH-AIM, both the SB field and the Lagrange multiplier should be in general complex-valued due to non-Hermiticity.

We proceed with the calculation by applying the path-integral formalism to the NH-AIM with the Lagrange multiplier in Eq.~\eqref{eq_NHAnderson_vtilde}. To analyze the effective ground state of $H_\mathrm{eff}$ with the smallest real part of energy, let us define the partition function in NH systems as 
\begin{align}
Z=\sum_n e^{-\beta E_n}=\sum_n {}_L\langle E_n|e^{-\beta H_\mathrm{eff}}|E_n\rangle_R,
\end{align}
where $|E_n\rangle_R$ and $|E_n\rangle_L$ are the right and left eigenstates of $H_\mathrm{eff}$ with eigenenergy $E_n$ satisfying the biorthonormal relation ${}_L\langle E_m|E_n\rangle_R=\delta_{mn}$, and $\beta$ is the \cred{real} parameter that characterizes the statistical weight of eigenstates.
\cred{We take the $\beta\to\infty$ limit in the end of the calculation, to focus on the quantum phase transition and not to suffer from the problem that arises from finite $\beta$. The partition function considered in our study describes the energy eigenstate with the smallest real part of the energy in the $\beta\to\infty$ limit, which can experimentally be realized by adiabatically introducing dissipation to the ground state in the Hermitian limit.}
Then, we express the partition function with a constraint and the action as
\begin{align}
Z=&\int \mathcal D[\bar \psi, \psi, \bar b, b, \tilde \lambda]e^{-S},\\
S=&\int_0^\beta d\tau \Big\{\sum_{\bm k\sigma}\bar c_{\bm k \sigma}(\tau)(\partial_\tau+\epsilon_{\bm k})c_{\bm k \sigma}(\tau)\notag\\
&+\sum_{\sigma}\bar d_{\sigma}(\tau)(\partial_\tau+E_d+\tilde \lambda)d_{\sigma}(\tau)\notag\\
&+\sum_{\bm k \sigma}\big[\tilde V_{\bm k d} \bar c_{\bm k \sigma}(\tau) \bar b(\tau) d_\sigma(\tau) + \tilde V_{d\bm k} \bar d_\sigma(\tau) b(\tau) c_{\bm k \sigma}(\tau)\big]\notag\\
&+\bar b(\tau) (\partial_\tau + \tilde \lambda)b(\tau) - \tilde \lambda\Big\}.
\label{eq_pathintegral}
\end{align}
Here, $\bar \psi$ and $\psi$ are the Grassmann variables for the set of Fermi fields. We note that we should take the limit $\beta\to\infty$ in the end of the calculation because we analyze the effective ground state, which is defined by the smallest real part of the energy. To obtain the saddle-point solution with respect to the Bose fields $\bar b$ and $\tilde \lambda$, we first integrate out the Fermi fields and get
\begin{align}
Z=\int\mathcal D[\bar b, b, \tilde \lambda]e^{-S_\mathrm{eff}}.
\end{align}
By taking the derivative of the effective action $S_\mathrm{eff}$ with respect to $\bar b$ and $\tilde \lambda$, \cred{the saddle-point conditions are given by
\begin{align}
&\frac{\delta S_\mathrm{eff}}{\delta \bar b(\tau)}=\bigg._{L}{\left\langle \frac{\delta S}{\delta \bar b(\tau)}\right\rangle}\bigg._{R}\notag\\
&=(\partial_\tau+\tilde \lambda) b(\tau) +\sum_{\bm k \sigma}\tilde V_{\bm k d}{}_L\langle c_{\bm k \sigma}^\dag d_\sigma\rangle_R=0,\label{eq_saddle_b}\\
&\frac{\delta S_\mathrm{eff}}{\delta \tilde \lambda}=\bigg._{L}{\left\langle \frac{\delta S}{\delta \tilde \lambda}\right\rangle}\bigg._{R}=\sum_\sigma{}_L\langle d_\sigma^\dag d_\sigma\rangle_R+\bar {b}(\tau) b(\tau)-1=0.\label{eq_saddle_lambda}
\end{align}
We emphasize that, due to the complex parameters $b$, $\bar b$, $\tilde \lambda$, and $\tilde V_{\bm k d}$, the expectation value ${}_L\langle\cdots\rangle_R\equiv \mathrm{Tr}[\cdots e^{-\beta H_\mathrm{eff}}]/Z$ for the fixed $b$, $\bar b$, and $\tilde \lambda$ with the use of Eq.~\eqref{eq_pathintegral} takes a complex value in general.} Though the total particle number at the impurity site is conserved due to the constraint in Eq.~\eqref{eq_saddle_lambda}, a local operator such as $d_\sigma^\dag d_\sigma$ does not commute with $H_\mathrm{eff}$, and therefore its expectation value ${}_L\langle d_\sigma^\dag d_\sigma\rangle_R$ can mathematically become complex.

\cred{Hereafter, we focus on the effective ground state of the NH-AIM, where the mean field described by the Bose field is static. When the $\tau$-dependence of the Bose field is considered, the mean field shows nontrivial imaginary-time dynamics, but we do not consider such a situation.} We note that the expectation values ${}_L\langle c_{\bm k \sigma}^\dag d_\sigma\rangle_R$ and ${}_L\langle d_\sigma^\dag d_\sigma\rangle_R$ are independent of $\tau$ because they are described by the equal-time limit of the Green functions, which have the time translation symmetry even in NH systems as explained in Appendix~\ref{sec_Lehmann}. Then, the SCEs are given by
\begin{gather}
\tilde \lambda b +\sum_{\bm k \sigma} \tilde V_{\bm k d} {}_L\langle c_{\bm k \sigma}^\dag d_\sigma\rangle_R=0,\label{eq_self1_vtilde}\\
\sum_\sigma{}_L\langle d_\sigma^\dag d_\sigma\rangle_R+b_0^2=1.\label{eq_self2_vtilde}
\end{gather}
\cred{Here, we remark that the above mean-field treatment spontaneously breaks the U(1) symmetry of the Hamiltonian generated by the transformation $b\to b e^{i\theta}$ and $d_\sigma \to d_\sigma e^{i\theta}$, and accordingly, the Bose fields are written as
\begin{align}
b &=b_0 e^{i\theta}, \label{eq_b1}\\
\bar{b}&=b_0 e^{-i\theta}.\label{eq_b2}
\end{align}
Though we are not studying $b_0$ itself in the current study, this is introduced to emphasize that $b$ and $\bar b$ are not complex conjugate to each other for $b_0 \in\mathbb C$ in NH systems. We note that Eqs.~\eqref{eq_b1} and \eqref{eq_b2} are reminiscent of the superfluid order parameters in the NH BCS theory \cite{Yamamoto19, Takemori24, Takemori24B}, and we indeed see some similarities in the behavior between Eqs.~\eqref{eq_b1} and \eqref{eq_b2} and superfluid gaps in the phase transition as discussed in Sec.~\ref{sec_numerics}.
}

Next, we evaluate the expectation values given in Eqs.~\eqref{eq_self1_vtilde} and \eqref{eq_self2_vtilde}. We here explain the calculation of ${}_L\langle d_\sigma^\dag d_\sigma\rangle_R$, and that of ${}_L\langle c_{\bm k \sigma}^\dag d_\sigma\rangle_R$ will be performed in a similar manner. With the use of the path integrals, we find that ${}_L\langle d_\sigma^\dag d_\sigma\rangle_R$ is rewritten as
\begin{align}
{}_L\langle d_\sigma^\dag d_\sigma\rangle_R=\beta^{-1}\sum_{\omega_n} e^{i\omega_n\eta}G_d^\sigma(i\omega_n),\label{eq_Green_d}
\end{align}
where $\eta\to+0$ limit is implicitly indicated, $\omega_n=(2n+1)\pi/\beta$ with $n\in\mathbb Z$ is the Matsubara frequency for fermions, and the NH Matsubara Green function $G_d^\sigma(i\omega_n)$ reads
\begin{align}
G_d^\sigma(i\omega_n)=[i\omega_n-E_d-\tilde \lambda-\Sigma_d^\sigma(i\omega_n)]^{-1}.\label{eq_Green_d2}
\end{align}
Here, the self-energy of the impurity fermion is given by 
\begin{align}
\Sigma_d^\sigma(i\omega_n)=\tilde V^2 b_0^2\sum_{\bm k} [i\omega_n-\epsilon_{\bm k}]^{-1}.\label{eq_SelfEnergy_d}
\end{align}
To proceed, we have to take the sum over the Matsubara frequency in Eq.~\eqref{eq_Green_d} with the contour integrations, for which we have to analytically continue the impurity self-energy to the complex-$\omega$ plane as
\begin{align}
\Sigma_d^\sigma(\omega)=-i\Delta_b \mathrm{sgn}(\mathrm{Im}\omega),
\label{eq_SelfEnergy2}
\end{align}
where $\Delta_b\equiv b_0^2 \tilde\Delta$, and we have performed the summation over $\bm k$ by assuming a constant density of states of reservoir fermions. Importantly, in Eq.~\eqref{eq_SelfEnergy2}, we find that the non-Hermiticity originates from both the complex SB field $b_0$ and the complex hybridization $\tilde \Delta$, the latter of which is assumed to take a real value in Ref.~\cite{Yamamoto24}. We arrive at the retarded (advanced) NH Green function given in Eq.~\eqref{eq_greenimpurity} by the analytical continuation $i\omega_n\to\omega\pm i\eta$ in Eq.~\eqref{eq_Green_d2}. In Sec.~\ref{sec_NHSB}, we assume that 
\begin{align}
\Delta_b^\mathrm{Re}\mp\mathrm{Im}\tilde\lambda>0,
\label{eq_analyticity}
\end{align}
which is smoothly connected to the Hermitian case in the $V\to0$ limit \cite{Coleman87}. If either one of Eq.~\eqref{eq_analyticity} is not preserved with increasing non-Hermiticity, the phase transition with the breakdown of the Kondo effect occurs \cite{Yamamoto24}. Finally, with the use of the detailed form of ${}_L\langle d_\sigma^\dag d_\sigma\rangle_R$ and ${}_L\langle c_{\bm k \sigma}^\dag d_\sigma\rangle_R$ after the contour integrations, we arrive at the SCEs \eqref{eq_self} in the $\beta\to\infty$ limit. We note that, due to the non-Hermiticity in $\tilde \Delta$, both the $\log$ term and the $\tan^{-1}$ term in Eq.~\eqref{eq_self} contribute to the real and imaginary parts of the SCEs \cite{SCE}.

\section{Lehmann representation for the non-Hermitian Green function}
\label{sec_Lehmann}
In Sec.~\ref{sec_NHSB}, we have analyzed the NH Green functions in many-body systems by using the path-integral formalism on the basis of the analytic continuation of parameters. However, in NH systems, such analytic continuation sometimes breaks down as seen in previous studies, e.g., beyond the weak dissipation regime upon phase transitions \cite{Kawakami98, Yamamoto19, Nakagawa21, Yamamoto24}. In Hermitian quantum many-body systems, the Lehmann representation of the Green function plays a fundamental role in relating the imaginary-time Green function to the retarded and advanced Green functions via analytic continuation \cite{Fetter71}. In this appendix, we construct the Lehmann representation for the NH Green function. We then demonstrate that the analytic continuation from the Hermitian case does not hold in the retarded and advanced Green functions when their analyticity in the half-complex-$\omega$ plane breaks down due to the Kondo breakdown. We note that the NH Lehmann representation has been recently obtained in Ref.~\cite{Coveney25} for the coupled-cluster Green function, but we rather focus on the breakdown of analytic continuation of the conventional Lehmann representation \cite{Fetter71}.

We first introduce the single-particle Green function for an impurity fermion in the NH-AIM \eqref{eq_NHAnderson} as
\begin{align}
G_\sigma(\tau, \tau^\prime)=-{}_L\langle T_\tau[c_{d\sigma}(\tau) c_{d\sigma}^\dag(\tau^\prime)]\rangle_R,
\label{eq_GreenMatsubara}
\end{align}
where $\tau$ is the imaginary time, $T_\tau$ stands for the $\tau$-product given by
\begin{align}
&T_\tau[c_{d\sigma}(\tau)c_{d\sigma}^\dag(\tau^\prime)]\notag\displaybreak[2]\\
&=\theta_H(\tau-\tau^\prime)c_{d\sigma}(\tau)c_{d\sigma}^\dag(\tau^\prime)-\theta_H(\tau^\prime-\tau)c_{d\sigma}^\dag(\tau^\prime)c_{d\sigma}(\tau).
\end{align}
Here, $\theta_H(x)$ is the Heaviside unit-step function. The expectation value is defined as
\begin{align}
{}_L\langle\cdots\rangle_R
&=\frac{1}{\Xi}\mathrm{Tr}[e^{-\beta(H_\mathrm{eff}-\tilde \mu \hat M)}\cdots]\notag\\
&=\frac{1}{\Xi}\sum_n {}_L\langle E_n|\cdots|E_n\rangle_R e^{-\beta (E_n-\tilde\mu M_n)},
\label{eq_expectationgrand}
\end{align}
where $|E_n\rangle_R$ and $|E_n\rangle_L$ are the right and left eigenstates of $H_\mathrm{eff}$ with eigenenergy $E_n$ and satisfy the biorthonormal relation as ${}_L\langle E_m|E_n\rangle_R=\delta_{mn}$. Also,
\begin{align}
\Xi=\mathrm{Tr}[e^{-\beta(H_\mathrm{eff}-\tilde \mu \hat M)}],
\label{eq_Xi}
\end{align}
is the partition function with the real statistical weight parameter $\beta$ and the complex parameter $\tilde \mu$, which is introduced as an analytical continuation of the equilibrium chemical potential. In Eq.~\eqref{eq_Xi}, $\hat M$ is the particle-number operator with eigenvalue $M_n$ (note that the NH-AIM conserves the total particle number and that $H_\mathrm{eff}$ and $\hat M$ have simultaneous eigenstates). In addition, we have introduced the Heisenberg representation for the NH system as
\begin{gather}
c_{d\sigma}(\tau)=e^{\tau(H_\mathrm{eff}-\tilde \mu \hat M)} c_{d\sigma} e^{-\tau(H_\mathrm{eff}-\tilde \mu \hat M)},\label{eq_Heisan}\displaybreak[2]\\
c_{d\sigma}^\dag(\tau)=e^{\tau(H_\mathrm{eff}-\tilde \mu \hat M)} c_{d\sigma}^\dag e^{-\tau(H_\mathrm{eff}-\tilde \mu \hat M)}.\label{eq_Heiscr}\displaybreak[2]
\end{gather}
By employing the cyclic property of the trace together with Eqs.~\eqref{eq_expectationgrand}, \eqref{eq_Heisan}, and \eqref{eq_Heiscr}, we find that the NH Green function \eqref{eq_GreenMatsubara} has the time translational symmetry as $G_\sigma(\tau,\tau^\prime)=G_\sigma(\tau-\tau^\prime, 0)\equiv G_\sigma(\tau-\tau^\prime)$.

\begin{widetext}
Next, by inserting the biorthogonal completeness relation $\sum_n|E_n\rangle_R {}_L\langle E_n|=1$ into Eq.~\eqref{eq_GreenMatsubara}, we obtain the Lehmann representation for the NH Green function as 
\begin{align}
G_\sigma(\tau-\tau^\prime)=-\frac{1}{\Xi}\sum_{m,n}e^{-\beta (E_n-\tilde \mu M_n)}\big[&\theta_H(\tau-\tau^\prime){}_L\langle E_n|c_{d\sigma}(\tau)|E_m\rangle_R{}_L\langle E_m|c_{d\sigma}^\dagger(\tau^\prime)|E_n\rangle_R\notag\\
&-\theta_H(\tau^\prime-\tau){}_L\langle E_n|c_{d\sigma}^\dag(\tau^\prime)|E_m\rangle_R{}_L\langle E_m|c_{d\sigma}(\tau)|E_n\rangle_R\big]\notag\displaybreak[2]\\
=-\frac{1}{\Xi}\sum_{m,n}e^{-\beta (E_n-\tilde \mu M_n)}\big[&\theta_H(\tau-\tau^\prime)e^{(E_n-E_m+\tilde\mu)(\tau-\tau^\prime)}{}_L\langle E_n|c_{d\sigma}|E_m\rangle_R{}_L\langle E_m|c_{d\sigma}^\dagger|E_n\rangle_R\notag\\
&-\theta_H(\tau^\prime-\tau)e^{(E_n-E_m-\tilde\mu)(\tau^\prime-\tau)}{}_L\langle E_n|c_{d\sigma}^\dag|E_m\rangle_R{}_L\langle E_m|c_{d\sigma}|E_n\rangle_R\big].\displaybreak[2]
\end{align}
The Fourier transform of the NH Green function is defined by 
\begin{gather}
G_\sigma(\tau)=\frac{1}{\beta}\sum_{\omega_l} e^{-i\omega_l\tau}G_\sigma(i\omega_l),\displaybreak[2]\\
G_\sigma(i\omega_l)=\int_0^\beta e^{i\omega_l\tau} G_\sigma(\tau)d\tau,\displaybreak[2]
\end{gather}
which leads to
\begin{align}
G_\sigma(i\omega_l)=&-\frac{1}{\Xi}\sum_{m,n}e^{-\beta (E_n-\tilde \mu M_n)}\int_0^\beta d\tau e^{i\omega_l\tau}e^{(E_n-E_m+\tilde\mu)\tau}{}_L\langle E_n|c_{d\sigma}|E_m\rangle_R{}_L\langle E_m|c_{d\sigma}^\dagger|E_n\rangle_R\notag\displaybreak[2]\\
=&-\frac{1}{\Xi}\sum_{m,n}e^{-\beta (E_n-\tilde \mu M_n)}\frac{e^{(i\omega_l+E_n-E_m+\tilde\mu)\beta}-1}{i\omega_l+E_n-E_m+\tilde \mu}{}_L\langle E_n|c_{d\sigma}|E_m\rangle_R{}_L\langle E_m|c_{d\sigma}^\dagger|E_n\rangle_R\notag\displaybreak[2]\\
=&\frac{1}{\Xi}\sum_{m,n}\frac{e^{-\beta(E_m-\tilde\mu M_m)}+e^{-\beta(E_n-\tilde\mu M_n)}}{i\omega_l+E_n-E_m+\tilde \mu}{}_L\langle E_n|c_{d\sigma}|E_m\rangle_R{}_L\langle E_m|c_{d\sigma}^\dagger|E_n\rangle_R,\displaybreak[2]
\label{eq_LehmannMatsubara}
\end{align}
where we have used $M_m=M_n+1$ and $\omega_l=(2l+1)\pi/\beta$ ($l \in \mathbb{Z}$). Here, we note that $\int_0^\beta e^{\alpha x}dx=(e^{\alpha \beta}-1)/\alpha$ holds for arbitrary complex number $\alpha$. 

At first sight, it seems that no issue occurs in the Lehmann representation for the NH Green function, but a problem arises when it comes to the real-time representation. Let us define the retarded NH Green function as
\begin{align}
G_\sigma^{R}(t, t^\prime)=&-i\theta_H(t-t^\prime){}_L\langle\{c_{d\sigma}(t), c_{d\sigma}^\dag(t^\prime) \}\rangle_R\notag\\
=&-\frac{i\theta_H(t-t^\prime)}{\Xi}\sum_{m,n}e^{-\beta (E_n-\tilde \mu M_n)}\big[e^{-i(E_m-E_n-\tilde \mu)(t-t^\prime)}{}_L\langle E_n|c_{d\sigma}|E_m\rangle_R{}_L\langle E_m|c_{d\sigma}^\dagger|E_n\rangle_R\notag\\
&\hspace{4cm}+e^{-i(E_m-E_n+\tilde \mu)(t^\prime-t)}{}_L\langle E_n|c_{d\sigma}^\dag|E_m\rangle_R{}_L\langle E_m|c_{d\sigma}|E_n\rangle_R\big],
\label{eq_LehmannR}
\end{align}
where we have used the definition of the real-time Heisenberg representation given by
\begin{gather}
c_{d\sigma}(t)=e^{i(H_\mathrm{eff}-\tilde \mu \hat M) t}c_{d\sigma} e^{-i (H_\mathrm{eff}-\tilde \mu \hat M) t},\\
c_{d\sigma}^\dag(t)=e^{i(H_\mathrm{eff}-\tilde \mu \hat M) t}c_{d\sigma}^\dag e^{-i (H_\mathrm{eff}-\tilde \mu \hat M) t}.
\label{eq_Heisenberg}
\end{gather}
If the Lehmann representation in the frequency space were given by the analytic continuation as in the Hermitian system as
\begin{align}
G_\sigma(i\omega_l)\xrightarrow[i\omega_l\to\omega+i\eta]{?}G_\sigma^{R}(\omega),
\label{eq_ACMatsubara}
\end{align}
we could obtain
\begin{align}
G_\sigma^{R}(\omega)
\overset{?}{=}&\frac{1}{\Xi}\sum_{m,n}\frac{e^{-\beta (E_m-\tilde \mu M_m)}+e^{-\beta (E_n-\tilde \mu M_n)}}{\omega+E_n-E_m+\tilde \mu+i\eta}{}_L\langle E_n|c_{d\sigma}|E_m\rangle_R{}_L\langle E_m|c^\dagger_{d\sigma}|E_n\rangle_R,
\label{eq_LehmannRetarded}
\end{align}
where $i\eta=+i0$ in the denominator would be omitted due to the complex quantities $E_n$, $E_m$, and $\tilde \mu$. However, this procedure contains a problem coming from the analyticity of the retarded and advanced Green functions. We recall that, in the Hermitian system, the retarded (advanced) Green function is analytic in the half-upper (lower) complex-$\omega$ plane. However, the complex-valued energy $E_n$ can break this analyticity in the strong dissipation regime beyond the Kondo breakdown in the NH-AIM \cite{Yamamoto24}, where the resonance width becomes negative. Then, the right-hand side of Eq.~\eqref{eq_LehmannRetarded} cannot be identified as the well-defined retarded Green function that reflects the analyticity of the Fourier transform of Eq.~\eqref{eq_LehmannR}. Mathematically, this means that the complex path in the integration of the Fourier transformation cannot be deformed onto the real axis as the path crosses the poles of the integrand. Therefore, though the Lehmann representation for $G_\sigma(i\omega_l)$ is well-defined in NH systems, that for $G_\sigma^{R(A)}(\omega)$ is not given by the analytic continuation \eqref{eq_ACMatsubara} in general, and Eq.~\eqref{eq_LehmannRetarded}, which is the analytic continuation of the conventional Hermitian Lehmann representation to complex parameters, no longer holds in NH systems. To formulate the complete Lehmann representation in NH systems, it seems that we need a different concept to recover the analyticity of retarded and advanced Green functions, but it is beyond the scope of this paper.

\section{Calculations of the nested Bethe equations}
\label{App_bethe}
In this appendix, we give the detailed calculations of the nested Bethe equations and obtain the condition for the Kondo breakdown. First, we separate the formulae ~\eqref{eq_bethespinthermo} and \eqref{eq_bethechargethermo}, \cred{where $\rho(k)$ [$\sigma(\lambda)$] represents the distribution function for spin (charge) modes,} into the host part denoted by the superscript (c) and the impurity part represented by the superscript (i). By introducing
\begin{gather}
\rho_0^{(\mathrm c)}(k)=\frac{1}{2\pi},\\
\rho_0^{(\mathrm i)}(k)=\frac{\tilde \Delta}{\pi[(k-E_d)^2+\tilde \Delta^2]},\displaybreak[2]\\
\sigma_0^{(\mathrm c)}(\lambda)=\frac{1}{\pi},\displaybreak[2]\\
\sigma_0^{(\mathrm i)}(\lambda)=\frac{2\tilde \Delta}{\pi[(\lambda-E_d)^2+(2\tilde \Delta)^2]},\displaybreak[2]
\end{gather}
and defining the integral of the function $f(k)$ multiplied by the Lorentzian function as
\begin{gather}
\int [\:n\:]_{\lambda, \lambda^\prime}f(\lambda^\prime)d\lambda^\prime =\int \frac{n\tilde \Delta}{\pi[(\lambda-\lambda^\prime)^2+(n\tilde \Delta)^2]}f(\lambda^\prime)d\lambda^\prime,\displaybreak[2]
\end{gather}
where $n$ is a positive integer, Eqs.~\eqref{eq_bethespinthermo} and \eqref{eq_bethechargethermo} read
\begin{gather}
\rho_0^{(\mathrm c)}(k) + \frac{1}{L}\rho_0^{(\mathrm i)}(k)=\rho(k) + \int_{-\infty-iB_-}^{\infty+iB_+} [\:1\:]_{k,\lambda}\sigma(\lambda)d\lambda- \int_B^{\infty+iB_+} [\:1\:]_{k,\lambda}\sigma(\lambda)d\lambda,\label{eq_bethespinthermo1}\displaybreak[2]\\
\sigma_0^{(\mathrm c)}(\lambda) + \frac{1}{L}\sigma_0^{(\mathrm i)}(\lambda)=\sigma(\lambda) + \int_{-\infty-iB_-}^{\infty+iB_+} [\:2\:]_{\lambda,\lambda^\prime}\sigma(\lambda^\prime)d\lambda^\prime- \int_B^{\infty+iB_+} [\:2\:]_{\lambda,\lambda^\prime}\sigma(\lambda^\prime)d\lambda^\prime+\int_{-\infty-iQ_-}^Q [\:1\:]_{\lambda,k}\rho(k)dk,\label{eq_bethechargethermo1}\displaybreak[2]
\end{gather}
\cred{where quasimomenta $k$ for spin modes are distributed over the region $(-\infty-iQ_-,Q)_c$, and rapidities $\lambda$ for charge modes span the range $(-\infty-iB_-,B)_c$. Here, we have rewritten the integral by separating the large $|B|$ part by introducing the complex quantity $B_+$ to focus on the Kondo limit later.}

In the following, we simplify Eqs.~\eqref{eq_bethespinthermo1} and \eqref{eq_bethechargethermo1} by using the Fourier transform. The Fourier transform for a function $f(k)$ of complex quasimomentum $k$ is introduced as
\begin{gather}
\tilde f(\omega) =\int_{\mathcal C}e^{i\omega k}f(k) dk = \int_{-\infty}^\infty e^{i\omega k}f(k)dk,
\end{gather}
where we have used that the complex path $\mathcal C=(-\infty-ia_-,\infty+ia_+)_c$ can continuously be deformed onto the real axis if it does not cross the poles of the integrand. We also note that, as the imaginary part of $k$ is finite, the convergence of the Fourier transform is guaranteed. By performing the Fourier transformation in Eq.~\eqref{eq_bethechargethermo1}, we obtain 
\begin{align}
\tilde \sigma_0^{(\mathrm c)}(\omega) + \frac{1}{L}\tilde \sigma_0^{(\mathrm i)}(\omega)=\tilde \sigma(\omega) + e^{-2\tilde \Delta |\omega|}\tilde \sigma(\omega) -\int_B^{\infty+iB_+}e^{-2\tilde \Delta |\omega|}e^{i\lambda^\prime \omega}\sigma(\lambda^\prime)d\lambda^\prime+\int_{-\infty-iQ_-}^Q e^{-\tilde \Delta |\omega|}e^{i k \omega}\rho(k)dk,\label{eq_bethechargethermo2}
\end{align}
where $\tilde \sigma_0^{(\mathrm c)}(\omega)=2\delta(\omega)$ and $\tilde \sigma_0^{(\mathrm i)}(\omega)=e^{iE_d\omega}e^{-2\tilde\Delta|\omega|}$. We rewrite Eq.~\eqref{eq_bethechargethermo2} as
\begin{align}
\tilde \sigma(\omega) =\tilde \sigma_s^{(\mathrm c)}(\omega) + \frac{1}{L}\tilde \sigma_s^{(\mathrm i)}(\omega) +\int_B^{\infty+iB_+}\tilde R(\omega)e^{i\lambda^\prime \omega}\sigma(\lambda^\prime)d\lambda^\prime-\int_{-\infty-iQ_-}^Q \tilde S(\omega)e^{i k \omega}\rho(k)dk.
\label{eq_bethechargethermo3}
\end{align}
Here, the following quantities are introduced for readability: $\tilde \sigma_s^{(\mathrm c)}(\omega)=\tilde \sigma_0^{(\mathrm c)}(\omega) / (1+e^{-2\tilde\Delta|\omega|})$, $\tilde \sigma_s^{(\mathrm i)}(\omega)=\tilde \sigma_0^{(\mathrm i)}(\omega) / (1+e^{-2\tilde\Delta|\omega|})=e^{iE_d\omega}\tilde R(\omega)$, $\tilde R(\omega)=e^{-2\tilde \Delta |\omega|}/(1+e^{-2\tilde\Delta|\omega|})$, and $\tilde S(\omega)=e^{-\tilde \Delta |\omega|}/(1+e^{-2\tilde\Delta|\omega|})$. Finally, by performing the inverse Fourier transform in Eq.~\eqref{eq_bethechargethermo3}, we arrive at the Bethe equation for the charge degrees of freedom as
\begin{align}
\sigma(\lambda) + \int_{-\infty-iQ_-}^Q S(\lambda-k)\rho(k)dk - \int_B^{\infty+iB_+} R(\lambda-\lambda^\prime)\sigma(\lambda^\prime)d\lambda^\prime=\sigma_s^{(\mathrm c)}(\lambda) + \frac{1}{L} \sigma_s^{(\mathrm i)}(\lambda),
\label{eq_bethechargethermo4}
\end{align}
where $\sigma_s^{(\mathrm c)}(\lambda)=1/2\pi$ and $\sigma_s^{(\mathrm i)}(\lambda)=R(\lambda-E_d)$.

To calculate Eq.~\eqref{eq_bethespinthermo1}, we substitute Eq.~\eqref{eq_bethechargethermo4} into the second term in the right hand side of Eq.~\eqref{eq_bethespinthermo1} and proceed with the calculation with the help of the Fourier transform, obtaining
\begin{align}
\rho(k) - \int_{-\infty-iQ_-}^Q R(k-k^\prime)\rho(k^\prime)dk^\prime - \int_B^{\infty+iB_+} S(k-\lambda)\sigma(\lambda)d\lambda=\rho_s^{(\mathrm c)}(k) + \frac{1}{L} \rho_s^{(\mathrm i)}(k),
\label{eq_bethespinthermo4}
\end{align}
where we find that $\rho_s^{(\mathrm c)}(k)=0$ and $\rho_s^{(\mathrm i)}(k)=S(k-E_d)$. By decomposing Eqs.~\eqref{eq_bethechargethermo4} and \eqref{eq_bethespinthermo4} into the host part and the impurity part, we arrive at
\begin{gather}
\rho^{(\mathrm i)}(k) - \int_{-\infty-iQ_-}^Q R(k-k^\prime)\rho^{(\mathrm i)}(k^\prime)dk^\prime - \int_B^{\infty+iB_+} S(k-\lambda)\sigma^{(\mathrm i)}(\lambda)d\lambda=S(k-E_d),
\label{eq_bethespinthermoi}\\
\rho^{(\mathrm c)}(k) - \int_{-\infty-iQ_-}^Q R(k-k^\prime)\rho^{(\mathrm c)}(k^\prime)dk^\prime - \int_B^{\infty+iB_+} S(k-\lambda)\sigma^{(\mathrm c)}(\lambda)d\lambda=0,
\label{eq_bethespinthermoc}\\
\sigma^{(\mathrm i)}(\lambda) + \int_{-\infty-iQ_-}^Q S(\lambda-k)\rho^{(\mathrm i)}(k)dk - \int_B^{\infty+iB_+} R(\lambda-\lambda^\prime)\sigma^{(\mathrm i)}(\lambda^\prime)d\lambda^\prime=R(\lambda-E_d),
\label{eq_bethechargethermoi}\\
\sigma^{(\mathrm c)}(\lambda) + \int_{-\infty-iQ_-}^Q S(\lambda-k)\rho^{(\mathrm c)}(k)dk - \int_B^{\infty+iB_+} R(\lambda-\lambda^\prime)\sigma^{(\mathrm c)}(\lambda^\prime)d\lambda^\prime=\frac{1}{2\pi}.
\label{eq_bethechargethermoi}
\end{gather}

Equations \eqref{eq_bethespinthermoi}-\eqref{eq_bethechargethermoi} are useful to describe key physical quantities for the impurity physics. For example, the particle number density of the impurity is given by
\begin{align}
\frac{N_d}{L} = 1-\int_B^{\infty+iB_+}\sigma^{(\mathrm i)}(\lambda)d\lambda,
\end{align}
and the magnetization density of the impurity and that of the host are respectively represented by
\begin{gather}
\frac{M_z^{(\mathrm i)}}{L}=\frac{1}{2} \int_{-\infty-iQ_-}^Q\rho^{(\mathrm i)}(k)dk,\label{eq_magi}\\
\frac{M_z^{(\mathrm c)}}{L}=\frac{1}{2} \int_{-\infty-iQ_-}^Q\rho^{(\mathrm c)}(k)dk.\label{eq_magc}
\end{gather}
Importantly, $D_s^{(\mathrm i)}[\epsilon(Q)]$ given in Eq.~\eqref{eq_susceptibility}, which is obtained from Eq.~\eqref{eq_DOS}, is calculated with the use of Eqs.~\eqref{eq_bethespinthermoi}-\eqref{eq_bethechargethermoi}. To see this, we first show the relation $\epsilon^\prime(k)=2\pi\rho^{(\mathrm c)}(k)$. The formal solution of the distribution function $\rho(k)$ is written as
\begin{align}
\rho(k)&=\rho_0(k) + \int_{-\infty - iQ_-}^Q R^{(\rho)}(k,k^\prime)\rho_0(k^\prime)dk^\prime\notag\\
&=\int_{-\infty - iQ_-}^Q [\delta(k-k^\prime)+R^{(\rho)}(k,k^\prime)]\rho_0(k^\prime)dk^\prime,
\label{eq_rhoformal}
\end{align}
and similarly, $\sigma(\lambda)$ is represented as
\begin{align}
\sigma(\lambda)&=\sigma_0(\lambda) + \int_{-\infty - iB_-}^B R^{(\sigma)}(\lambda,\lambda^\prime)\sigma_0(\lambda^\prime)d\lambda^\prime\notag\displaybreak[2]\\
&=\int_{-\infty - iB_-}^B [\delta(\lambda-\lambda^\prime)+R^{(\sigma)}(\lambda,\lambda^\prime)]\sigma_0(\lambda^\prime)d\lambda^\prime,
\label{eq_sigmaformal}
\end{align}
where $\rho_0(k)=\rho_0^{(\mathrm c)}(k) + \rho_0^{(\mathrm i)}(k)/L$, $\sigma_0(\lambda)=\sigma_0^{(\mathrm c)}(\lambda) + \sigma_0^{(\mathrm i)}(\lambda)/L$, and the kernels $R^{(\rho)}(k,k^\prime)$ and $R^{(\sigma)}(\lambda,\lambda^\prime)$ are introduced. By substituting Eqs.~\eqref{eq_rhoformal} and \eqref{eq_sigmaformal} into Eq.~\eqref{eq_energy}, the energy density of the system is rewritten by incorporating terms coming from the chemical potential $\tilde \mu$ and the magnetic field $\tilde H$ as
\begin{align}
\frac{E -\tilde \mu N - g\mu_B \tilde H M_z}{L}&= \int_{-\infty-iQ_-}^Q \epsilon_0(k)\rho(k) dk + \int_{-\infty-iB_-}^B \kappa_0(\lambda)\sigma(\lambda) d\lambda\notag\displaybreak[2]\\
&= \int_{-\infty-iQ_-}^Q dk \int_{-\infty-iQ_-}^Q dk^\prime \epsilon_0(k)[\delta(k-k^\prime)+R^{(\rho)}(k,k^\prime)]\rho_0(k^\prime) \notag\\
&\quad+ \int_{-\infty-iB_-}^B d\lambda \int_{-\infty-iB_-}^B d\lambda^\prime \kappa_0(\lambda)[\delta(\lambda-\lambda^\prime)+R^{(\sigma)}(\lambda,\lambda^\prime)] \sigma_0(\lambda^\prime)\notag\displaybreak[2]\\
&= \int_{-\infty-iQ_-}^Q \epsilon(k)\rho_0(k) dk + \int_{-\infty-iB_-}^B \kappa(\lambda)\sigma_0(\lambda) d\lambda,
\end{align}
where we have used Eqs.~\eqref{eq_ndensity} and \eqref{eq_magnetizationdensity}. Here, $\epsilon_0(k)\equiv k-\tilde \mu - \frac{1}{2}g\mu_B\tilde H$ and $\kappa_0(\lambda)\equiv 2\lambda-2\tilde \mu$ are defined by using the $g$-factor and the Bohr magneton $\mu_B$. Dressed energies $\epsilon(k)$ for spin excitations and $\kappa(\lambda)$ for charge excitations are introduced to satisfy
\begin{align}
\epsilon(k)
&=\int_{-\infty - iQ_-}^Q [\delta(k-k^\prime)+R^{(\rho)}(k,k^\prime)]\epsilon_0(k^\prime)dk^\prime,\label{eq_epsilonformal}\\
\kappa(\lambda)
&=\int_{-\infty - iB_-}^B [\delta(\lambda-\lambda^\prime)+R^{(\sigma)}(\lambda,\lambda^\prime)]\kappa_0(\lambda^\prime)d\lambda^\prime.
\label{eq_kappaformal}
\end{align}
We find that $\epsilon(k)$ and $\kappa(\lambda)$ obey the equations of the same form as $\rho(k)$ in Eq.~\eqref{eq_rhoformal} and $\sigma(\lambda)$ in Eq.~\eqref{eq_sigmaformal}, respectively. This means that nested Bethe equations for $\epsilon(k)$ and $\kappa(\lambda)$ are obtained by replacing $\rho_0(k), \sigma_0(\lambda), \rho(k), \sigma(\lambda)$ by $\epsilon_0(k), \kappa_0(\lambda), \epsilon(k), \kappa(\lambda)$ in Eqs.~\eqref{eq_bethespinthermo} and \eqref{eq_bethechargethermo} as
\begin{gather}
\epsilon_0(k)=\epsilon(k) + \int_{-\infty-iB_-}^B \frac{\tilde \Delta}{\pi[(k-\lambda)^2+\tilde \Delta^2]}\kappa(\lambda)d\lambda,\label{eq_betheepsilon}\displaybreak[2]\\
\kappa_0(\lambda)=\kappa(\lambda) + \int_{-\infty-iB_-}^B \frac{2\tilde \Delta}{\pi[(\lambda-\lambda^\prime)^2+(2\tilde \Delta)^2]}\kappa(\lambda^\prime)d\lambda^\prime+\int_{-\infty-iQ_-}^Q \frac{\tilde \Delta}{\pi[(\lambda-k)^2+\tilde \Delta^2]}\epsilon(k)dk.\label{eq_bethekappa}
\end{gather}
By differentiating Eqs.~\eqref{eq_betheepsilon} and \eqref{eq_bethekappa} with respect to $k$ and $\lambda$ with the use of $\epsilon_0(k)=k-\tilde \mu - \frac{1}{2}g\mu_B\tilde H$ and $\kappa_0(\lambda)=2\lambda-2\tilde \mu$, we obtain
\begin{gather}
\frac{1}{2\pi}=\frac{\epsilon^\prime(k)}{2\pi} + \int_{-\infty-iB_-}^B \frac{\tilde \Delta}{\pi[(k-\lambda)^2+\tilde \Delta^2]}\frac{\kappa^\prime(\lambda)}{2\pi}d\lambda,\label{eq_betheepsilonD}\displaybreak[2]\\
\frac{1}{\pi}=\frac{\kappa^\prime(\lambda)}{2\pi} + \int_{-\infty-iB_-}^B \frac{2\tilde \Delta}{\pi[(\lambda-\lambda^\prime)^2+(2\tilde \Delta)^2]}\frac{\kappa^\prime(\lambda^\prime)}{2\pi}d\lambda^\prime+\int_{-\infty-iQ_-}^Q \frac{\tilde \Delta}{\pi[(\lambda-k)^2+\tilde \Delta^2]}\frac{\epsilon^\prime(k)}{2\pi}dk,\label{eq_bethekappaD}
\end{gather}
where we have performed the partial integration with the help of the fact that excitation energies are zero at quasi-Fermi points as $\epsilon(Q)=0$ and $\kappa(B)=0$. Then, we find that Eqs.~\eqref{eq_betheepsilonD} and \eqref{eq_bethekappaD} are the same form as the host part of Eqs.~\eqref{eq_bethespinthermo} and \eqref{eq_bethechargethermo} including the driving terms. Thus, we obtain $\epsilon^\prime(k)=2\pi\rho^{(\mathrm c)}(k)$ and $\kappa^\prime(\lambda)=2\pi\sigma^{(\mathrm c)}(\lambda)$, which reflect that itinerant fermions are free fermions.

We now analyze Eq.~\eqref{eq_susceptibility} with Eqs.~\eqref{eq_bethespinthermoi}-\eqref{eq_bethechargethermoi}, which is calculated as
\begin{align}
D_s^{(\mathrm i)}[\epsilon(Q)]
&=\lim_{Q\to-\infty-iQ_-}\frac{\frac{1}{2\tilde\Delta}e^{\frac{\pi(Q-E_d)}{2\tilde\Delta}}+\int_B^{\infty+iB_+}d\lambda\frac{1}{2\tilde\Delta}e^{\frac{\pi(Q-\lambda)}{2\tilde\Delta}}\sigma^{(\mathrm i)}(\lambda)}{2\pi \int_B^{\infty+iB_+}d\lambda\frac{1}{2\tilde\Delta}e^{\frac{\pi(Q-\lambda)}{2\tilde\Delta}}\sigma^{(\mathrm c)}(\lambda)}\notag\displaybreak[2]\\
&=\frac{e^{-\frac{\pi E_d}{2\tilde\Delta}}+\int_B^{\infty+iB_+}d\lambda e^{-\frac{\pi\lambda}{2\tilde\Delta}}\sigma^{(\mathrm i)}(\lambda)}{2\pi \int_B^{\infty+iB_+}d\lambda e^{-\frac{\pi\lambda}{2\tilde\Delta}}\sigma^{(\mathrm c)}(\lambda)},
\label{eq_susceptibility1}
\end{align}
where we have used the asymptotic form $S(Q)=\mathrm{sech}(\pi Q / 2\tilde \Delta)/4\tilde \Delta\sim \exp(\pi Q /2\tilde \Delta)/2\tilde \Delta$ for $Q\to-\infty-iQ_-$, and the distribution function for the charge excitations satisfies
\begin{gather}
\sigma^{(\mathrm i)}(\lambda)- \int_B^{\infty+iB_+} R(\lambda-\lambda^\prime)\sigma^{(\mathrm i)}(\lambda^\prime)d\lambda^\prime=R(\lambda-E_d),
\label{eq_bethechargethermoi_approx}\\
\sigma^{(\mathrm c)}(\lambda)- \int_B^{\infty+iB_+} R(\lambda-\lambda^\prime)\sigma^{(\mathrm c)}(\lambda^\prime)d\lambda^\prime=\frac{1}{2\pi}.
\label{eq_bethespinthermoi_approx}
\end{gather}
We now evaluate the NH Kondo scale, which is proportional to $D_s^{(\mathrm i)}[\epsilon(Q)]^{-1}$ in the Kondo limit. As the second term in Eq.~\eqref{eq_susceptibility1} can be ignored compared to the first term in the Kondo limit, Eq.~\eqref{eq_susceptibility1} is evaluated as
\begin{align}
D_s^{(\mathrm i)}[\epsilon(Q)]=\frac{e^{-\frac{\pi \tilde E_d}{2\tilde\Delta}}}{2\pi\int_0^{\infty+iB_+^\prime}d\lambda e^{-\frac{\pi\lambda}{2\tilde\Delta}}\sigma^{(\mathrm c)}(\lambda+B)}
=\alpha e^{-\frac{\pi \tilde E_d}{2\tilde\Delta}},
\label{eq_KondoScaleEApp}
\end{align}
where we have introduced the shifted impurity level as $\tilde E_d = E_d - B$ and $\alpha$ as the normalization constant, respectively. Equation \eqref{eq_KondoScaleEApp} is nothing but the inverse NH Kondo scale given in Eq.~\eqref{eq_KondoScalE}. We remark that, in the condition for the Kondo breakdown $\mathrm{Re}[D_s^{(\mathrm i)}[\epsilon(Q)]^{-1}]=0$, the contribution coming from $\alpha^{-1}$ can be ignored compared to that from $e^{\frac{\pi \tilde E_d}{2\tilde\Delta}}$ for the large value of $|\tilde E_d^{\mathrm R}|$. Thus, we obtain Eq.~\eqref{eq_breakdownexact} as the condition for the Kondo breakdown.
\end{widetext}

\nocite{apsrev42Control}
\bibliographystyle{apsrev4-2}
\bibliography{NH_Anderson_VComplex.bib}

\begin{thebibliography}{122}%
\makeatletter
\providecommand \@ifxundefined [1]{%
 \@ifx{#1\undefined}
}%
\providecommand \@ifnum [1]{%
 \ifnum #1\expandafter \@firstoftwo
 \else \expandafter \@secondoftwo
 \fi
}%
\providecommand \@ifx [1]{%
 \ifx #1\expandafter \@firstoftwo
 \else \expandafter \@secondoftwo
 \fi
}%
\providecommand \natexlab [1]{#1}%
\providecommand \enquote  [1]{``#1''}%
\providecommand \bibnamefont  [1]{#1}%
\providecommand \bibfnamefont [1]{#1}%
\providecommand \citenamefont [1]{#1}%
\providecommand \href@noop [0]{\@secondoftwo}%
\providecommand \href [0]{\begingroup \@sanitize@url \@href}%
\providecommand \@href[1]{\@@startlink{#1}\@@href}%
\providecommand \@@href[1]{\endgroup#1\@@endlink}%
\providecommand \@sanitize@url [0]{\catcode `\\12\catcode `\$12\catcode
  `\&12\catcode `\#12\catcode `\^12\catcode `\_12\catcode `\%12\relax}%
\providecommand \@@startlink[1]{}%
\providecommand \@@endlink[0]{}%
\providecommand \url  [0]{\begingroup\@sanitize@url \@url }%
\providecommand \@url [1]{\endgroup\@href {#1}{\urlprefix }}%
\providecommand \urlprefix  [0]{URL }%
\providecommand \Eprint [0]{\href }%
\providecommand \doibase [0]{https://doi.org/}%
\providecommand \selectlanguage [0]{\@gobble}%
\providecommand \bibinfo  [0]{\@secondoftwo}%
\providecommand \bibfield  [0]{\@secondoftwo}%
\providecommand \translation [1]{[#1]}%
\providecommand \BibitemOpen [0]{}%
\providecommand \bibitemStop [0]{}%
\providecommand \bibitemNoStop [0]{.\EOS\space}%
\providecommand \EOS [0]{\spacefactor3000\relax}%
\providecommand \BibitemShut  [1]{\csname bibitem#1\endcsname}%
\let\auto@bib@innerbib\@empty
\bibitem [{\citenamefont {Kondo}(1964)}]{Kondo64}%
  \BibitemOpen
  \bibfield  {author} {\bibinfo {author} {\bibfnamefont {J.}~\bibnamefont
  {Kondo}},\ }\bibfield  {title} {\bibinfo {title} {Resistance minimum in
  dilute magnetic alloys},\ }\href@noop {} {\bibfield  {journal} {\bibinfo
  {journal} {Prog. Theor. Phys.}\ }\textbf {\bibinfo {volume} {32}},\ \bibinfo
  {pages} {37} (\bibinfo {year} {1964})}\BibitemShut {NoStop}%
\bibitem [{\citenamefont {Hewson}(1997)}]{Hewson97}%
  \BibitemOpen
  \bibfield  {author} {\bibinfo {author} {\bibfnamefont {A.~C.}\ \bibnamefont
  {Hewson}},\ }\href@noop {} {\emph {\bibinfo {title} {{The Kondo problem to
  heavy fermions}}}}\ (\bibinfo  {publisher} {Cambridge university press},\
  \bibinfo {year} {1997})\BibitemShut {NoStop}%
\bibitem [{\citenamefont {Coleman}(2015)}]{Coleman15}%
  \BibitemOpen
  \bibfield  {author} {\bibinfo {author} {\bibfnamefont {P.}~\bibnamefont
  {Coleman}},\ }\href@noop {} {\emph {\bibinfo {title} {Introduction to
  many-body physics}}}\ (\bibinfo  {publisher} {Cambridge University Press},\
  \bibinfo {year} {2015})\BibitemShut {NoStop}%
\bibitem [{\citenamefont {Anderson}(1961)}]{Anderson61}%
  \BibitemOpen
  \bibfield  {author} {\bibinfo {author} {\bibfnamefont {P.~W.}\ \bibnamefont
  {Anderson}},\ }\bibfield  {title} {\bibinfo {title} {Localized magnetic
  states in metals},\ }\href {https://doi.org/10.1103/PhysRev.124.41}
  {\bibfield  {journal} {\bibinfo  {journal} {Phys. Rev.}\ }\textbf {\bibinfo
  {volume} {124}},\ \bibinfo {pages} {41} (\bibinfo {year} {1961})}\BibitemShut
  {NoStop}%
\bibitem [{\citenamefont {Tsvelick}\ and\ \citenamefont
  {Wiegmann}(1983{\natexlab{a}})}]{Tsvelick83}%
  \BibitemOpen
  \bibfield  {author} {\bibinfo {author} {\bibfnamefont {A.}~\bibnamefont
  {Tsvelick}}\ and\ \bibinfo {author} {\bibfnamefont {P.}~\bibnamefont
  {Wiegmann}},\ }\bibfield  {title} {\bibinfo {title} {Exact results in the
  theory of magnetic alloys},\ }\href@noop {} {\bibfield  {journal} {\bibinfo
  {journal} {Adv. Phys.}\ }\textbf {\bibinfo {volume} {32}},\ \bibinfo {pages}
  {453} (\bibinfo {year} {1983}{\natexlab{a}})}\BibitemShut {NoStop}%
\bibitem [{\citenamefont {Okiji}\ and\ \citenamefont
  {Kawakami}(1984)}]{Okiji84}%
  \BibitemOpen
  \bibfield  {author} {\bibinfo {author} {\bibfnamefont {A.}~\bibnamefont
  {Okiji}}\ and\ \bibinfo {author} {\bibfnamefont {N.}~\bibnamefont
  {Kawakami}},\ }\bibfield  {title} {\bibinfo {title} {{Thermodynamic
  properties of the Anderson model}},\ }\href@noop {} {\bibfield  {journal}
  {\bibinfo  {journal} {J. App. Phys.}\ }\textbf {\bibinfo {volume} {55}},\
  \bibinfo {pages} {1931} (\bibinfo {year} {1984})}\BibitemShut {NoStop}%
\bibitem [{\citenamefont {Meir}\ \emph {et~al.}(1991)\citenamefont {Meir},
  \citenamefont {Wingreen},\ and\ \citenamefont {Lee}}]{Meir91}%
  \BibitemOpen
  \bibfield  {author} {\bibinfo {author} {\bibfnamefont {Y.}~\bibnamefont
  {Meir}}, \bibinfo {author} {\bibfnamefont {N.~S.}\ \bibnamefont {Wingreen}},\
  and\ \bibinfo {author} {\bibfnamefont {P.~A.}\ \bibnamefont {Lee}},\
  }\bibfield  {title} {\bibinfo {title} {{Transport through a strongly
  interacting electron system: Theory of periodic conductance oscillations}},\
  }\href {https://doi.org/10.1103/PhysRevLett.66.3048} {\bibfield  {journal}
  {\bibinfo  {journal} {Phys. Rev. Lett.}\ }\textbf {\bibinfo {volume} {66}},\
  \bibinfo {pages} {3048} (\bibinfo {year} {1991})}\BibitemShut {NoStop}%
\bibitem [{\citenamefont {Beenakker}(1991)}]{Beenakker91}%
  \BibitemOpen
  \bibfield  {author} {\bibinfo {author} {\bibfnamefont {C.~W.~J.}\
  \bibnamefont {Beenakker}},\ }\bibfield  {title} {\bibinfo {title} {Theory of
  coulomb-blockade oscillations in the conductance of a quantum dot},\ }\href
  {https://doi.org/10.1103/PhysRevB.44.1646} {\bibfield  {journal} {\bibinfo
  {journal} {Phys. Rev. B}\ }\textbf {\bibinfo {volume} {44}},\ \bibinfo
  {pages} {1646} (\bibinfo {year} {1991})}\BibitemShut {NoStop}%
\bibitem [{\citenamefont {Meir}\ and\ \citenamefont {Wingreen}(1992)}]{Meir92}%
  \BibitemOpen
  \bibfield  {author} {\bibinfo {author} {\bibfnamefont {Y.}~\bibnamefont
  {Meir}}\ and\ \bibinfo {author} {\bibfnamefont {N.~S.}\ \bibnamefont
  {Wingreen}},\ }\bibfield  {title} {\bibinfo {title} {Landauer formula for the
  current through an interacting electron region},\ }\href
  {https://doi.org/10.1103/PhysRevLett.68.2512} {\bibfield  {journal} {\bibinfo
   {journal} {Phys. Rev. Lett.}\ }\textbf {\bibinfo {volume} {68}},\ \bibinfo
  {pages} {2512} (\bibinfo {year} {1992})}\BibitemShut {NoStop}%
\bibitem [{\citenamefont {Meir}\ \emph {et~al.}(1993)\citenamefont {Meir},
  \citenamefont {Wingreen},\ and\ \citenamefont {Lee}}]{Meir93}%
  \BibitemOpen
  \bibfield  {author} {\bibinfo {author} {\bibfnamefont {Y.}~\bibnamefont
  {Meir}}, \bibinfo {author} {\bibfnamefont {N.~S.}\ \bibnamefont {Wingreen}},\
  and\ \bibinfo {author} {\bibfnamefont {P.~A.}\ \bibnamefont {Lee}},\
  }\bibfield  {title} {\bibinfo {title} {{Low-temperature transport through a
  quantum dot: The Anderson model out of equilibrium}},\ }\href
  {https://doi.org/10.1103/PhysRevLett.70.2601} {\bibfield  {journal} {\bibinfo
   {journal} {Phys. Rev. Lett.}\ }\textbf {\bibinfo {volume} {70}},\ \bibinfo
  {pages} {2601} (\bibinfo {year} {1993})}\BibitemShut {NoStop}%
\bibitem [{\citenamefont {Ralph}\ and\ \citenamefont
  {Buhrman}(1994)}]{Ralph94}%
  \BibitemOpen
  \bibfield  {author} {\bibinfo {author} {\bibfnamefont {D.~C.}\ \bibnamefont
  {Ralph}}\ and\ \bibinfo {author} {\bibfnamefont {R.~A.}\ \bibnamefont
  {Buhrman}},\ }\bibfield  {title} {\bibinfo {title} {{Kondo-assisted and
  resonant tunneling via a single charge trap: A realization of the Anderson
  model out of equilibrium}},\ }\href
  {https://doi.org/10.1103/PhysRevLett.72.3401} {\bibfield  {journal} {\bibinfo
   {journal} {Phys. Rev. Lett.}\ }\textbf {\bibinfo {volume} {72}},\ \bibinfo
  {pages} {3401} (\bibinfo {year} {1994})}\BibitemShut {NoStop}%
\bibitem [{\citenamefont {L\'opez}\ and\ \citenamefont
  {S\'anchez}(2003)}]{Lopez03}%
  \BibitemOpen
  \bibfield  {author} {\bibinfo {author} {\bibfnamefont {R.}~\bibnamefont
  {L\'opez}}\ and\ \bibinfo {author} {\bibfnamefont {D.}~\bibnamefont
  {S\'anchez}},\ }\bibfield  {title} {\bibinfo {title} {{Nonequilibrium
  Spintronic Transport through an Artificial Kondo Impurity: Conductance,
  Magnetoresistance, and Shot Noise}},\ }\href
  {https://doi.org/10.1103/PhysRevLett.90.116602} {\bibfield  {journal}
  {\bibinfo  {journal} {Phys. Rev. Lett.}\ }\textbf {\bibinfo {volume} {90}},\
  \bibinfo {pages} {116602} (\bibinfo {year} {2003})}\BibitemShut {NoStop}%
\bibitem [{\citenamefont {Coleman}(1984)}]{Coleman84}%
  \BibitemOpen
  \bibfield  {author} {\bibinfo {author} {\bibfnamefont {P.}~\bibnamefont
  {Coleman}},\ }\bibfield  {title} {\bibinfo {title} {New approach to the
  mixed-valence problem},\ }\href {https://doi.org/10.1103/PhysRevB.29.3035}
  {\bibfield  {journal} {\bibinfo  {journal} {Phys. Rev. B}\ }\textbf {\bibinfo
  {volume} {29}},\ \bibinfo {pages} {3035} (\bibinfo {year}
  {1984})}\BibitemShut {NoStop}%
\bibitem [{\citenamefont {Coleman}(1987)}]{Coleman87}%
  \BibitemOpen
  \bibfield  {author} {\bibinfo {author} {\bibfnamefont {P.}~\bibnamefont
  {Coleman}},\ }\bibfield  {title} {\bibinfo {title} {Mixed valence as an
  almost broken symmetry},\ }\href {https://doi.org/10.1103/PhysRevB.35.5072}
  {\bibfield  {journal} {\bibinfo  {journal} {Phys. Rev. B}\ }\textbf {\bibinfo
  {volume} {35}},\ \bibinfo {pages} {5072} (\bibinfo {year}
  {1987})}\BibitemShut {NoStop}%
\bibitem [{\citenamefont {Newns}\ and\ \citenamefont {Read}(1987)}]{Newns87}%
  \BibitemOpen
  \bibfield  {author} {\bibinfo {author} {\bibfnamefont {D.}~\bibnamefont
  {Newns}}\ and\ \bibinfo {author} {\bibfnamefont {N.}~\bibnamefont {Read}},\
  }\bibfield  {title} {\bibinfo {title} {Mean-field theory of intermediate
  valence/heavy fermion systems},\ }\href@noop {} {\bibfield  {journal}
  {\bibinfo  {journal} {Adv. Phys.}\ }\textbf {\bibinfo {volume} {36}},\
  \bibinfo {pages} {799} (\bibinfo {year} {1987})}\BibitemShut {NoStop}%
\bibitem [{\citenamefont {Schr{\"o}der}\ \emph {et~al.}(2000)\citenamefont
  {Schr{\"o}der}, \citenamefont {Aeppli}, \citenamefont {Coldea}, \citenamefont
  {Adams}, \citenamefont {Stockert}, \citenamefont {L{\"o}hneysen},
  \citenamefont {Bucher}, \citenamefont {Ramazashvili},\ and\ \citenamefont
  {Coleman}}]{Schroder00}%
  \BibitemOpen
  \bibfield  {author} {\bibinfo {author} {\bibfnamefont {A.}~\bibnamefont
  {Schr{\"o}der}}, \bibinfo {author} {\bibfnamefont {G.}~\bibnamefont
  {Aeppli}}, \bibinfo {author} {\bibfnamefont {R.}~\bibnamefont {Coldea}},
  \bibinfo {author} {\bibfnamefont {M.}~\bibnamefont {Adams}}, \bibinfo
  {author} {\bibfnamefont {O.}~\bibnamefont {Stockert}}, \bibinfo {author}
  {\bibfnamefont {H.}~\bibnamefont {L{\"o}hneysen}}, \bibinfo {author}
  {\bibfnamefont {E.}~\bibnamefont {Bucher}}, \bibinfo {author} {\bibfnamefont
  {R.}~\bibnamefont {Ramazashvili}},\ and\ \bibinfo {author} {\bibfnamefont
  {P.}~\bibnamefont {Coleman}},\ }\bibfield  {title} {\bibinfo {title} {Onset
  of antiferromagnetism in heavy-fermion metals},\ }\href@noop {} {\bibfield
  {journal} {\bibinfo  {journal} {Nature}\ }\textbf {\bibinfo {volume} {407}},\
  \bibinfo {pages} {351} (\bibinfo {year} {2000})}\BibitemShut {NoStop}%
\bibitem [{\citenamefont {Aynajian}\ \emph {et~al.}(2012)\citenamefont
  {Aynajian}, \citenamefont {da~Silva~Neto}, \citenamefont {Gyenis},
  \citenamefont {Baumbach}, \citenamefont {Thompson}, \citenamefont {Fisk},
  \citenamefont {Bauer},\ and\ \citenamefont {Yazdani}}]{Aynajian12}%
  \BibitemOpen
  \bibfield  {author} {\bibinfo {author} {\bibfnamefont {P.}~\bibnamefont
  {Aynajian}}, \bibinfo {author} {\bibfnamefont {E.~H.}\ \bibnamefont
  {da~Silva~Neto}}, \bibinfo {author} {\bibfnamefont {A.}~\bibnamefont
  {Gyenis}}, \bibinfo {author} {\bibfnamefont {R.~E.}\ \bibnamefont
  {Baumbach}}, \bibinfo {author} {\bibfnamefont {J.~D.}\ \bibnamefont
  {Thompson}}, \bibinfo {author} {\bibfnamefont {Z.}~\bibnamefont {Fisk}},
  \bibinfo {author} {\bibfnamefont {E.~D.}\ \bibnamefont {Bauer}},\ and\
  \bibinfo {author} {\bibfnamefont {A.}~\bibnamefont {Yazdani}},\ }\bibfield
  {title} {\bibinfo {title} {{Visualizing heavy fermions emerging in a quantum
  critical Kondo lattice}},\ }\href@noop {} {\bibfield  {journal} {\bibinfo
  {journal} {Nature}\ }\textbf {\bibinfo {volume} {486}},\ \bibinfo {pages}
  {201} (\bibinfo {year} {2012})}\BibitemShut {NoStop}%
\bibitem [{\citenamefont {Goldhaber-Gordon}\ \emph {et~al.}(1998)\citenamefont
  {Goldhaber-Gordon}, \citenamefont {Shtrikman}, \citenamefont {Mahalu},
  \citenamefont {Abusch-Magder}, \citenamefont {Meirav},\ and\ \citenamefont
  {Kastner}}]{Gordon98}%
  \BibitemOpen
  \bibfield  {author} {\bibinfo {author} {\bibfnamefont {D.}~\bibnamefont
  {Goldhaber-Gordon}}, \bibinfo {author} {\bibfnamefont {H.}~\bibnamefont
  {Shtrikman}}, \bibinfo {author} {\bibfnamefont {D.}~\bibnamefont {Mahalu}},
  \bibinfo {author} {\bibfnamefont {D.}~\bibnamefont {Abusch-Magder}}, \bibinfo
  {author} {\bibfnamefont {U.}~\bibnamefont {Meirav}},\ and\ \bibinfo {author}
  {\bibfnamefont {M.}~\bibnamefont {Kastner}},\ }\bibfield  {title} {\bibinfo
  {title} {Kondo effect in a single-electron transistor},\ }\href@noop {}
  {\bibfield  {journal} {\bibinfo  {journal} {Nature}\ }\textbf {\bibinfo
  {volume} {391}},\ \bibinfo {pages} {156} (\bibinfo {year}
  {1998})}\BibitemShut {NoStop}%
\bibitem [{\citenamefont {Cronenwett}\ \emph {et~al.}(1998)\citenamefont
  {Cronenwett}, \citenamefont {Oosterkamp},\ and\ \citenamefont
  {Kouwenhoven}}]{Cronenwett98}%
  \BibitemOpen
  \bibfield  {author} {\bibinfo {author} {\bibfnamefont {S.~M.}\ \bibnamefont
  {Cronenwett}}, \bibinfo {author} {\bibfnamefont {T.~H.}\ \bibnamefont
  {Oosterkamp}},\ and\ \bibinfo {author} {\bibfnamefont {L.~P.}\ \bibnamefont
  {Kouwenhoven}},\ }\bibfield  {title} {\bibinfo {title} {{A tunable Kondo
  effect in quantum dots}},\ }\href@noop {} {\bibfield  {journal} {\bibinfo
  {journal} {Science}\ }\textbf {\bibinfo {volume} {281}},\ \bibinfo {pages}
  {540} (\bibinfo {year} {1998})}\BibitemShut {NoStop}%
\bibitem [{\citenamefont {V.~Borzenets}\ \emph {et~al.}(2020)\citenamefont
  {V.~Borzenets}, \citenamefont {Shim}, \citenamefont {Chen}, \citenamefont
  {Ludwig}, \citenamefont {Wieck}, \citenamefont {Tarucha}, \citenamefont
  {Sim},\ and\ \citenamefont {Yamamoto}}]{Borzenets20}%
  \BibitemOpen
  \bibfield  {author} {\bibinfo {author} {\bibfnamefont {I.}~\bibnamefont
  {V.~Borzenets}}, \bibinfo {author} {\bibfnamefont {J.}~\bibnamefont {Shim}},
  \bibinfo {author} {\bibfnamefont {J.~C.}\ \bibnamefont {Chen}}, \bibinfo
  {author} {\bibfnamefont {A.}~\bibnamefont {Ludwig}}, \bibinfo {author}
  {\bibfnamefont {A.~D.}\ \bibnamefont {Wieck}}, \bibinfo {author}
  {\bibfnamefont {S.}~\bibnamefont {Tarucha}}, \bibinfo {author} {\bibfnamefont
  {H.-S.}\ \bibnamefont {Sim}},\ and\ \bibinfo {author} {\bibfnamefont
  {M.}~\bibnamefont {Yamamoto}},\ }\bibfield  {title} {\bibinfo {title}
  {{Observation of the Kondo screening cloud}},\ }\href@noop {} {\bibfield
  {journal} {\bibinfo  {journal} {Nature}\ }\textbf {\bibinfo {volume} {579}},\
  \bibinfo {pages} {210} (\bibinfo {year} {2020})}\BibitemShut {NoStop}%
\bibitem [{\citenamefont {Scazza}\ \emph {et~al.}(2014)\citenamefont {Scazza},
  \citenamefont {Hofrichter}, \citenamefont {H{\"o}fer}, \citenamefont
  {De~Groot}, \citenamefont {Bloch},\ and\ \citenamefont
  {F{\"o}lling}}]{Scazza14}%
  \BibitemOpen
  \bibfield  {author} {\bibinfo {author} {\bibfnamefont {F.}~\bibnamefont
  {Scazza}}, \bibinfo {author} {\bibfnamefont {C.}~\bibnamefont {Hofrichter}},
  \bibinfo {author} {\bibfnamefont {M.}~\bibnamefont {H{\"o}fer}}, \bibinfo
  {author} {\bibfnamefont {P.}~\bibnamefont {De~Groot}}, \bibinfo {author}
  {\bibfnamefont {I.}~\bibnamefont {Bloch}},\ and\ \bibinfo {author}
  {\bibfnamefont {S.}~\bibnamefont {F{\"o}lling}},\ }\bibfield  {title}
  {\bibinfo {title} {{Observation of two-orbital spin-exchange interactions
  with ultracold SU(N)-symmetric fermions}},\ }\href@noop {} {\bibfield
  {journal} {\bibinfo  {journal} {Nat. Phys.}\ }\textbf {\bibinfo {volume}
  {10}},\ \bibinfo {pages} {779} (\bibinfo {year} {2014})}\BibitemShut
  {NoStop}%
\bibitem [{\citenamefont {Riegger}\ \emph {et~al.}(2018)\citenamefont
  {Riegger}, \citenamefont {Darkwah~Oppong}, \citenamefont {H\"ofer},
  \citenamefont {Fernandes}, \citenamefont {Bloch},\ and\ \citenamefont
  {F\"olling}}]{Riegger18}%
  \BibitemOpen
  \bibfield  {author} {\bibinfo {author} {\bibfnamefont {L.}~\bibnamefont
  {Riegger}}, \bibinfo {author} {\bibfnamefont {N.}~\bibnamefont
  {Darkwah~Oppong}}, \bibinfo {author} {\bibfnamefont {M.}~\bibnamefont
  {H\"ofer}}, \bibinfo {author} {\bibfnamefont {D.~R.}\ \bibnamefont
  {Fernandes}}, \bibinfo {author} {\bibfnamefont {I.}~\bibnamefont {Bloch}},\
  and\ \bibinfo {author} {\bibfnamefont {S.}~\bibnamefont {F\"olling}},\
  }\bibfield  {title} {\bibinfo {title} {{Localized Magnetic Moments with
  Tunable Spin Exchange in a Gas of Ultracold Fermions}},\ }\href
  {https://doi.org/10.1103/PhysRevLett.120.143601} {\bibfield  {journal}
  {\bibinfo  {journal} {Phys. Rev. Lett.}\ }\textbf {\bibinfo {volume} {120}},\
  \bibinfo {pages} {143601} (\bibinfo {year} {2018})}\BibitemShut {NoStop}%
\bibitem [{\citenamefont {Ono}\ \emph {et~al.}(2021)\citenamefont {Ono},
  \citenamefont {Amano}, \citenamefont {Higomoto}, \citenamefont {Saito},\ and\
  \citenamefont {Takahashi}}]{Ono21}%
  \BibitemOpen
  \bibfield  {author} {\bibinfo {author} {\bibfnamefont {K.}~\bibnamefont
  {Ono}}, \bibinfo {author} {\bibfnamefont {Y.}~\bibnamefont {Amano}}, \bibinfo
  {author} {\bibfnamefont {T.}~\bibnamefont {Higomoto}}, \bibinfo {author}
  {\bibfnamefont {Y.}~\bibnamefont {Saito}},\ and\ \bibinfo {author}
  {\bibfnamefont {Y.}~\bibnamefont {Takahashi}},\ }\bibfield  {title} {\bibinfo
  {title} {{Observation of spin-exchange dynamics between itinerant and
  localized $^{171}\mathrm{Yb}$ atoms}},\ }\href
  {https://doi.org/10.1103/PhysRevA.103.L041303} {\bibfield  {journal}
  {\bibinfo  {journal} {Phys. Rev. A}\ }\textbf {\bibinfo {volume} {103}},\
  \bibinfo {pages} {L041303} (\bibinfo {year} {2021})}\BibitemShut {NoStop}%
\bibitem [{\citenamefont {Daley}(2014)}]{Daley14}%
  \BibitemOpen
  \bibfield  {author} {\bibinfo {author} {\bibfnamefont {A.~J.}\ \bibnamefont
  {Daley}},\ }\bibfield  {title} {\bibinfo {title} {Quantum trajectories and
  open many-body quantum systems},\ }\href@noop {} {\bibfield  {journal}
  {\bibinfo  {journal} {Adv. Phys.}\ }\textbf {\bibinfo {volume} {63}},\
  \bibinfo {pages} {77} (\bibinfo {year} {2014})}\BibitemShut {NoStop}%
\bibitem [{\citenamefont {Ashida}\ \emph {et~al.}(2020)\citenamefont {Ashida},
  \citenamefont {Gong},\ and\ \citenamefont {Ueda}}]{Ashida20}%
  \BibitemOpen
  \bibfield  {author} {\bibinfo {author} {\bibfnamefont {Y.}~\bibnamefont
  {Ashida}}, \bibinfo {author} {\bibfnamefont {Z.}~\bibnamefont {Gong}},\ and\
  \bibinfo {author} {\bibfnamefont {M.}~\bibnamefont {Ueda}},\ }\bibfield
  {title} {\bibinfo {title} {Non-hermitian physics},\ }\href@noop {} {\bibfield
   {journal} {\bibinfo  {journal} {Adv. Phys.}\ }\textbf {\bibinfo {volume}
  {69}},\ \bibinfo {pages} {249} (\bibinfo {year} {2020})}\BibitemShut
  {NoStop}%
\bibitem [{\citenamefont {Fazio}\ \emph {et~al.}()\citenamefont {Fazio},
  \citenamefont {Keeling}, \citenamefont {Mazza},\ and\ \citenamefont
  {Schiro}}]{Fazio24}%
  \BibitemOpen
  \bibfield  {author} {\bibinfo {author} {\bibfnamefont {R.}~\bibnamefont
  {Fazio}}, \bibinfo {author} {\bibfnamefont {J.}~\bibnamefont {Keeling}},
  \bibinfo {author} {\bibfnamefont {L.}~\bibnamefont {Mazza}},\ and\ \bibinfo
  {author} {\bibfnamefont {M.}~\bibnamefont {Schiro}},\ }\bibfield  {title}
  {\bibinfo {title} {Many-body open quantum systems},\ }\href@noop {} {\bibinfo
   {journal} {arXiv:2409.10300}\ }\BibitemShut {NoStop}%
\bibitem [{\citenamefont {Syassen}\ \emph {et~al.}(2008)\citenamefont
  {Syassen}, \citenamefont {Bauer}, \citenamefont {Lettner}, \citenamefont
  {Volz}, \citenamefont {Dietze}, \citenamefont {Garcia-Ripoll}, \citenamefont
  {Cirac}, \citenamefont {Rempe},\ and\ \citenamefont {D{\"u}rr}}]{Syassen08}%
  \BibitemOpen
\bibfield  {journal} {  }\bibfield  {author} {\bibinfo {author} {\bibfnamefont
  {N.}~\bibnamefont {Syassen}}, \bibinfo {author} {\bibfnamefont {D.~M.}\
  \bibnamefont {Bauer}}, \bibinfo {author} {\bibfnamefont {M.}~\bibnamefont
  {Lettner}}, \bibinfo {author} {\bibfnamefont {T.}~\bibnamefont {Volz}},
  \bibinfo {author} {\bibfnamefont {D.}~\bibnamefont {Dietze}}, \bibinfo
  {author} {\bibfnamefont {J.~J.}\ \bibnamefont {Garcia-Ripoll}}, \bibinfo
  {author} {\bibfnamefont {J.~I.}\ \bibnamefont {Cirac}}, \bibinfo {author}
  {\bibfnamefont {G.}~\bibnamefont {Rempe}},\ and\ \bibinfo {author}
  {\bibfnamefont {S.}~\bibnamefont {D{\"u}rr}},\ }\bibfield  {title} {\bibinfo
  {title} {Strong dissipation inhibits losses and induces correlations in cold
  molecular gases},\ }\href@noop {} {\bibfield  {journal} {\bibinfo  {journal}
  {Science}\ }\textbf {\bibinfo {volume} {320}},\ \bibinfo {pages} {1329}
  (\bibinfo {year} {2008})}\BibitemShut {NoStop}%
\bibitem [{\citenamefont {Yan}\ \emph {et~al.}(2013)\citenamefont {Yan},
  \citenamefont {Moses}, \citenamefont {Gadway}, \citenamefont {Covey},
  \citenamefont {Hazzard}, \citenamefont {Rey}, \citenamefont {Jin},\ and\
  \citenamefont {Ye}}]{Yan13}%
  \BibitemOpen
  \bibfield  {author} {\bibinfo {author} {\bibfnamefont {B.}~\bibnamefont
  {Yan}}, \bibinfo {author} {\bibfnamefont {S.~A.}\ \bibnamefont {Moses}},
  \bibinfo {author} {\bibfnamefont {B.}~\bibnamefont {Gadway}}, \bibinfo
  {author} {\bibfnamefont {J.~P.}\ \bibnamefont {Covey}}, \bibinfo {author}
  {\bibfnamefont {K.~R.}\ \bibnamefont {Hazzard}}, \bibinfo {author}
  {\bibfnamefont {A.~M.}\ \bibnamefont {Rey}}, \bibinfo {author} {\bibfnamefont
  {D.~S.}\ \bibnamefont {Jin}},\ and\ \bibinfo {author} {\bibfnamefont
  {J.}~\bibnamefont {Ye}},\ }\bibfield  {title} {\bibinfo {title} {Observation
  of dipolar spin-exchange interactions with lattice-confined polar
  molecules},\ }\href@noop {} {\bibfield  {journal} {\bibinfo  {journal}
  {Nature (London)}\ }\textbf {\bibinfo {volume} {501}},\ \bibinfo {pages}
  {521} (\bibinfo {year} {2013})}\BibitemShut {NoStop}%
\bibitem [{\citenamefont {Patil}\ \emph {et~al.}(2015)\citenamefont {Patil},
  \citenamefont {Chakram},\ and\ \citenamefont {Vengalattore}}]{Patil15}%
  \BibitemOpen
  \bibfield  {author} {\bibinfo {author} {\bibfnamefont {Y.~S.}\ \bibnamefont
  {Patil}}, \bibinfo {author} {\bibfnamefont {S.}~\bibnamefont {Chakram}},\
  and\ \bibinfo {author} {\bibfnamefont {M.}~\bibnamefont {Vengalattore}},\
  }\bibfield  {title} {\bibinfo {title} {{Measurement-Induced Localization of
  an Ultracold Lattice Gas}},\ }\href
  {https://doi.org/10.1103/PhysRevLett.115.140402} {\bibfield  {journal}
  {\bibinfo  {journal} {Phys. Rev. Lett.}\ }\textbf {\bibinfo {volume} {115}},\
  \bibinfo {pages} {140402} (\bibinfo {year} {2015})}\BibitemShut {NoStop}%
\bibitem [{\citenamefont {L\"uschen}\ \emph {et~al.}(2017)\citenamefont
  {L\"uschen}, \citenamefont {Bordia}, \citenamefont {Hodgman}, \citenamefont
  {Schreiber}, \citenamefont {Sarkar}, \citenamefont {Daley}, \citenamefont
  {Fischer}, \citenamefont {Altman}, \citenamefont {Bloch},\ and\ \citenamefont
  {Schneider}}]{Schneider17}%
  \BibitemOpen
  \bibfield  {author} {\bibinfo {author} {\bibfnamefont {H.~P.}\ \bibnamefont
  {L\"uschen}}, \bibinfo {author} {\bibfnamefont {P.}~\bibnamefont {Bordia}},
  \bibinfo {author} {\bibfnamefont {S.~S.}\ \bibnamefont {Hodgman}}, \bibinfo
  {author} {\bibfnamefont {M.}~\bibnamefont {Schreiber}}, \bibinfo {author}
  {\bibfnamefont {S.}~\bibnamefont {Sarkar}}, \bibinfo {author} {\bibfnamefont
  {A.~J.}\ \bibnamefont {Daley}}, \bibinfo {author} {\bibfnamefont {M.~H.}\
  \bibnamefont {Fischer}}, \bibinfo {author} {\bibfnamefont {E.}~\bibnamefont
  {Altman}}, \bibinfo {author} {\bibfnamefont {I.}~\bibnamefont {Bloch}},\ and\
  \bibinfo {author} {\bibfnamefont {U.}~\bibnamefont {Schneider}},\ }\bibfield
  {title} {\bibinfo {title} {{Signatures of Many-Body Localization in a
  Controlled Open Quantum System}},\ }\href
  {https://doi.org/10.1103/PhysRevX.7.011034} {\bibfield  {journal} {\bibinfo
  {journal} {Phys. Rev. X}\ }\textbf {\bibinfo {volume} {7}},\ \bibinfo {pages}
  {011034} (\bibinfo {year} {2017})}\BibitemShut {NoStop}%
\bibitem [{\citenamefont {Tomita}\ \emph {et~al.}(2017)\citenamefont {Tomita},
  \citenamefont {Nakajima}, \citenamefont {Danshita}, \citenamefont {Takasu},\
  and\ \citenamefont {Takahashi}}]{Tomita17}%
  \BibitemOpen
  \bibfield  {author} {\bibinfo {author} {\bibfnamefont {T.}~\bibnamefont
  {Tomita}}, \bibinfo {author} {\bibfnamefont {S.}~\bibnamefont {Nakajima}},
  \bibinfo {author} {\bibfnamefont {I.}~\bibnamefont {Danshita}}, \bibinfo
  {author} {\bibfnamefont {Y.}~\bibnamefont {Takasu}},\ and\ \bibinfo {author}
  {\bibfnamefont {Y.}~\bibnamefont {Takahashi}},\ }\bibfield  {title} {\bibinfo
  {title} {{Observation of the Mott insulator to superfluid crossover of a
  driven-dissipative Bose-Hubbard system}},\ }\href@noop {} {\bibfield
  {journal} {\bibinfo  {journal} {Sci. Adv.}\ }\textbf {\bibinfo {volume}
  {3}},\ \bibinfo {pages} {e1701513} (\bibinfo {year} {2017})}\BibitemShut
  {NoStop}%
\bibitem [{\citenamefont {Sponselee}\ \emph {et~al.}(2018)\citenamefont
  {Sponselee}, \citenamefont {Freystatzky}, \citenamefont {Abeln},
  \citenamefont {Diem}, \citenamefont {Hundt}, \citenamefont {Kochanke},
  \citenamefont {Ponath}, \citenamefont {Santra}, \citenamefont {Mathey},
  \citenamefont {Sengstock},\ and\ \citenamefont {Becker}}]{Spon18}%
  \BibitemOpen
  \bibfield  {author} {\bibinfo {author} {\bibfnamefont {K.}~\bibnamefont
  {Sponselee}}, \bibinfo {author} {\bibfnamefont {L.}~\bibnamefont
  {Freystatzky}}, \bibinfo {author} {\bibfnamefont {B.}~\bibnamefont {Abeln}},
  \bibinfo {author} {\bibfnamefont {M.}~\bibnamefont {Diem}}, \bibinfo {author}
  {\bibfnamefont {B.}~\bibnamefont {Hundt}}, \bibinfo {author} {\bibfnamefont
  {A.}~\bibnamefont {Kochanke}}, \bibinfo {author} {\bibfnamefont
  {T.}~\bibnamefont {Ponath}}, \bibinfo {author} {\bibfnamefont
  {B.}~\bibnamefont {Santra}}, \bibinfo {author} {\bibfnamefont
  {L.}~\bibnamefont {Mathey}}, \bibinfo {author} {\bibfnamefont
  {K.}~\bibnamefont {Sengstock}},\ and\ \bibinfo {author} {\bibfnamefont
  {C.}~\bibnamefont {Becker}},\ }\bibfield  {title} {\bibinfo {title}
  {{Dynamics of ultracold quantum gases in the dissipative Fermi--Hubbard
  model}},\ }\href@noop {} {\bibfield  {journal} {\bibinfo  {journal} {Quantum
  Sci. Technol.}\ }\textbf {\bibinfo {volume} {4}},\ \bibinfo {pages} {014002}
  (\bibinfo {year} {2018})}\BibitemShut {NoStop}%
\bibitem [{\citenamefont {Bouganne}\ \emph {et~al.}(2020)\citenamefont
  {Bouganne}, \citenamefont {Aguilera}, \citenamefont {Ghermaoui},
  \citenamefont {Beugnon},\ and\ \citenamefont {Gerbier}}]{Gerbier20}%
  \BibitemOpen
  \bibfield  {author} {\bibinfo {author} {\bibfnamefont {R.}~\bibnamefont
  {Bouganne}}, \bibinfo {author} {\bibfnamefont {M.~B.}\ \bibnamefont
  {Aguilera}}, \bibinfo {author} {\bibfnamefont {A.}~\bibnamefont {Ghermaoui}},
  \bibinfo {author} {\bibfnamefont {J.}~\bibnamefont {Beugnon}},\ and\ \bibinfo
  {author} {\bibfnamefont {F.}~\bibnamefont {Gerbier}},\ }\bibfield  {title}
  {\bibinfo {title} {Anomalous decay of coherence in a dissipative many-body
  system},\ }\href@noop {} {\bibfield  {journal} {\bibinfo  {journal} {Nat.
  Phys.}\ }\textbf {\bibinfo {volume} {16}},\ \bibinfo {pages} {21} (\bibinfo
  {year} {2020})}\BibitemShut {NoStop}%
\bibitem [{\citenamefont {Honda}\ \emph {et~al.}(2023)\citenamefont {Honda},
  \citenamefont {Taie}, \citenamefont {Takasu}, \citenamefont {Nishizawa},
  \citenamefont {Nakagawa},\ and\ \citenamefont {Takahashi}}]{Honda23}%
  \BibitemOpen
  \bibfield  {author} {\bibinfo {author} {\bibfnamefont {K.}~\bibnamefont
  {Honda}}, \bibinfo {author} {\bibfnamefont {S.}~\bibnamefont {Taie}},
  \bibinfo {author} {\bibfnamefont {Y.}~\bibnamefont {Takasu}}, \bibinfo
  {author} {\bibfnamefont {N.}~\bibnamefont {Nishizawa}}, \bibinfo {author}
  {\bibfnamefont {M.}~\bibnamefont {Nakagawa}},\ and\ \bibinfo {author}
  {\bibfnamefont {Y.}~\bibnamefont {Takahashi}},\ }\bibfield  {title} {\bibinfo
  {title} {{Observation of the Sign Reversal of the Magnetic Correlation in a
  Driven-Dissipative Fermi Gas in Double Wells}},\ }\href
  {https://doi.org/10.1103/PhysRevLett.130.063001} {\bibfield  {journal}
  {\bibinfo  {journal} {Phys. Rev. Lett.}\ }\textbf {\bibinfo {volume} {130}},\
  \bibinfo {pages} {063001} (\bibinfo {year} {2023})}\BibitemShut {NoStop}%
\bibitem [{\citenamefont {Barontini}\ \emph {et~al.}(2013)\citenamefont
  {Barontini}, \citenamefont {Labouvie}, \citenamefont {Stubenrauch},
  \citenamefont {Vogler}, \citenamefont {Guarrera},\ and\ \citenamefont
  {Ott}}]{Ott13}%
  \BibitemOpen
  \bibfield  {author} {\bibinfo {author} {\bibfnamefont {G.}~\bibnamefont
  {Barontini}}, \bibinfo {author} {\bibfnamefont {R.}~\bibnamefont {Labouvie}},
  \bibinfo {author} {\bibfnamefont {F.}~\bibnamefont {Stubenrauch}}, \bibinfo
  {author} {\bibfnamefont {A.}~\bibnamefont {Vogler}}, \bibinfo {author}
  {\bibfnamefont {V.}~\bibnamefont {Guarrera}},\ and\ \bibinfo {author}
  {\bibfnamefont {H.}~\bibnamefont {Ott}},\ }\bibfield  {title} {\bibinfo
  {title} {{Controlling the Dynamics of an Open Many-Body Quantum System with
  Localized Dissipation}},\ }\href
  {https://doi.org/10.1103/PhysRevLett.110.035302} {\bibfield  {journal}
  {\bibinfo  {journal} {Phys. Rev. Lett.}\ }\textbf {\bibinfo {volume} {110}},\
  \bibinfo {pages} {035302} (\bibinfo {year} {2013})}\BibitemShut {NoStop}%
\bibitem [{\citenamefont {Labouvie}\ \emph {et~al.}(2016)\citenamefont
  {Labouvie}, \citenamefont {Santra}, \citenamefont {Heun},\ and\ \citenamefont
  {Ott}}]{Labouvie16}%
  \BibitemOpen
  \bibfield  {author} {\bibinfo {author} {\bibfnamefont {R.}~\bibnamefont
  {Labouvie}}, \bibinfo {author} {\bibfnamefont {B.}~\bibnamefont {Santra}},
  \bibinfo {author} {\bibfnamefont {S.}~\bibnamefont {Heun}},\ and\ \bibinfo
  {author} {\bibfnamefont {H.}~\bibnamefont {Ott}},\ }\bibfield  {title}
  {\bibinfo {title} {{Bistability in a Driven-Dissipative Superfluid}},\ }\href
  {https://doi.org/10.1103/PhysRevLett.116.235302} {\bibfield  {journal}
  {\bibinfo  {journal} {Phys. Rev. Lett.}\ }\textbf {\bibinfo {volume} {116}},\
  \bibinfo {pages} {235302} (\bibinfo {year} {2016})}\BibitemShut {NoStop}%
\bibitem [{\citenamefont {Benary}\ \emph {et~al.}(2022)\citenamefont {Benary},
  \citenamefont {Baals}, \citenamefont {Bernhart}, \citenamefont {Jiang},
  \citenamefont {R{\"o}hrle},\ and\ \citenamefont {Ott}}]{Benary22}%
  \BibitemOpen
  \bibfield  {author} {\bibinfo {author} {\bibfnamefont {J.}~\bibnamefont
  {Benary}}, \bibinfo {author} {\bibfnamefont {C.}~\bibnamefont {Baals}},
  \bibinfo {author} {\bibfnamefont {E.}~\bibnamefont {Bernhart}}, \bibinfo
  {author} {\bibfnamefont {J.}~\bibnamefont {Jiang}}, \bibinfo {author}
  {\bibfnamefont {M.}~\bibnamefont {R{\"o}hrle}},\ and\ \bibinfo {author}
  {\bibfnamefont {H.}~\bibnamefont {Ott}},\ }\bibfield  {title} {\bibinfo
  {title} {Experimental observation of a dissipative phase transition in a
  multi-mode many-body quantum system},\ }\href@noop {} {\bibfield  {journal}
  {\bibinfo  {journal} {New J. Phys.}\ }\textbf {\bibinfo {volume} {24}},\
  \bibinfo {pages} {103034} (\bibinfo {year} {2022})}\BibitemShut {NoStop}%
\bibitem [{\citenamefont {Takasu}\ \emph {et~al.}(2020)\citenamefont {Takasu},
  \citenamefont {Yagami}, \citenamefont {Ashida}, \citenamefont {Hamazaki},
  \citenamefont {Kuno},\ and\ \citenamefont {Takahashi}}]{Takasu20}%
  \BibitemOpen
  \bibfield  {author} {\bibinfo {author} {\bibfnamefont {Y.}~\bibnamefont
  {Takasu}}, \bibinfo {author} {\bibfnamefont {T.}~\bibnamefont {Yagami}},
  \bibinfo {author} {\bibfnamefont {Y.}~\bibnamefont {Ashida}}, \bibinfo
  {author} {\bibfnamefont {R.}~\bibnamefont {Hamazaki}}, \bibinfo {author}
  {\bibfnamefont {Y.}~\bibnamefont {Kuno}},\ and\ \bibinfo {author}
  {\bibfnamefont {Y.}~\bibnamefont {Takahashi}},\ }\bibfield  {title} {\bibinfo
  {title} {{PT-symmetric non-Hermitian quantum many-body system using ultracold
  atoms in an optical lattice with controlled dissipation}},\ }\href@noop {}
  {\bibfield  {journal} {\bibinfo  {journal} {Prog. Theor. Exp. Phys.}\
  }\textbf {\bibinfo {volume} {2020}},\ \bibinfo {pages} {12A110} (\bibinfo
  {year} {2020})}\BibitemShut {NoStop}%
\bibitem [{\citenamefont {Yamamoto}\ \emph {et~al.}(2021)\citenamefont
  {Yamamoto}, \citenamefont {Nakagawa}, \citenamefont {Tsuji}, \citenamefont
  {Ueda},\ and\ \citenamefont {Kawakami}}]{Yamamoto21}%
  \BibitemOpen
  \bibfield  {author} {\bibinfo {author} {\bibfnamefont {K.}~\bibnamefont
  {Yamamoto}}, \bibinfo {author} {\bibfnamefont {M.}~\bibnamefont {Nakagawa}},
  \bibinfo {author} {\bibfnamefont {N.}~\bibnamefont {Tsuji}}, \bibinfo
  {author} {\bibfnamefont {M.}~\bibnamefont {Ueda}},\ and\ \bibinfo {author}
  {\bibfnamefont {N.}~\bibnamefont {Kawakami}},\ }\bibfield  {title} {\bibinfo
  {title} {{Collective Excitations and Nonequilibrium Phase Transition in
  Dissipative Fermionic Superfluids}},\ }\href
  {https://doi.org/10.1103/PhysRevLett.127.055301} {\bibfield  {journal}
  {\bibinfo  {journal} {Phys. Rev. Lett.}\ }\textbf {\bibinfo {volume} {127}},\
  \bibinfo {pages} {055301} (\bibinfo {year} {2021})}\BibitemShut {NoStop}%
\bibitem [{\citenamefont {Ren}\ \emph {et~al.}(2022)\citenamefont {Ren},
  \citenamefont {Liu}, \citenamefont {Zhao}, \citenamefont {He}, \citenamefont
  {Pak}, \citenamefont {Li},\ and\ \citenamefont {Jo}}]{Ren22}%
  \BibitemOpen
  \bibfield  {author} {\bibinfo {author} {\bibfnamefont {Z.}~\bibnamefont
  {Ren}}, \bibinfo {author} {\bibfnamefont {D.}~\bibnamefont {Liu}}, \bibinfo
  {author} {\bibfnamefont {E.}~\bibnamefont {Zhao}}, \bibinfo {author}
  {\bibfnamefont {C.}~\bibnamefont {He}}, \bibinfo {author} {\bibfnamefont
  {K.~K.}\ \bibnamefont {Pak}}, \bibinfo {author} {\bibfnamefont
  {J.}~\bibnamefont {Li}},\ and\ \bibinfo {author} {\bibfnamefont {G.-B.}\
  \bibnamefont {Jo}},\ }\bibfield  {title} {\bibinfo {title} {{Chiral control
  of quantum states in non-Hermitian spin--orbit-coupled fermions}},\
  }\href@noop {} {\bibfield  {journal} {\bibinfo  {journal} {Nat. Phys.}\
  }\textbf {\bibinfo {volume} {18}},\ \bibinfo {pages} {385} (\bibinfo {year}
  {2022})}\BibitemShut {NoStop}%
\bibitem [{\citenamefont {Liang}\ \emph {et~al.}(2022)\citenamefont {Liang},
  \citenamefont {Xie}, \citenamefont {Dong}, \citenamefont {Li}, \citenamefont
  {Li}, \citenamefont {Gadway}, \citenamefont {Yi},\ and\ \citenamefont
  {Yan}}]{Liang22}%
  \BibitemOpen
  \bibfield  {author} {\bibinfo {author} {\bibfnamefont {Q.}~\bibnamefont
  {Liang}}, \bibinfo {author} {\bibfnamefont {D.}~\bibnamefont {Xie}}, \bibinfo
  {author} {\bibfnamefont {Z.}~\bibnamefont {Dong}}, \bibinfo {author}
  {\bibfnamefont {H.}~\bibnamefont {Li}}, \bibinfo {author} {\bibfnamefont
  {H.}~\bibnamefont {Li}}, \bibinfo {author} {\bibfnamefont {B.}~\bibnamefont
  {Gadway}}, \bibinfo {author} {\bibfnamefont {W.}~\bibnamefont {Yi}},\ and\
  \bibinfo {author} {\bibfnamefont {B.}~\bibnamefont {Yan}},\ }\bibfield
  {title} {\bibinfo {title} {{Dynamic Signatures of Non-Hermitian Skin Effect
  and Topology in Ultracold Atoms}},\ }\href
  {https://doi.org/10.1103/PhysRevLett.129.070401} {\bibfield  {journal}
  {\bibinfo  {journal} {Phys. Rev. Lett.}\ }\textbf {\bibinfo {volume} {129}},\
  \bibinfo {pages} {070401} (\bibinfo {year} {2022})}\BibitemShut {NoStop}%
\bibitem [{\citenamefont {Tsuno}\ \emph {et~al.}()\citenamefont {Tsuno},
  \citenamefont {Taie}, \citenamefont {Takasu}, \citenamefont {Yamashita},
  \citenamefont {Ozawa},\ and\ \citenamefont {Takahashi}}]{Tsuno24}%
  \BibitemOpen
  \bibfield  {author} {\bibinfo {author} {\bibfnamefont {T.}~\bibnamefont
  {Tsuno}}, \bibinfo {author} {\bibfnamefont {S.}~\bibnamefont {Taie}},
  \bibinfo {author} {\bibfnamefont {Y.}~\bibnamefont {Takasu}}, \bibinfo
  {author} {\bibfnamefont {K.}~\bibnamefont {Yamashita}}, \bibinfo {author}
  {\bibfnamefont {T.}~\bibnamefont {Ozawa}},\ and\ \bibinfo {author}
  {\bibfnamefont {Y.}~\bibnamefont {Takahashi}},\ }\bibfield  {title} {\bibinfo
  {title} {Gain engineering and topological atom laser in synthetic
  dimensions},\ }\href@noop {} {\bibinfo  {journal} {arXiv:2404.13769}\
  }\BibitemShut {NoStop}%
\bibitem [{\citenamefont {Zhao}\ \emph {et~al.}(2025)\citenamefont {Zhao},
  \citenamefont {Wang}, \citenamefont {He}, \citenamefont {Poon}, \citenamefont
  {Pak}, \citenamefont {Liu}, \citenamefont {Ren}, \citenamefont {Liu},\ and\
  \citenamefont {Jo}}]{Jo25}%
  \BibitemOpen
\bibfield  {journal} {  }\bibfield  {author} {\bibinfo {author} {\bibfnamefont
  {E.}~\bibnamefont {Zhao}}, \bibinfo {author} {\bibfnamefont {Z.}~\bibnamefont
  {Wang}}, \bibinfo {author} {\bibfnamefont {C.}~\bibnamefont {He}}, \bibinfo
  {author} {\bibfnamefont {T.~F.~J.}\ \bibnamefont {Poon}}, \bibinfo {author}
  {\bibfnamefont {K.~K.}\ \bibnamefont {Pak}}, \bibinfo {author} {\bibfnamefont
  {Y.-J.}\ \bibnamefont {Liu}}, \bibinfo {author} {\bibfnamefont
  {P.}~\bibnamefont {Ren}}, \bibinfo {author} {\bibfnamefont {X.-J.}\
  \bibnamefont {Liu}},\ and\ \bibinfo {author} {\bibfnamefont {G.-B.}\
  \bibnamefont {Jo}},\ }\bibfield  {title} {\bibinfo {title} {{Two-dimensional
  non-Hermitian skin effect in an ultracold Fermi gas}},\ }\href@noop {}
  {\bibfield  {journal} {\bibinfo  {journal} {Nature}\ }\textbf {\bibinfo
  {volume} {637}},\ \bibinfo {pages} {565} (\bibinfo {year}
  {2025})}\BibitemShut {NoStop}%
\bibitem [{\citenamefont {Yamamoto}\ \emph {et~al.}(2019)\citenamefont
  {Yamamoto}, \citenamefont {Nakagawa}, \citenamefont {Adachi}, \citenamefont
  {Takasan}, \citenamefont {Ueda},\ and\ \citenamefont
  {Kawakami}}]{Yamamoto19}%
  \BibitemOpen
  \bibfield  {author} {\bibinfo {author} {\bibfnamefont {K.}~\bibnamefont
  {Yamamoto}}, \bibinfo {author} {\bibfnamefont {M.}~\bibnamefont {Nakagawa}},
  \bibinfo {author} {\bibfnamefont {K.}~\bibnamefont {Adachi}}, \bibinfo
  {author} {\bibfnamefont {K.}~\bibnamefont {Takasan}}, \bibinfo {author}
  {\bibfnamefont {M.}~\bibnamefont {Ueda}},\ and\ \bibinfo {author}
  {\bibfnamefont {N.}~\bibnamefont {Kawakami}},\ }\bibfield  {title} {\bibinfo
  {title} {{Theory of Non-Hermitian Fermionic Superfluidity with a
  Complex-Valued Interaction}},\ }\href
  {https://doi.org/10.1103/PhysRevLett.123.123601} {\bibfield  {journal}
  {\bibinfo  {journal} {Phys. Rev. Lett.}\ }\textbf {\bibinfo {volume} {123}},\
  \bibinfo {pages} {123601} (\bibinfo {year} {2019})}\BibitemShut {NoStop}%
\bibitem [{\citenamefont {Hamazaki}\ \emph {et~al.}(2019)\citenamefont
  {Hamazaki}, \citenamefont {Kawabata},\ and\ \citenamefont
  {Ueda}}]{Hamazaki19}%
  \BibitemOpen
  \bibfield  {author} {\bibinfo {author} {\bibfnamefont {R.}~\bibnamefont
  {Hamazaki}}, \bibinfo {author} {\bibfnamefont {K.}~\bibnamefont {Kawabata}},\
  and\ \bibinfo {author} {\bibfnamefont {M.}~\bibnamefont {Ueda}},\ }\bibfield
  {title} {\bibinfo {title} {{Non-Hermitian Many-Body Localization}},\ }\href
  {https://doi.org/10.1103/PhysRevLett.123.090603} {\bibfield  {journal}
  {\bibinfo  {journal} {Phys. Rev. Lett.}\ }\textbf {\bibinfo {volume} {123}},\
  \bibinfo {pages} {090603} (\bibinfo {year} {2019})}\BibitemShut {NoStop}%
\bibitem [{\citenamefont {Ashida}\ \emph {et~al.}(2016)\citenamefont {Ashida},
  \citenamefont {Furukawa},\ and\ \citenamefont {Ueda}}]{Ashida16}%
  \BibitemOpen
  \bibfield  {author} {\bibinfo {author} {\bibfnamefont {Y.}~\bibnamefont
  {Ashida}}, \bibinfo {author} {\bibfnamefont {S.}~\bibnamefont {Furukawa}},\
  and\ \bibinfo {author} {\bibfnamefont {M.}~\bibnamefont {Ueda}},\ }\bibfield
  {title} {\bibinfo {title} {Quantum critical behavior influenced by
  measurement backaction in ultracold gases},\ }\href
  {https://doi.org/10.1103/PhysRevA.94.053615} {\bibfield  {journal} {\bibinfo
  {journal} {Phys. Rev. A}\ }\textbf {\bibinfo {volume} {94}},\ \bibinfo
  {pages} {053615} (\bibinfo {year} {2016})}\BibitemShut {NoStop}%
\bibitem [{\citenamefont {Ashida}\ \emph {et~al.}(2017)\citenamefont {Ashida},
  \citenamefont {Furukawa},\ and\ \citenamefont {Ueda}}]{Ashida17}%
  \BibitemOpen
  \bibfield  {author} {\bibinfo {author} {\bibfnamefont {Y.}~\bibnamefont
  {Ashida}}, \bibinfo {author} {\bibfnamefont {S.}~\bibnamefont {Furukawa}},\
  and\ \bibinfo {author} {\bibfnamefont {M.}~\bibnamefont {Ueda}},\ }\bibfield
  {title} {\bibinfo {title} {Parity-time-symmetric quantum critical
  phenomena},\ }\href {https://doi.org/10.1038/ncomms15791} {\bibfield
  {journal} {\bibinfo  {journal} {Nat. Commun.}\ }\textbf {\bibinfo {volume}
  {8}},\ \bibinfo {pages} {15791} (\bibinfo {year} {2017})}\BibitemShut
  {NoStop}%
\bibitem [{\citenamefont {Xu}\ and\ \citenamefont {Chen}(2020)}]{Xu20}%
  \BibitemOpen
  \bibfield  {author} {\bibinfo {author} {\bibfnamefont {Z.}~\bibnamefont
  {Xu}}\ and\ \bibinfo {author} {\bibfnamefont {S.}~\bibnamefont {Chen}},\
  }\bibfield  {title} {\bibinfo {title} {{Topological Bose-Mott insulators in
  one-dimensional non-Hermitian superlattices}},\ }\href
  {https://doi.org/10.1103/PhysRevB.102.035153} {\bibfield  {journal} {\bibinfo
   {journal} {Phys. Rev. B}\ }\textbf {\bibinfo {volume} {102}},\ \bibinfo
  {pages} {035153} (\bibinfo {year} {2020})}\BibitemShut {NoStop}%
\bibitem [{\citenamefont {Zhang}\ \emph {et~al.}(2020)\citenamefont {Zhang},
  \citenamefont {Chen}, \citenamefont {Zhang}, \citenamefont {Lang},
  \citenamefont {Li},\ and\ \citenamefont {Zhu}}]{Zhang20}%
  \BibitemOpen
  \bibfield  {author} {\bibinfo {author} {\bibfnamefont {D.-W.}\ \bibnamefont
  {Zhang}}, \bibinfo {author} {\bibfnamefont {Y.-L.}\ \bibnamefont {Chen}},
  \bibinfo {author} {\bibfnamefont {G.-Q.}\ \bibnamefont {Zhang}}, \bibinfo
  {author} {\bibfnamefont {L.-J.}\ \bibnamefont {Lang}}, \bibinfo {author}
  {\bibfnamefont {Z.}~\bibnamefont {Li}},\ and\ \bibinfo {author}
  {\bibfnamefont {S.-L.}\ \bibnamefont {Zhu}},\ }\bibfield  {title} {\bibinfo
  {title} {{Skin superfluid, topological Mott insulators, and asymmetric
  dynamics in an interacting non-Hermitian Aubry-Andr\'e-Harper model}},\
  }\href {https://doi.org/10.1103/PhysRevB.101.235150} {\bibfield  {journal}
  {\bibinfo  {journal} {Phys. Rev. B}\ }\textbf {\bibinfo {volume} {101}},\
  \bibinfo {pages} {235150} (\bibinfo {year} {2020})}\BibitemShut {NoStop}%
\bibitem [{\citenamefont {Liu}\ \emph {et~al.}(2020)\citenamefont {Liu},
  \citenamefont {He}, \citenamefont {Yoshida}, \citenamefont {Xiang},\ and\
  \citenamefont {Nori}}]{Liu20}%
  \BibitemOpen
  \bibfield  {author} {\bibinfo {author} {\bibfnamefont {T.}~\bibnamefont
  {Liu}}, \bibinfo {author} {\bibfnamefont {J.~J.}\ \bibnamefont {He}},
  \bibinfo {author} {\bibfnamefont {T.}~\bibnamefont {Yoshida}}, \bibinfo
  {author} {\bibfnamefont {Z.-L.}\ \bibnamefont {Xiang}},\ and\ \bibinfo
  {author} {\bibfnamefont {F.}~\bibnamefont {Nori}},\ }\bibfield  {title}
  {\bibinfo {title} {{Non-Hermitian topological Mott insulators in
  one-dimensional fermionic superlattices}},\ }\href
  {https://doi.org/10.1103/PhysRevB.102.235151} {\bibfield  {journal} {\bibinfo
   {journal} {Phys. Rev. B}\ }\textbf {\bibinfo {volume} {102}},\ \bibinfo
  {pages} {235151} (\bibinfo {year} {2020})}\BibitemShut {NoStop}%
\bibitem [{\citenamefont {Yamamoto}\ \emph {et~al.}(2022)\citenamefont
  {Yamamoto}, \citenamefont {Nakagawa}, \citenamefont {Tezuka}, \citenamefont
  {Ueda},\ and\ \citenamefont {Kawakami}}]{Yamamoto22}%
  \BibitemOpen
  \bibfield  {author} {\bibinfo {author} {\bibfnamefont {K.}~\bibnamefont
  {Yamamoto}}, \bibinfo {author} {\bibfnamefont {M.}~\bibnamefont {Nakagawa}},
  \bibinfo {author} {\bibfnamefont {M.}~\bibnamefont {Tezuka}}, \bibinfo
  {author} {\bibfnamefont {M.}~\bibnamefont {Ueda}},\ and\ \bibinfo {author}
  {\bibfnamefont {N.}~\bibnamefont {Kawakami}},\ }\bibfield  {title} {\bibinfo
  {title} {{Universal properties of dissipative Tomonaga-Luttinger liquids:
  Case study of a non-Hermitian XXZ spin chain}},\ }\href
  {https://doi.org/10.1103/PhysRevB.105.205125} {\bibfield  {journal} {\bibinfo
   {journal} {Phys. Rev. B}\ }\textbf {\bibinfo {volume} {105}},\ \bibinfo
  {pages} {205125} (\bibinfo {year} {2022})}\BibitemShut {NoStop}%
\bibitem [{\citenamefont {Yamamoto}\ and\ \citenamefont
  {Kawakami}(2023)}]{Yamamoto23}%
  \BibitemOpen
  \bibfield  {author} {\bibinfo {author} {\bibfnamefont {K.}~\bibnamefont
  {Yamamoto}}\ and\ \bibinfo {author} {\bibfnamefont {N.}~\bibnamefont
  {Kawakami}},\ }\bibfield  {title} {\bibinfo {title} {{Universal description
  of dissipative Tomonaga-Luttinger liquids with $\mathrm{SU}(N)$ spin
  symmetry: Exact spectrum and critical exponents}},\ }\href
  {https://doi.org/10.1103/PhysRevB.107.045110} {\bibfield  {journal} {\bibinfo
   {journal} {Phys. Rev. B}\ }\textbf {\bibinfo {volume} {107}},\ \bibinfo
  {pages} {045110} (\bibinfo {year} {2023})}\BibitemShut {NoStop}%
\bibitem [{\citenamefont {Kawabata}\ \emph {et~al.}(2022)\citenamefont
  {Kawabata}, \citenamefont {Shiozaki},\ and\ \citenamefont
  {Ryu}}]{Kawabata22}%
  \BibitemOpen
  \bibfield  {author} {\bibinfo {author} {\bibfnamefont {K.}~\bibnamefont
  {Kawabata}}, \bibinfo {author} {\bibfnamefont {K.}~\bibnamefont {Shiozaki}},\
  and\ \bibinfo {author} {\bibfnamefont {S.}~\bibnamefont {Ryu}},\ }\bibfield
  {title} {\bibinfo {title} {{Many-body topology of non-Hermitian systems}},\
  }\href {https://doi.org/10.1103/PhysRevB.105.165137} {\bibfield  {journal}
  {\bibinfo  {journal} {Phys. Rev. B}\ }\textbf {\bibinfo {volume} {105}},\
  \bibinfo {pages} {165137} (\bibinfo {year} {2022})}\BibitemShut {NoStop}%
\bibitem [{\citenamefont {Yoshida}\ \emph {et~al.}(2024)\citenamefont
  {Yoshida}, \citenamefont {Zhang}, \citenamefont {Neupert},\ and\
  \citenamefont {Kawakami}}]{TYoshida24}%
  \BibitemOpen
  \bibfield  {author} {\bibinfo {author} {\bibfnamefont {T.}~\bibnamefont
  {Yoshida}}, \bibinfo {author} {\bibfnamefont {S.-B.}\ \bibnamefont {Zhang}},
  \bibinfo {author} {\bibfnamefont {T.}~\bibnamefont {Neupert}},\ and\ \bibinfo
  {author} {\bibfnamefont {N.}~\bibnamefont {Kawakami}},\ }\bibfield  {title}
  {\bibinfo {title} {{Non-Hermitian Mott Skin Effect}},\ }\href
  {https://doi.org/10.1103/PhysRevLett.133.076502} {\bibfield  {journal}
  {\bibinfo  {journal} {Phys. Rev. Lett.}\ }\textbf {\bibinfo {volume} {133}},\
  \bibinfo {pages} {076502} (\bibinfo {year} {2024})}\BibitemShut {NoStop}%
\bibitem [{\citenamefont {Kim}\ \emph {et~al.}(2024)\citenamefont {Kim},
  \citenamefont {Han},\ and\ \citenamefont {Park}}]{Kim24}%
  \BibitemOpen
  \bibfield  {author} {\bibinfo {author} {\bibfnamefont {B.~H.}\ \bibnamefont
  {Kim}}, \bibinfo {author} {\bibfnamefont {J.-H.}\ \bibnamefont {Han}},\ and\
  \bibinfo {author} {\bibfnamefont {M.~J.}\ \bibnamefont {Park}},\ }\bibfield
  {title} {\bibinfo {title} {{Collective non-Hermitian skin effect: point-gap
  topology and the doublon-holon excitations in non-reciprocal many-body
  systems}},\ }\href@noop {} {\bibfield  {journal} {\bibinfo  {journal}
  {Commun. Phys.}\ }\textbf {\bibinfo {volume} {7}},\ \bibinfo {pages} {73}
  (\bibinfo {year} {2024})}\BibitemShut {NoStop}%
\bibitem [{\citenamefont {Takemori}\ \emph {et~al.}(2025)\citenamefont
  {Takemori}, \citenamefont {Yamamoto},\ and\ \citenamefont
  {Koga}}]{Takemori25}%
  \BibitemOpen
  \bibfield  {author} {\bibinfo {author} {\bibfnamefont {S.}~\bibnamefont
  {Takemori}}, \bibinfo {author} {\bibfnamefont {K.}~\bibnamefont {Yamamoto}},\
  and\ \bibinfo {author} {\bibfnamefont {A.}~\bibnamefont {Koga}},\ }\bibfield
  {title} {\bibinfo {title} {{Spin-Depairing-Induced Exceptional Fermionic
  Superfluidity}},\ }\href {https://doi.org/10.1103/ntjf-zb2v} {\bibfield
  {journal} {\bibinfo  {journal} {Phys. Rev. Lett.}\ }\textbf {\bibinfo
  {volume} {135}},\ \bibinfo {pages} {266002} (\bibinfo {year}
  {2025})}\BibitemShut {NoStop}%
\bibitem [{\citenamefont {Takemori}\ \emph {et~al.}(2026)\citenamefont
  {Takemori}, \citenamefont {Yamamoto},\ and\ \citenamefont
  {Koga}}]{Takemori25L}%
  \BibitemOpen
  \bibfield  {author} {\bibinfo {author} {\bibfnamefont {S.}~\bibnamefont
  {Takemori}}, \bibinfo {author} {\bibfnamefont {K.}~\bibnamefont {Yamamoto}},\
  and\ \bibinfo {author} {\bibfnamefont {A.}~\bibnamefont {Koga}},\ }\bibfield
  {title} {\bibinfo {title} {Dimensionality effect on exceptional fermionic
  superfluidity with spin-dependent asymmetric hopping},\ }\href@noop {}
  {\bibfield  {journal} {\bibinfo  {journal} {J. Low Temp. Phys.}\ }\textbf
  {\bibinfo {volume} {222}},\ \bibinfo {pages} {37} (\bibinfo {year}
  {2026})}\BibitemShut {NoStop}%
\bibitem [{\citenamefont {Hanai}\ \emph {et~al.}(2019)\citenamefont {Hanai},
  \citenamefont {Edelman}, \citenamefont {Ohashi},\ and\ \citenamefont
  {Littlewood}}]{Hanai19}%
  \BibitemOpen
  \bibfield  {author} {\bibinfo {author} {\bibfnamefont {R.}~\bibnamefont
  {Hanai}}, \bibinfo {author} {\bibfnamefont {A.}~\bibnamefont {Edelman}},
  \bibinfo {author} {\bibfnamefont {Y.}~\bibnamefont {Ohashi}},\ and\ \bibinfo
  {author} {\bibfnamefont {P.~B.}\ \bibnamefont {Littlewood}},\ }\bibfield
  {title} {\bibinfo {title} {{Non-Hermitian Phase Transition from a Polariton
  Bose-Einstein Condensate to a Photon Laser}},\ }\href
  {https://doi.org/10.1103/PhysRevLett.122.185301} {\bibfield  {journal}
  {\bibinfo  {journal} {Phys. Rev. Lett.}\ }\textbf {\bibinfo {volume} {122}},\
  \bibinfo {pages} {185301} (\bibinfo {year} {2019})}\BibitemShut {NoStop}%
\bibitem [{\citenamefont {Gopalakrishnan}\ and\ \citenamefont
  {Gullans}(2021)}]{Gopa21}%
  \BibitemOpen
  \bibfield  {author} {\bibinfo {author} {\bibfnamefont {S.}~\bibnamefont
  {Gopalakrishnan}}\ and\ \bibinfo {author} {\bibfnamefont {M.~J.}\
  \bibnamefont {Gullans}},\ }\bibfield  {title} {\bibinfo {title}
  {{Entanglement and Purification Transitions in Non-Hermitian Quantum
  Mechanics}},\ }\href {https://doi.org/10.1103/PhysRevLett.126.170503}
  {\bibfield  {journal} {\bibinfo  {journal} {Phys. Rev. Lett.}\ }\textbf
  {\bibinfo {volume} {126}},\ \bibinfo {pages} {170503} (\bibinfo {year}
  {2021})}\BibitemShut {NoStop}%
\bibitem [{\citenamefont {Yang}\ \emph {et~al.}(2021)\citenamefont {Yang},
  \citenamefont {Morampudi},\ and\ \citenamefont {Bergholtz}}]{Yang21}%
  \BibitemOpen
  \bibfield  {author} {\bibinfo {author} {\bibfnamefont {K.}~\bibnamefont
  {Yang}}, \bibinfo {author} {\bibfnamefont {S.~C.}\ \bibnamefont
  {Morampudi}},\ and\ \bibinfo {author} {\bibfnamefont {E.~J.}\ \bibnamefont
  {Bergholtz}},\ }\bibfield  {title} {\bibinfo {title} {{Exceptional Spin
  Liquids from Couplings to the Environment}},\ }\href
  {https://doi.org/10.1103/PhysRevLett.126.077201} {\bibfield  {journal}
  {\bibinfo  {journal} {Phys. Rev. Lett.}\ }\textbf {\bibinfo {volume} {126}},\
  \bibinfo {pages} {077201} (\bibinfo {year} {2021})}\BibitemShut {NoStop}%
\bibitem [{\citenamefont {Sun}\ \emph {et~al.}(2023)\citenamefont {Sun},
  \citenamefont {Shi}, \citenamefont {Liu}, \citenamefont {Zhang},
  \citenamefont {Xiao}, \citenamefont {Jia},\ and\ \citenamefont {Hu}}]{Sun23}%
  \BibitemOpen
  \bibfield  {author} {\bibinfo {author} {\bibfnamefont {Y.}~\bibnamefont
  {Sun}}, \bibinfo {author} {\bibfnamefont {T.}~\bibnamefont {Shi}}, \bibinfo
  {author} {\bibfnamefont {Z.}~\bibnamefont {Liu}}, \bibinfo {author}
  {\bibfnamefont {Z.}~\bibnamefont {Zhang}}, \bibinfo {author} {\bibfnamefont
  {L.}~\bibnamefont {Xiao}}, \bibinfo {author} {\bibfnamefont {S.}~\bibnamefont
  {Jia}},\ and\ \bibinfo {author} {\bibfnamefont {Y.}~\bibnamefont {Hu}},\
  }\bibfield  {title} {\bibinfo {title} {{Fractional Quantum Zeno Effect
  Emerging from Non-Hermitian Physics}},\ }\href
  {https://doi.org/10.1103/PhysRevX.13.031009} {\bibfield  {journal} {\bibinfo
  {journal} {Phys. Rev. X}\ }\textbf {\bibinfo {volume} {13}},\ \bibinfo
  {pages} {031009} (\bibinfo {year} {2023})}\BibitemShut {NoStop}%
\bibitem [{\citenamefont {Yu}\ \emph {et~al.}(2024)\citenamefont {Yu},
  \citenamefont {Pan}, \citenamefont {Xu},\ and\ \citenamefont {Li}}]{Yu24}%
  \BibitemOpen
  \bibfield  {author} {\bibinfo {author} {\bibfnamefont {X.-J.}\ \bibnamefont
  {Yu}}, \bibinfo {author} {\bibfnamefont {Z.}~\bibnamefont {Pan}}, \bibinfo
  {author} {\bibfnamefont {L.}~\bibnamefont {Xu}},\ and\ \bibinfo {author}
  {\bibfnamefont {Z.-X.}\ \bibnamefont {Li}},\ }\bibfield  {title} {\bibinfo
  {title} {{Non-Hermitian Strongly Interacting Dirac Fermions}},\ }\href
  {https://doi.org/10.1103/PhysRevLett.132.116503} {\bibfield  {journal}
  {\bibinfo  {journal} {Phys. Rev. Lett.}\ }\textbf {\bibinfo {volume} {132}},\
  \bibinfo {pages} {116503} (\bibinfo {year} {2024})}\BibitemShut {NoStop}%
\bibitem [{\citenamefont {Yamamoto}\ and\ \citenamefont
  {Hamazaki}()}]{Yamamoto26}%
  \BibitemOpen
  \bibfield  {author} {\bibinfo {author} {\bibfnamefont {K.}~\bibnamefont
  {Yamamoto}}\ and\ \bibinfo {author} {\bibfnamefont {R.}~\bibnamefont
  {Hamazaki}},\ }\bibfield  {title} {\bibinfo {title} {Anomalous waiting-time
  distributions in postselection-free quantum many-body dynamics under
  continuous monitoring},\ }\href@noop {} {\bibinfo  {journal}
  {arXiv:2604.00358}\ }\BibitemShut {NoStop}%
\bibitem [{\citenamefont {Yamamoto}\ \emph {et~al.}(2025)\citenamefont
  {Yamamoto}, \citenamefont {Nakagawa},\ and\ \citenamefont
  {Kawakami}}]{Yamamoto24}%
  \BibitemOpen
\bibfield  {journal} {  }\bibfield  {author} {\bibinfo {author} {\bibfnamefont
  {K.}~\bibnamefont {Yamamoto}}, \bibinfo {author} {\bibfnamefont
  {M.}~\bibnamefont {Nakagawa}},\ and\ \bibinfo {author} {\bibfnamefont
  {N.}~\bibnamefont {Kawakami}},\ }\bibfield  {title} {\bibinfo {title}
  {{Correlation versus dissipation in a non-Hermitian Anderson impurity
  model}},\ }\href {https://doi.org/10.1103/PhysRevB.111.125157} {\bibfield
  {journal} {\bibinfo  {journal} {Phys. Rev. B}\ }\textbf {\bibinfo {volume}
  {111}},\ \bibinfo {pages} {125157} (\bibinfo {year} {2025})}\BibitemShut
  {NoStop}%
\bibitem [{\citenamefont {Gorshkov}\ \emph {et~al.}(2010)\citenamefont
  {Gorshkov}, \citenamefont {Hermele}, \citenamefont {Gurarie}, \citenamefont
  {Xu}, \citenamefont {Julienne}, \citenamefont {Ye}, \citenamefont {Zoller},
  \citenamefont {Demler}, \citenamefont {Lukin},\ and\ \citenamefont
  {Rey}}]{Gorshkov10}%
  \BibitemOpen
  \bibfield  {author} {\bibinfo {author} {\bibfnamefont {A.~V.}\ \bibnamefont
  {Gorshkov}}, \bibinfo {author} {\bibfnamefont {M.}~\bibnamefont {Hermele}},
  \bibinfo {author} {\bibfnamefont {V.}~\bibnamefont {Gurarie}}, \bibinfo
  {author} {\bibfnamefont {C.}~\bibnamefont {Xu}}, \bibinfo {author}
  {\bibfnamefont {P.~S.}\ \bibnamefont {Julienne}}, \bibinfo {author}
  {\bibfnamefont {J.}~\bibnamefont {Ye}}, \bibinfo {author} {\bibfnamefont
  {P.}~\bibnamefont {Zoller}}, \bibinfo {author} {\bibfnamefont
  {E.}~\bibnamefont {Demler}}, \bibinfo {author} {\bibfnamefont {M.~D.}\
  \bibnamefont {Lukin}},\ and\ \bibinfo {author} {\bibfnamefont
  {A.}~\bibnamefont {Rey}},\ }\bibfield  {title} {\bibinfo {title}
  {{Two-orbital SU (N) magnetism with ultracold alkaline-earth atoms}},\
  }\href@noop {} {\bibfield  {journal} {\bibinfo  {journal} {Nat. Phys.}\
  }\textbf {\bibinfo {volume} {6}},\ \bibinfo {pages} {289} (\bibinfo {year}
  {2010})}\BibitemShut {NoStop}%
\bibitem [{\citenamefont {Foss-Feig}\ \emph
  {et~al.}(2010{\natexlab{a}})\citenamefont {Foss-Feig}, \citenamefont
  {Hermele},\ and\ \citenamefont {Rey}}]{ARey10}%
  \BibitemOpen
  \bibfield  {author} {\bibinfo {author} {\bibfnamefont {M.}~\bibnamefont
  {Foss-Feig}}, \bibinfo {author} {\bibfnamefont {M.}~\bibnamefont {Hermele}},\
  and\ \bibinfo {author} {\bibfnamefont {A.~M.}\ \bibnamefont {Rey}},\
  }\bibfield  {title} {\bibinfo {title} {{Probing the Kondo lattice model with
  alkaline-earth-metal atoms}},\ }\href
  {https://doi.org/10.1103/PhysRevA.81.051603} {\bibfield  {journal} {\bibinfo
  {journal} {Phys. Rev. A}\ }\textbf {\bibinfo {volume} {81}},\ \bibinfo
  {pages} {051603} (\bibinfo {year} {2010}{\natexlab{a}})}\BibitemShut
  {NoStop}%
\bibitem [{\citenamefont {Foss-Feig}\ \emph
  {et~al.}(2010{\natexlab{b}})\citenamefont {Foss-Feig}, \citenamefont
  {Hermele}, \citenamefont {Gurarie},\ and\ \citenamefont {Rey}}]{ARey10A}%
  \BibitemOpen
  \bibfield  {author} {\bibinfo {author} {\bibfnamefont {M.}~\bibnamefont
  {Foss-Feig}}, \bibinfo {author} {\bibfnamefont {M.}~\bibnamefont {Hermele}},
  \bibinfo {author} {\bibfnamefont {V.}~\bibnamefont {Gurarie}},\ and\ \bibinfo
  {author} {\bibfnamefont {A.~M.}\ \bibnamefont {Rey}},\ }\bibfield  {title}
  {\bibinfo {title} {Heavy fermions in an optical lattice},\ }\href
  {https://doi.org/10.1103/PhysRevA.82.053624} {\bibfield  {journal} {\bibinfo
  {journal} {Phys. Rev. A}\ }\textbf {\bibinfo {volume} {82}},\ \bibinfo
  {pages} {053624} (\bibinfo {year} {2010}{\natexlab{b}})}\BibitemShut
  {NoStop}%
\bibitem [{\citenamefont {Bauer}\ \emph {et~al.}(2013)\citenamefont {Bauer},
  \citenamefont {Salomon},\ and\ \citenamefont {Demler}}]{Demler13}%
  \BibitemOpen
  \bibfield  {author} {\bibinfo {author} {\bibfnamefont {J.}~\bibnamefont
  {Bauer}}, \bibinfo {author} {\bibfnamefont {C.}~\bibnamefont {Salomon}},\
  and\ \bibinfo {author} {\bibfnamefont {E.}~\bibnamefont {Demler}},\
  }\bibfield  {title} {\bibinfo {title} {{Realizing a Kondo-Correlated State
  with Ultracold Atoms}},\ }\href
  {https://doi.org/10.1103/PhysRevLett.111.215304} {\bibfield  {journal}
  {\bibinfo  {journal} {Phys. Rev. Lett.}\ }\textbf {\bibinfo {volume} {111}},\
  \bibinfo {pages} {215304} (\bibinfo {year} {2013})}\BibitemShut {NoStop}%
\bibitem [{\citenamefont {Isaev}\ and\ \citenamefont {Rey}(2015)}]{ARey15}%
  \BibitemOpen
  \bibfield  {author} {\bibinfo {author} {\bibfnamefont {L.}~\bibnamefont
  {Isaev}}\ and\ \bibinfo {author} {\bibfnamefont {A.~M.}\ \bibnamefont
  {Rey}},\ }\bibfield  {title} {\bibinfo {title} {{Heavy-Fermion Valence-Bond
  Liquids in Ultracold Atoms: Cooperation of the Kondo Effect and Geometric
  Frustration}},\ }\href {https://doi.org/10.1103/PhysRevLett.115.165302}
  {\bibfield  {journal} {\bibinfo  {journal} {Phys. Rev. Lett.}\ }\textbf
  {\bibinfo {volume} {115}},\ \bibinfo {pages} {165302} (\bibinfo {year}
  {2015})}\BibitemShut {NoStop}%
\bibitem [{\citenamefont {Nakagawa}\ and\ \citenamefont
  {Kawakami}(2015)}]{Nakagawa15}%
  \BibitemOpen
  \bibfield  {author} {\bibinfo {author} {\bibfnamefont {M.}~\bibnamefont
  {Nakagawa}}\ and\ \bibinfo {author} {\bibfnamefont {N.}~\bibnamefont
  {Kawakami}},\ }\bibfield  {title} {\bibinfo {title} {{Laser-Induced Kondo
  Effect in Ultracold Alkaline-Earth Fermions}},\ }\href
  {https://doi.org/10.1103/PhysRevLett.115.165303} {\bibfield  {journal}
  {\bibinfo  {journal} {Phys. Rev. Lett.}\ }\textbf {\bibinfo {volume} {115}},\
  \bibinfo {pages} {165303} (\bibinfo {year} {2015})}\BibitemShut {NoStop}%
\bibitem [{\citenamefont {Nishida}(2016)}]{Nishida16}%
  \BibitemOpen
  \bibfield  {author} {\bibinfo {author} {\bibfnamefont {Y.}~\bibnamefont
  {Nishida}},\ }\bibfield  {title} {\bibinfo {title} {{Transport measurement of
  the orbital Kondo effect with ultracold atoms}},\ }\href
  {https://doi.org/10.1103/PhysRevA.93.011606} {\bibfield  {journal} {\bibinfo
  {journal} {Phys. Rev. A}\ }\textbf {\bibinfo {volume} {93}},\ \bibinfo
  {pages} {011606} (\bibinfo {year} {2016})}\BibitemShut {NoStop}%
\bibitem [{\citenamefont {Zhang}\ \emph {et~al.}(2016)\citenamefont {Zhang},
  \citenamefont {Zhang}, \citenamefont {Cheng}, \citenamefont {Chen},
  \citenamefont {Zhang},\ and\ \citenamefont {Zhai}}]{Zhai16}%
  \BibitemOpen
  \bibfield  {author} {\bibinfo {author} {\bibfnamefont {R.}~\bibnamefont
  {Zhang}}, \bibinfo {author} {\bibfnamefont {D.}~\bibnamefont {Zhang}},
  \bibinfo {author} {\bibfnamefont {Y.}~\bibnamefont {Cheng}}, \bibinfo
  {author} {\bibfnamefont {W.}~\bibnamefont {Chen}}, \bibinfo {author}
  {\bibfnamefont {P.}~\bibnamefont {Zhang}},\ and\ \bibinfo {author}
  {\bibfnamefont {H.}~\bibnamefont {Zhai}},\ }\bibfield  {title} {\bibinfo
  {title} {Kondo effect in alkaline-earth-metal atomic gases with
  confinement-induced resonances},\ }\href
  {https://doi.org/10.1103/PhysRevA.93.043601} {\bibfield  {journal} {\bibinfo
  {journal} {Phys. Rev. A}\ }\textbf {\bibinfo {volume} {93}},\ \bibinfo
  {pages} {043601} (\bibinfo {year} {2016})}\BibitemShut {NoStop}%
\bibitem [{\citenamefont {Amaricci}\ \emph {et~al.}(2025)\citenamefont
  {Amaricci}, \citenamefont {Richaud}, \citenamefont {Capone}, \citenamefont
  {Darkwah~Oppong},\ and\ \citenamefont {Scazza}}]{Scazza25}%
  \BibitemOpen
  \bibfield  {author} {\bibinfo {author} {\bibfnamefont {A.}~\bibnamefont
  {Amaricci}}, \bibinfo {author} {\bibfnamefont {A.}~\bibnamefont {Richaud}},
  \bibinfo {author} {\bibfnamefont {M.}~\bibnamefont {Capone}}, \bibinfo
  {author} {\bibfnamefont {N.}~\bibnamefont {Darkwah~Oppong}},\ and\ \bibinfo
  {author} {\bibfnamefont {F.}~\bibnamefont {Scazza}},\ }\bibfield  {title}
  {\bibinfo {title} {Engineering the kondo impurity problem with
  alkaline-earth-atom arrays},\ }\href {https://doi.org/10.1103/m7l3-y2f8}
  {\bibfield  {journal} {\bibinfo  {journal} {Phys. Rev. A}\ }\textbf {\bibinfo
  {volume} {112}},\ \bibinfo {pages} {043301} (\bibinfo {year}
  {2025})}\BibitemShut {NoStop}%
\bibitem [{\citenamefont {Kan\'asz-Nagy}\ \emph {et~al.}(2018)\citenamefont
  {Kan\'asz-Nagy}, \citenamefont {Ashida}, \citenamefont {Shi}, \citenamefont
  {Moca}, \citenamefont {Ikeda}, \citenamefont {F\"olling}, \citenamefont
  {Cirac}, \citenamefont {Zar\'and},\ and\ \citenamefont {Demler}}]{Ashida18}%
  \BibitemOpen
  \bibfield  {author} {\bibinfo {author} {\bibfnamefont {M.}~\bibnamefont
  {Kan\'asz-Nagy}}, \bibinfo {author} {\bibfnamefont {Y.}~\bibnamefont
  {Ashida}}, \bibinfo {author} {\bibfnamefont {T.}~\bibnamefont {Shi}},
  \bibinfo {author} {\bibfnamefont {C.~P.}\ \bibnamefont {Moca}}, \bibinfo
  {author} {\bibfnamefont {T.~N.}\ \bibnamefont {Ikeda}}, \bibinfo {author}
  {\bibfnamefont {S.}~\bibnamefont {F\"olling}}, \bibinfo {author}
  {\bibfnamefont {J.~I.}\ \bibnamefont {Cirac}}, \bibinfo {author}
  {\bibfnamefont {G.}~\bibnamefont {Zar\'and}},\ and\ \bibinfo {author}
  {\bibfnamefont {E.~A.}\ \bibnamefont {Demler}},\ }\bibfield  {title}
  {\bibinfo {title} {{Exploring the anisotropic Kondo model in and out of
  equilibrium with alkaline-earth atoms}},\ }\href
  {https://doi.org/10.1103/PhysRevB.97.155156} {\bibfield  {journal} {\bibinfo
  {journal} {Phys. Rev. B}\ }\textbf {\bibinfo {volume} {97}},\ \bibinfo
  {pages} {155156} (\bibinfo {year} {2018})}\BibitemShut {NoStop}%
\bibitem [{\citenamefont {Vanhoecke}\ and\ \citenamefont
  {Schir\`o}(2024)}]{Schiro24}%
  \BibitemOpen
  \bibfield  {author} {\bibinfo {author} {\bibfnamefont {M.}~\bibnamefont
  {Vanhoecke}}\ and\ \bibinfo {author} {\bibfnamefont {M.}~\bibnamefont
  {Schir\`o}},\ }\bibfield  {title} {\bibinfo {title} {{Diagrammatic Monte
  Carlo for dissipative quantum impurity models}},\ }\href
  {https://doi.org/10.1103/PhysRevB.109.125125} {\bibfield  {journal} {\bibinfo
   {journal} {Phys. Rev. B}\ }\textbf {\bibinfo {volume} {109}},\ \bibinfo
  {pages} {125125} (\bibinfo {year} {2024})}\BibitemShut {NoStop}%
\bibitem [{\citenamefont {Vanhoecke}\ and\ \citenamefont
  {Schir\'o}()}]{Schiro24arXiv}%
  \BibitemOpen
  \bibfield  {author} {\bibinfo {author} {\bibfnamefont {M.}~\bibnamefont
  {Vanhoecke}}\ and\ \bibinfo {author} {\bibfnamefont {M.}~\bibnamefont
  {Schir\'o}},\ }\bibfield  {title} {\bibinfo {title} {Kondo-zeno crossover in
  the dynamics of a monitored quantum dot},\ }\href@noop {} {\bibinfo
  {journal} {arXiv:2405.17348}\ }\BibitemShut {NoStop}%
\bibitem [{\citenamefont {Stefanini}\ \emph {et~al.}(2025)\citenamefont
  {Stefanini}, \citenamefont {Qu}, \citenamefont {Esslinger}, \citenamefont
  {Gopalakrishnan}, \citenamefont {Demler},\ and\ \citenamefont
  {Marino}}]{Stefanini25}%
  \BibitemOpen
\bibfield  {journal} {  }\bibfield  {author} {\bibinfo {author} {\bibfnamefont
  {M.}~\bibnamefont {Stefanini}}, \bibinfo {author} {\bibfnamefont {Y.-F.}\
  \bibnamefont {Qu}}, \bibinfo {author} {\bibfnamefont {T.}~\bibnamefont
  {Esslinger}}, \bibinfo {author} {\bibfnamefont {S.}~\bibnamefont
  {Gopalakrishnan}}, \bibinfo {author} {\bibfnamefont {E.}~\bibnamefont
  {Demler}},\ and\ \bibinfo {author} {\bibfnamefont {J.}~\bibnamefont
  {Marino}},\ }\bibfield  {title} {\bibinfo {title} {{Dissipative realization
  of Kondo models}},\ }\href@noop {} {\bibfield  {journal} {\bibinfo  {journal}
  {Commun. Phys.}\ }\textbf {\bibinfo {volume} {8}},\ \bibinfo {pages} {1}
  (\bibinfo {year} {2025})}\BibitemShut {NoStop}%
\bibitem [{\citenamefont {Vanhoecke}\ \emph {et~al.}()\citenamefont
  {Vanhoecke}, \citenamefont {Tsuji},\ and\ \citenamefont
  {Schir{\`o}}}]{Vanh25}%
  \BibitemOpen
  \bibfield  {author} {\bibinfo {author} {\bibfnamefont {M.}~\bibnamefont
  {Vanhoecke}}, \bibinfo {author} {\bibfnamefont {N.}~\bibnamefont {Tsuji}},\
  and\ \bibinfo {author} {\bibfnamefont {M.}~\bibnamefont {Schir{\`o}}},\
  }\bibfield  {title} {\bibinfo {title} {{Dissipative Kondo physics in the
  Anderson Impurity Model with two-body losses}},\ }\href@noop {} {\bibinfo
  {journal} {arXiv:2506.22302}\ }\BibitemShut {NoStop}%
\bibitem [{\citenamefont {Werner}\ \emph {et~al.}()\citenamefont {Werner},
  \citenamefont {Vanhoecke}, \citenamefont {Schir{\`o}},\ and\ \citenamefont
  {Arrigoni}}]{Werner25}%
  \BibitemOpen
\bibfield  {journal} {  }\bibfield  {author} {\bibinfo {author} {\bibfnamefont
  {D.}~\bibnamefont {Werner}}, \bibinfo {author} {\bibfnamefont
  {M.}~\bibnamefont {Vanhoecke}}, \bibinfo {author} {\bibfnamefont
  {M.}~\bibnamefont {Schir{\`o}}},\ and\ \bibinfo {author} {\bibfnamefont
  {E.}~\bibnamefont {Arrigoni}},\ }\bibfield  {title} {\bibinfo {title}
  {Nonequilibrium transport through an interacting monitored quantum dot},\
  }\href@noop {} {\bibinfo  {journal} {arXiv:2507.20779}\ }\BibitemShut
  {NoStop}%
\bibitem [{\citenamefont {Qu}\ \emph {et~al.}(2025)\citenamefont {Qu},
  \citenamefont {Stefanini}, \citenamefont {Shi}, \citenamefont {Esslinger},
  \citenamefont {Gopalakrishnan}, \citenamefont {Marino},\ and\ \citenamefont
  {Demler}}]{Qu25}%
  \BibitemOpen
\bibfield  {journal} {  }\bibfield  {author} {\bibinfo {author} {\bibfnamefont
  {Y.-F.}\ \bibnamefont {Qu}}, \bibinfo {author} {\bibfnamefont
  {M.}~\bibnamefont {Stefanini}}, \bibinfo {author} {\bibfnamefont
  {T.}~\bibnamefont {Shi}}, \bibinfo {author} {\bibfnamefont {T.}~\bibnamefont
  {Esslinger}}, \bibinfo {author} {\bibfnamefont {S.}~\bibnamefont
  {Gopalakrishnan}}, \bibinfo {author} {\bibfnamefont {J.}~\bibnamefont
  {Marino}},\ and\ \bibinfo {author} {\bibfnamefont {E.}~\bibnamefont
  {Demler}},\ }\bibfield  {title} {\bibinfo {title} {Variational approach to
  the dynamics of dissipative quantum impurity models},\ }\href
  {https://doi.org/10.1103/PhysRevB.111.155113} {\bibfield  {journal} {\bibinfo
   {journal} {Phys. Rev. B}\ }\textbf {\bibinfo {volume} {111}},\ \bibinfo
  {pages} {155113} (\bibinfo {year} {2025})}\BibitemShut {NoStop}%
\bibitem [{\citenamefont {Hasegawa}\ \emph {et~al.}()\citenamefont {Hasegawa},
  \citenamefont {Nakagawa},\ and\ \citenamefont {Saito}}]{Hasegawa21}%
  \BibitemOpen
  \bibfield  {author} {\bibinfo {author} {\bibfnamefont {M.}~\bibnamefont
  {Hasegawa}}, \bibinfo {author} {\bibfnamefont {M.}~\bibnamefont {Nakagawa}},\
  and\ \bibinfo {author} {\bibfnamefont {K.}~\bibnamefont {Saito}},\ }\bibfield
   {title} {\bibinfo {title} {Kondo effect in a quantum dot under continuous
  quantum measurement},\ }\href@noop {} {\bibinfo  {journal}
  {arXiv:2111.07771}\ }\BibitemShut {NoStop}%
\bibitem [{\citenamefont {Han}\ \emph {et~al.}(2023)\citenamefont {Han},
  \citenamefont {Schultz},\ and\ \citenamefont {Kim}}]{Han23}%
  \BibitemOpen
\bibfield  {journal} {  }\bibfield  {author} {\bibinfo {author} {\bibfnamefont
  {S.}~\bibnamefont {Han}}, \bibinfo {author} {\bibfnamefont {D.~J.}\
  \bibnamefont {Schultz}},\ and\ \bibinfo {author} {\bibfnamefont {Y.~B.}\
  \bibnamefont {Kim}},\ }\bibfield  {title} {\bibinfo {title} {{Complex fixed
  points of the non-Hermitian Kondo model in a Luttinger liquid}},\ }\href
  {https://doi.org/10.1103/PhysRevB.107.235153} {\bibfield  {journal} {\bibinfo
   {journal} {Phys. Rev. B}\ }\textbf {\bibinfo {volume} {107}},\ \bibinfo
  {pages} {235153} (\bibinfo {year} {2023})}\BibitemShut {NoStop}%
\bibitem [{\citenamefont {Kulkarni}\ and\ \citenamefont
  {Vidhyadhiraja}()}]{Kulkarni25}%
  \BibitemOpen
  \bibfield  {author} {\bibinfo {author} {\bibfnamefont {V.~M.}\ \bibnamefont
  {Kulkarni}}\ and\ \bibinfo {author} {\bibfnamefont {N.}~\bibnamefont
  {Vidhyadhiraja}},\ }\bibfield  {title} {\bibinfo {title} {Anderson impurities
  in edge states with nonlinear and dissipative perturbations},\ }\href
  {https://scipost.org/submissions/scipost_202502_00039v1} {\bibinfo  {journal}
  {submitted to Scipost Phys.}\ }\BibitemShut {NoStop}%
\bibitem [{\citenamefont {Kulkarni}()}]{Kulkarni25light}%
  \BibitemOpen
\bibfield  {journal} {  }\bibfield  {author} {\bibinfo {author} {\bibfnamefont
  {V.~M.}\ \bibnamefont {Kulkarni}},\ }\bibfield  {title} {\bibinfo {title}
  {Light-driven bound state of interacting impurities in a dirac-like bath},\
  }\href@noop {} {\bibinfo  {journal} {arXiv:2505.17811}\ }\BibitemShut
  {NoStop}%
\bibitem [{\citenamefont {Kattel}\ \emph
  {et~al.}(2025{\natexlab{a}})\citenamefont {Kattel}, \citenamefont {Pasnoori},
  \citenamefont {Pixley},\ and\ \citenamefont {Andrei}}]{Kattel25}%
  \BibitemOpen
\bibfield  {journal} {  }\bibfield  {author} {\bibinfo {author} {\bibfnamefont
  {P.}~\bibnamefont {Kattel}}, \bibinfo {author} {\bibfnamefont {P.~R.}\
  \bibnamefont {Pasnoori}}, \bibinfo {author} {\bibfnamefont {J.~H.}\
  \bibnamefont {Pixley}},\ and\ \bibinfo {author} {\bibfnamefont
  {N.}~\bibnamefont {Andrei}},\ }\bibfield  {title} {\bibinfo {title} {{Spin
  chain with non-Hermitian $\mathcal{PT}$-symmetric boundary couplings: Exact
  solution, dissipative Kondo effect, and phase transitions on the edge}},\
  }\href {https://doi.org/10.1103/PhysRevB.111.224407} {\bibfield  {journal}
  {\bibinfo  {journal} {Phys. Rev. B}\ }\textbf {\bibinfo {volume} {111}},\
  \bibinfo {pages} {224407} (\bibinfo {year} {2025}{\natexlab{a}})}\BibitemShut
  {NoStop}%
\bibitem [{\citenamefont {Louren\ifmmode~\mbox{\c{c}}\else \c{c}\fi{}o}\ \emph
  {et~al.}(2018)\citenamefont {Louren\ifmmode~\mbox{\c{c}}\else \c{c}\fi{}o},
  \citenamefont {Eneias},\ and\ \citenamefont {Pereira}}]{Lourenco18}%
  \BibitemOpen
  \bibfield  {author} {\bibinfo {author} {\bibfnamefont {J.~A.~S.}\
  \bibnamefont {Louren\ifmmode~\mbox{\c{c}}\else \c{c}\fi{}o}}, \bibinfo
  {author} {\bibfnamefont {R.~L.}\ \bibnamefont {Eneias}},\ and\ \bibinfo
  {author} {\bibfnamefont {R.~G.}\ \bibnamefont {Pereira}},\ }\bibfield
  {title} {\bibinfo {title} {Kondo effect in a $\mathcal{PT}$-symmetric
  non-hermitian hamiltonian},\ }\href
  {https://doi.org/10.1103/PhysRevB.98.085126} {\bibfield  {journal} {\bibinfo
  {journal} {Phys. Rev. B}\ }\textbf {\bibinfo {volume} {98}},\ \bibinfo
  {pages} {085126} (\bibinfo {year} {2018})}\BibitemShut {NoStop}%
\bibitem [{\citenamefont {Kulkarni}\ \emph {et~al.}(2022)\citenamefont
  {Kulkarni}, \citenamefont {Gupta},\ and\ \citenamefont
  {Vidhyadhiraja}}]{Kulkarni22}%
  \BibitemOpen
  \bibfield  {author} {\bibinfo {author} {\bibfnamefont {V.~M.}\ \bibnamefont
  {Kulkarni}}, \bibinfo {author} {\bibfnamefont {A.}~\bibnamefont {Gupta}},\
  and\ \bibinfo {author} {\bibfnamefont {N.~S.}\ \bibnamefont
  {Vidhyadhiraja}},\ }\bibfield  {title} {\bibinfo {title} {{Kondo effect in a
  non-Hermitian $\mathcal{PT}$-symmetric Anderson model with Rashba spin-orbit
  coupling}},\ }\href {https://doi.org/10.1103/PhysRevB.106.075113} {\bibfield
  {journal} {\bibinfo  {journal} {Phys. Rev. B}\ }\textbf {\bibinfo {volume}
  {106}},\ \bibinfo {pages} {075113} (\bibinfo {year} {2022})}\BibitemShut
  {NoStop}%
\bibitem [{\citenamefont {Nakagawa}\ \emph {et~al.}(2018)\citenamefont
  {Nakagawa}, \citenamefont {Kawakami},\ and\ \citenamefont
  {Ueda}}]{Nakagawa18}%
  \BibitemOpen
  \bibfield  {author} {\bibinfo {author} {\bibfnamefont {M.}~\bibnamefont
  {Nakagawa}}, \bibinfo {author} {\bibfnamefont {N.}~\bibnamefont {Kawakami}},\
  and\ \bibinfo {author} {\bibfnamefont {M.}~\bibnamefont {Ueda}},\ }\bibfield
  {title} {\bibinfo {title} {{Non-Hermitian Kondo Effect in Ultracold
  Alkaline-Earth Atoms}},\ }\href
  {https://doi.org/10.1103/PhysRevLett.121.203001} {\bibfield  {journal}
  {\bibinfo  {journal} {Phys. Rev. Lett.}\ }\textbf {\bibinfo {volume} {121}},\
  \bibinfo {pages} {203001} (\bibinfo {year} {2018})}\BibitemShut {NoStop}%
\bibitem [{\citenamefont {Kattel}\ \emph
  {et~al.}(2025{\natexlab{b}})\citenamefont {Kattel}, \citenamefont {Zhakenov},
  \citenamefont {Pasnoori}, \citenamefont {Azaria},\ and\ \citenamefont
  {Andrei}}]{Kattel25L}%
  \BibitemOpen
  \bibfield  {author} {\bibinfo {author} {\bibfnamefont {P.}~\bibnamefont
  {Kattel}}, \bibinfo {author} {\bibfnamefont {A.}~\bibnamefont {Zhakenov}},
  \bibinfo {author} {\bibfnamefont {P.~R.}\ \bibnamefont {Pasnoori}}, \bibinfo
  {author} {\bibfnamefont {P.}~\bibnamefont {Azaria}},\ and\ \bibinfo {author}
  {\bibfnamefont {N.}~\bibnamefont {Andrei}},\ }\bibfield  {title} {\bibinfo
  {title} {Dissipation driven phase transition in the non-hermitian kondo
  model},\ }\href {https://doi.org/10.1103/PhysRevB.111.L201106} {\bibfield
  {journal} {\bibinfo  {journal} {Phys. Rev. B}\ }\textbf {\bibinfo {volume}
  {111}},\ \bibinfo {pages} {L201106} (\bibinfo {year}
  {2025}{\natexlab{b}})}\BibitemShut {NoStop}%
\bibitem [{\citenamefont {Chen}(2025)}]{Chen25}%
  \BibitemOpen
  \bibfield  {author} {\bibinfo {author} {\bibfnamefont {J.}~\bibnamefont
  {Chen}},\ }\bibfield  {title} {\bibinfo {title} {{Critical behavior of
  non-Hermitian Kondo effect in a pseudogap system}},\ }\href
  {https://doi.org/10.1103/PhysRevB.111.045124} {\bibfield  {journal} {\bibinfo
   {journal} {Phys. Rev. B}\ }\textbf {\bibinfo {volume} {111}},\ \bibinfo
  {pages} {045124} (\bibinfo {year} {2025})}\BibitemShut {NoStop}%
\bibitem [{\citenamefont {Burke}\ and\ \citenamefont
  {Mitchell}(2025)}]{Burke25}%
  \BibitemOpen
  \bibfield  {author} {\bibinfo {author} {\bibfnamefont {P.~C.}\ \bibnamefont
  {Burke}}\ and\ \bibinfo {author} {\bibfnamefont {A.~K.}\ \bibnamefont
  {Mitchell}},\ }\bibfield  {title} {\bibinfo {title} {{Non-Hermitian Numerical
  Renormalization Group: Solution of the Non-Hermitian Kondo Model}},\ }\href
  {https://doi.org/10.1103/19td-1k9s} {\bibfield  {journal} {\bibinfo
  {journal} {Phys. Rev. Lett.}\ }\textbf {\bibinfo {volume} {135}},\ \bibinfo
  {pages} {206502} (\bibinfo {year} {2025})}\BibitemShut {NoStop}%
\bibitem [{PT()}]{PT}%
  \BibitemOpen
  \href@noop {} {}\bibinfo {note} {We note that complex hybridization in the
  NH-AIM has been also studied in a special $\mathcal{PT}$-symmetric setup with
  Rashba spin-orbit couplings in Ref.~\cite{Kulkarni22}.}\BibitemShut {Stop}%
\bibitem [{BCS()}]{BCS}%
  \BibitemOpen
  \href@noop {} {}\bibinfo {note} {This fact is reminiscent of the behavior of
  the superfluid order parameter in the NH BCS theory \cite{Yamamoto19,
  Takemori24, Takemori24B}.}\BibitemShut {Stop}%
\bibitem [{SW()}]{SW}%
  \BibitemOpen
  \href@noop {} {}\bibinfo {note} {We note that the Schrieffer-Wolff
  transformation of the NH-AIM with complex $U$ has been employed to obtain the
  NH Kondo model in Ref.~\cite{Chen25}.}\BibitemShut {Stop}%
\bibitem [{\citenamefont {Wiegmann}(1980)}]{Wiegmann80}%
  \BibitemOpen
  \bibfield  {author} {\bibinfo {author} {\bibfnamefont {P.}~\bibnamefont
  {Wiegmann}},\ }\bibfield  {title} {\bibinfo {title} {{Towards an exact
  solution of the Anderson model}},\ }\href@noop {} {\bibfield  {journal}
  {\bibinfo  {journal} {Phys. Lett. A}\ }\textbf {\bibinfo {volume} {80}},\
  \bibinfo {pages} {163} (\bibinfo {year} {1980})}\BibitemShut {NoStop}%
\bibitem [{\citenamefont {Kawakami}\ and\ \citenamefont
  {Okiji}(1981)}]{Kawakami81}%
  \BibitemOpen
  \bibfield  {author} {\bibinfo {author} {\bibfnamefont {N.}~\bibnamefont
  {Kawakami}}\ and\ \bibinfo {author} {\bibfnamefont {A.}~\bibnamefont
  {Okiji}},\ }\bibfield  {title} {\bibinfo {title} {{Exact expression of the
  ground-state energy for the symmetric Anderson model}},\ }\href@noop {}
  {\bibfield  {journal} {\bibinfo  {journal} {Phys. Lett. A}\ }\textbf
  {\bibinfo {volume} {86}},\ \bibinfo {pages} {483} (\bibinfo {year}
  {1981})}\BibitemShut {NoStop}%
\bibitem [{\citenamefont {Kawakami}\ and\ \citenamefont
  {Okiji}(1982{\natexlab{a}})}]{Kawakami82}%
  \BibitemOpen
  \bibfield  {author} {\bibinfo {author} {\bibfnamefont {N.}~\bibnamefont
  {Kawakami}}\ and\ \bibinfo {author} {\bibfnamefont {A.}~\bibnamefont
  {Okiji}},\ }\bibfield  {title} {\bibinfo {title} {{Ground State of Anderson
  Hamiltonian}},\ }\href@noop {} {\bibfield  {journal} {\bibinfo  {journal} {J.
  Phys. Soc. Jpn.}\ }\textbf {\bibinfo {volume} {51}},\ \bibinfo {pages} {1145}
  (\bibinfo {year} {1982}{\natexlab{a}})}\BibitemShut {NoStop}%
\bibitem [{\citenamefont {Kawakami}\ and\ \citenamefont
  {Okiji}(1982{\natexlab{b}})}]{Kawakami82lett}%
  \BibitemOpen
  \bibfield  {author} {\bibinfo {author} {\bibfnamefont {N.}~\bibnamefont
  {Kawakami}}\ and\ \bibinfo {author} {\bibfnamefont {A.}~\bibnamefont
  {Okiji}},\ }\bibfield  {title} {\bibinfo {title} {{Ground State of Asymmetric
  Anderson Hamiltonian}},\ }\href@noop {} {\bibfield  {journal} {\bibinfo
  {journal} {J. Phys. Soc. Jpn.}\ }\textbf {\bibinfo {volume} {51}},\ \bibinfo
  {pages} {2043} (\bibinfo {year} {1982}{\natexlab{b}})}\BibitemShut {NoStop}%
\bibitem [{\citenamefont {Okiji}\ and\ \citenamefont
  {Kawakami}(1982{\natexlab{a}})}]{Okiji82}%
  \BibitemOpen
  \bibfield  {author} {\bibinfo {author} {\bibfnamefont {A.}~\bibnamefont
  {Okiji}}\ and\ \bibinfo {author} {\bibfnamefont {N.}~\bibnamefont
  {Kawakami}},\ }\bibfield  {title} {\bibinfo {title} {{Magnetic properties of
  Anderson model at zero temperature}},\ }\href@noop {} {\bibfield  {journal}
  {\bibinfo  {journal} {J. Phys. Soc. Jpn.}\ }\textbf {\bibinfo {volume}
  {51}},\ \bibinfo {pages} {3192} (\bibinfo {year}
  {1982}{\natexlab{a}})}\BibitemShut {NoStop}%
\bibitem [{\citenamefont {Okiji}\ and\ \citenamefont
  {Kawakami}(1982{\natexlab{b}})}]{Okiji82solid}%
  \BibitemOpen
  \bibfield  {author} {\bibinfo {author} {\bibfnamefont {A.}~\bibnamefont
  {Okiji}}\ and\ \bibinfo {author} {\bibfnamefont {N.}~\bibnamefont
  {Kawakami}},\ }\bibfield  {title} {\bibinfo {title} {{Exact expression of
  magnetic susceptibility for the Anderson model}},\ }\href@noop {} {\bibfield
  {journal} {\bibinfo  {journal} {Solid State Commun.}\ }\textbf {\bibinfo
  {volume} {43}},\ \bibinfo {pages} {365} (\bibinfo {year}
  {1982}{\natexlab{b}})}\BibitemShut {NoStop}%
\bibitem [{\citenamefont {Wiegmann}\ and\ \citenamefont
  {Tsvelick}(1983)}]{Wiegmann83C}%
  \BibitemOpen
  \bibfield  {author} {\bibinfo {author} {\bibfnamefont {P.}~\bibnamefont
  {Wiegmann}}\ and\ \bibinfo {author} {\bibfnamefont {A.}~\bibnamefont
  {Tsvelick}},\ }\bibfield  {title} {\bibinfo {title} {{Exact solution of the
  Anderson model: I}},\ }\href@noop {} {\bibfield  {journal} {\bibinfo
  {journal} {J. Phys. C: Solid State Phys.}\ }\textbf {\bibinfo {volume}
  {16}},\ \bibinfo {pages} {2281} (\bibinfo {year} {1983})}\BibitemShut
  {NoStop}%
\bibitem [{\citenamefont {Tsvelick}\ and\ \citenamefont
  {Wiegmann}(1983{\natexlab{b}})}]{Tsvelick83C}%
  \BibitemOpen
  \bibfield  {author} {\bibinfo {author} {\bibfnamefont {A.}~\bibnamefont
  {Tsvelick}}\ and\ \bibinfo {author} {\bibfnamefont {P.}~\bibnamefont
  {Wiegmann}},\ }\bibfield  {title} {\bibinfo {title} {{Exact solution of the
  Anderson model. II. Thermodynamic properties at finite temperatures}},\
  }\href@noop {} {\bibfield  {journal} {\bibinfo  {journal} {J. Phys. C: Solid
  State Phys.}\ }\textbf {\bibinfo {volume} {16}},\ \bibinfo {pages} {2321}
  (\bibinfo {year} {1983}{\natexlab{b}})}\BibitemShut {NoStop}%
\bibitem [{\citenamefont {Schlottmann}(1983)}]{Schlottmann83}%
  \BibitemOpen
  \bibfield  {author} {\bibinfo {author} {\bibfnamefont {P.}~\bibnamefont
  {Schlottmann}},\ }\bibfield  {title} {\bibinfo {title} {{Bethe-Ansatz
  Solution of the Anderson Model of a Magnetic Impurity with Orbital
  Degeneracy}},\ }\href {https://doi.org/10.1103/PhysRevLett.50.1697}
  {\bibfield  {journal} {\bibinfo  {journal} {Phys. Rev. Lett.}\ }\textbf
  {\bibinfo {volume} {50}},\ \bibinfo {pages} {1697} (\bibinfo {year}
  {1983})}\BibitemShut {NoStop}%
\bibitem [{\citenamefont {Kawakami}\ and\ \citenamefont
  {Okiji}(1986)}]{Kawakami86}%
  \BibitemOpen
  \bibfield  {author} {\bibinfo {author} {\bibfnamefont {N.}~\bibnamefont
  {Kawakami}}\ and\ \bibinfo {author} {\bibfnamefont {A.}~\bibnamefont
  {Okiji}},\ }\bibfield  {title} {\bibinfo {title} {{Transport properties of
  the Anderson model at low temperatures}},\ }\href@noop {} {\bibfield
  {journal} {\bibinfo  {journal} {Phys. Lett.A}\ }\textbf {\bibinfo {volume}
  {118}},\ \bibinfo {pages} {301} (\bibinfo {year} {1986})}\BibitemShut
  {NoStop}%
\bibitem [{\citenamefont {Schlottmann}(1989)}]{Schlottmann89}%
  \BibitemOpen
  \bibfield  {author} {\bibinfo {author} {\bibfnamefont {P.}~\bibnamefont
  {Schlottmann}},\ }\bibfield  {title} {\bibinfo {title} {Some exact results
  for dilute mixed-valent and heavy-fermion systems},\ }\href@noop {}
  {\bibfield  {journal} {\bibinfo  {journal} {Phys. Rep.}\ }\textbf {\bibinfo
  {volume} {181}},\ \bibinfo {pages} {1} (\bibinfo {year} {1989})}\BibitemShut
  {NoStop}%
\bibitem [{\citenamefont {Fukui}\ and\ \citenamefont
  {Kawakami}(1998{\natexlab{a}})}]{Fukui98}%
  \BibitemOpen
  \bibfield  {author} {\bibinfo {author} {\bibfnamefont {T.}~\bibnamefont
  {Fukui}}\ and\ \bibinfo {author} {\bibfnamefont {N.}~\bibnamefont
  {Kawakami}},\ }\bibfield  {title} {\bibinfo {title} {{Breakdown of the Mott
  insulator: Exact solution of an asymmetric Hubbard model}},\ }\href
  {https://doi.org/10.1103/PhysRevB.58.16051} {\bibfield  {journal} {\bibinfo
  {journal} {Phys. Rev. B}\ }\textbf {\bibinfo {volume} {58}},\ \bibinfo
  {pages} {16051} (\bibinfo {year} {1998}{\natexlab{a}})}\BibitemShut {NoStop}%
\bibitem [{\citenamefont {Shibata}\ and\ \citenamefont
  {Katsura}(2019)}]{Shibata19}%
  \BibitemOpen
  \bibfield  {author} {\bibinfo {author} {\bibfnamefont {N.}~\bibnamefont
  {Shibata}}\ and\ \bibinfo {author} {\bibfnamefont {H.}~\bibnamefont
  {Katsura}},\ }\bibfield  {title} {\bibinfo {title} {{Dissipative quantum
  Ising chain as a non-Hermitian Ashkin-Teller model}},\ }\href
  {https://doi.org/10.1103/PhysRevB.99.224432} {\bibfield  {journal} {\bibinfo
  {journal} {Phys. Rev. B}\ }\textbf {\bibinfo {volume} {99}},\ \bibinfo
  {pages} {224432} (\bibinfo {year} {2019})}\BibitemShut {NoStop}%
\bibitem [{\citenamefont {Nakagawa}\ \emph {et~al.}(2021)\citenamefont
  {Nakagawa}, \citenamefont {Kawakami},\ and\ \citenamefont
  {Ueda}}]{Nakagawa21}%
  \BibitemOpen
  \bibfield  {author} {\bibinfo {author} {\bibfnamefont {M.}~\bibnamefont
  {Nakagawa}}, \bibinfo {author} {\bibfnamefont {N.}~\bibnamefont {Kawakami}},\
  and\ \bibinfo {author} {\bibfnamefont {M.}~\bibnamefont {Ueda}},\ }\bibfield
  {title} {\bibinfo {title} {{Exact Liouvillian Spectrum of a One-Dimensional
  Dissipative Hubbard Model}},\ }\href
  {https://doi.org/10.1103/PhysRevLett.126.110404} {\bibfield  {journal}
  {\bibinfo  {journal} {Phys. Rev. Lett.}\ }\textbf {\bibinfo {volume} {126}},\
  \bibinfo {pages} {110404} (\bibinfo {year} {2021})}\BibitemShut {NoStop}%
\bibitem [{\citenamefont {Bu{\v{c}}a}\ \emph {et~al.}(2020)\citenamefont
  {Bu{\v{c}}a}, \citenamefont {Booker}, \citenamefont {Medenjak},\ and\
  \citenamefont {Jaksch}}]{Buca20}%
  \BibitemOpen
  \bibfield  {author} {\bibinfo {author} {\bibfnamefont {B.}~\bibnamefont
  {Bu{\v{c}}a}}, \bibinfo {author} {\bibfnamefont {C.}~\bibnamefont {Booker}},
  \bibinfo {author} {\bibfnamefont {M.}~\bibnamefont {Medenjak}},\ and\
  \bibinfo {author} {\bibfnamefont {D.}~\bibnamefont {Jaksch}},\ }\bibfield
  {title} {\bibinfo {title} {Bethe ansatz approach for dissipation: exact
  solutions of quantum many-body dynamics under loss},\ }\href@noop {}
  {\bibfield  {journal} {\bibinfo  {journal} {New J. Phys.}\ }\textbf {\bibinfo
  {volume} {22}},\ \bibinfo {pages} {123040} (\bibinfo {year}
  {2020})}\BibitemShut {NoStop}%
\bibitem [{\citenamefont {Mao}\ \emph {et~al.}(2023)\citenamefont {Mao},
  \citenamefont {Hao},\ and\ \citenamefont {Pan}}]{Mao23}%
  \BibitemOpen
  \bibfield  {author} {\bibinfo {author} {\bibfnamefont {L.}~\bibnamefont
  {Mao}}, \bibinfo {author} {\bibfnamefont {Y.}~\bibnamefont {Hao}},\ and\
  \bibinfo {author} {\bibfnamefont {L.}~\bibnamefont {Pan}},\ }\bibfield
  {title} {\bibinfo {title} {{Non-Hermitian skin effect in a one-dimensional
  interacting Bose gas}},\ }\href {https://doi.org/10.1103/PhysRevA.107.043315}
  {\bibfield  {journal} {\bibinfo  {journal} {Phys. Rev. A}\ }\textbf {\bibinfo
  {volume} {107}},\ \bibinfo {pages} {043315} (\bibinfo {year}
  {2023})}\BibitemShut {NoStop}%
\bibitem [{\citenamefont {March\'e}\ \emph {et~al.}(2024)\citenamefont
  {March\'e}, \citenamefont {Yoshida}, \citenamefont {Nardin}, \citenamefont
  {Katsura},\ and\ \citenamefont {Mazza}}]{Marche24}%
  \BibitemOpen
  \bibfield  {author} {\bibinfo {author} {\bibfnamefont {A.}~\bibnamefont
  {March\'e}}, \bibinfo {author} {\bibfnamefont {H.}~\bibnamefont {Yoshida}},
  \bibinfo {author} {\bibfnamefont {A.}~\bibnamefont {Nardin}}, \bibinfo
  {author} {\bibfnamefont {H.}~\bibnamefont {Katsura}},\ and\ \bibinfo {author}
  {\bibfnamefont {L.}~\bibnamefont {Mazza}},\ }\bibfield  {title} {\bibinfo
  {title} {{Universality and two-body losses: Lessons from the effective
  non-Hermitian dynamics of two particles}},\ }\href
  {https://doi.org/10.1103/PhysRevA.110.033321} {\bibfield  {journal} {\bibinfo
   {journal} {Phys. Rev. A}\ }\textbf {\bibinfo {volume} {110}},\ \bibinfo
  {pages} {033321} (\bibinfo {year} {2024})}\BibitemShut {NoStop}%
\bibitem [{\citenamefont {Yang}(1967)}]{Yang67}%
  \BibitemOpen
  \bibfield  {author} {\bibinfo {author} {\bibfnamefont {C.~N.}\ \bibnamefont
  {Yang}},\ }\bibfield  {title} {\bibinfo {title} {{Some Exact Results for the
  Many-Body Problem in one Dimension with Repulsive Delta-Function
  Interaction}},\ }\href {https://doi.org/10.1103/PhysRevLett.19.1312}
  {\bibfield  {journal} {\bibinfo  {journal} {Phys. Rev. Lett.}\ }\textbf
  {\bibinfo {volume} {19}},\ \bibinfo {pages} {1312} (\bibinfo {year}
  {1967})}\BibitemShut {NoStop}%
\bibitem [{Zen()}]{Zenodo}%
  \BibitemOpen
  \href@noop {} {}\bibinfo {note} {Yamamoto, K., and Nakagawa, M., and Kawakami
  N. (2025). Dataset for "Kondo breakdown induced by non-Hermitian complex
  hybridization" [Data set]. Zenodo.
  https://doi.org/10.5281/zenodo.17421349}\BibitemShut {NoStop}%
\bibitem [{\citenamefont {Yamamoto}\ and\ \citenamefont
  {Hamazaki}(2023)}]{Yamamoto23L}%
  \BibitemOpen
  \bibfield  {author} {\bibinfo {author} {\bibfnamefont {K.}~\bibnamefont
  {Yamamoto}}\ and\ \bibinfo {author} {\bibfnamefont {R.}~\bibnamefont
  {Hamazaki}},\ }\bibfield  {title} {\bibinfo {title} {Localization properties
  in disordered quantum many-body dynamics under continuous measurement},\
  }\href {https://doi.org/10.1103/PhysRevB.107.L220201} {\bibfield  {journal}
  {\bibinfo  {journal} {Phys. Rev. B}\ }\textbf {\bibinfo {volume} {107}},\
  \bibinfo {pages} {L220201} (\bibinfo {year} {2023})}\BibitemShut {NoStop}%
\bibitem [{\citenamefont {Yamamoto}\ and\ \citenamefont
  {Hamazaki}(2026)}]{Yamamoto25}%
  \BibitemOpen
  \bibfield  {author} {\bibinfo {author} {\bibfnamefont {K.}~\bibnamefont
  {Yamamoto}}\ and\ \bibinfo {author} {\bibfnamefont {R.}~\bibnamefont
  {Hamazaki}},\ }\bibfield  {title} {\bibinfo {title} {Measurement-induced
  crossover of quantum jump statistics in postselection-free many-body
  dynamics},\ }\bibfield  {journal} {\bibinfo  {journal} {Phys. Rev. Lett.}\
  }\href {https://doi.org/10.1103/wv5b-r6sb} {10.1103/wv5b-r6sb} (\bibinfo
  {year} {2026}),\ \bibinfo {note} {to be published},\ \Eprint
  {https://arxiv.org/abs/2503.02418} {arXiv:2503.02418 [cond-mat.stat-mech]}
  \BibitemShut {NoStop}%
\bibitem [{\citenamefont {Gong}\ \emph {et~al.}(2018)\citenamefont {Gong},
  \citenamefont {Ashida}, \citenamefont {Kawabata}, \citenamefont {Takasan},
  \citenamefont {Higashikawa},\ and\ \citenamefont {Ueda}}]{Gong18}%
  \BibitemOpen
  \bibfield  {author} {\bibinfo {author} {\bibfnamefont {Z.}~\bibnamefont
  {Gong}}, \bibinfo {author} {\bibfnamefont {Y.}~\bibnamefont {Ashida}},
  \bibinfo {author} {\bibfnamefont {K.}~\bibnamefont {Kawabata}}, \bibinfo
  {author} {\bibfnamefont {K.}~\bibnamefont {Takasan}}, \bibinfo {author}
  {\bibfnamefont {S.}~\bibnamefont {Higashikawa}},\ and\ \bibinfo {author}
  {\bibfnamefont {M.}~\bibnamefont {Ueda}},\ }\bibfield  {title} {\bibinfo
  {title} {{Topological Phases of Non-Hermitian Systems}},\ }\href
  {https://doi.org/10.1103/PhysRevX.8.031079} {\bibfield  {journal} {\bibinfo
  {journal} {Phys. Rev. X}\ }\textbf {\bibinfo {volume} {8}},\ \bibinfo {pages}
  {031079} (\bibinfo {year} {2018})}\BibitemShut {NoStop}%
\bibitem [{\citenamefont {Takemori}\ \emph
  {et~al.}(2024{\natexlab{a}})\citenamefont {Takemori}, \citenamefont
  {Yamamoto},\ and\ \citenamefont {Koga}}]{Takemori24}%
  \BibitemOpen
  \bibfield  {author} {\bibinfo {author} {\bibfnamefont {S.}~\bibnamefont
  {Takemori}}, \bibinfo {author} {\bibfnamefont {K.}~\bibnamefont {Yamamoto}},\
  and\ \bibinfo {author} {\bibfnamefont {A.}~\bibnamefont {Koga}},\ }\bibfield
  {title} {\bibinfo {title} {{Theory of non-Hermitian fermionic superfluidity
  on a honeycomb lattice: Interplay between exceptional manifolds and Van Hove
  singularity}},\ }\href {https://doi.org/10.1103/PhysRevB.109.L060501}
  {\bibfield  {journal} {\bibinfo  {journal} {Phys. Rev. B}\ }\textbf {\bibinfo
  {volume} {109}},\ \bibinfo {pages} {L060501} (\bibinfo {year}
  {2024}{\natexlab{a}})}\BibitemShut {NoStop}%
\bibitem [{\citenamefont {Takemori}\ \emph
  {et~al.}(2024{\natexlab{b}})\citenamefont {Takemori}, \citenamefont
  {Yamamoto},\ and\ \citenamefont {Koga}}]{Takemori24B}%
  \BibitemOpen
  \bibfield  {author} {\bibinfo {author} {\bibfnamefont {S.}~\bibnamefont
  {Takemori}}, \bibinfo {author} {\bibfnamefont {K.}~\bibnamefont {Yamamoto}},\
  and\ \bibinfo {author} {\bibfnamefont {A.}~\bibnamefont {Koga}},\ }\bibfield
  {title} {\bibinfo {title} {{Phase diagram of non-Hermitian BCS superfluids in
  a dissipative asymmetric Hubbard model}},\ }\href
  {https://doi.org/10.1103/PhysRevB.110.184518} {\bibfield  {journal} {\bibinfo
   {journal} {Phys. Rev. B}\ }\textbf {\bibinfo {volume} {110}},\ \bibinfo
  {pages} {184518} (\bibinfo {year} {2024}{\natexlab{b}})}\BibitemShut
  {NoStop}%
\bibitem [{SCE()}]{SCE}%
  \BibitemOpen
  \href@noop {} {}\bibinfo {note} {Such a structure is different from the
  corresponding SCE in the Hermitian AIM \cite{Coleman87} and the NH-AIM with
  one-body loss \cite{Yamamoto24}.}\BibitemShut {Stop}%
\bibitem [{\citenamefont {Fukui}\ and\ \citenamefont
  {Kawakami}(1998{\natexlab{b}})}]{Kawakami98}%
  \BibitemOpen
  \bibfield  {author} {\bibinfo {author} {\bibfnamefont {T.}~\bibnamefont
  {Fukui}}\ and\ \bibinfo {author} {\bibfnamefont {N.}~\bibnamefont
  {Kawakami}},\ }\bibfield  {title} {\bibinfo {title} {{Breakdown of the Mott
  insulator: Exact solution of an asymmetric Hubbard model}},\ }\href
  {https://doi.org/10.1103/PhysRevB.58.16051} {\bibfield  {journal} {\bibinfo
  {journal} {Phys. Rev. B}\ }\textbf {\bibinfo {volume} {58}},\ \bibinfo
  {pages} {16051} (\bibinfo {year} {1998}{\natexlab{b}})}\BibitemShut {NoStop}%
\bibitem [{\citenamefont {Fetter}\ and\ \citenamefont
  {Walecka}(1971)}]{Fetter71}%
  \BibitemOpen
  \bibfield  {author} {\bibinfo {author} {\bibfnamefont {A.~L.}\ \bibnamefont
  {Fetter}}\ and\ \bibinfo {author} {\bibfnamefont {J.~D.}\ \bibnamefont
  {Walecka}},\ }\href@noop {} {\emph {\bibinfo {title} {Quantum theory of
  many-particle systems}}}\ (\bibinfo  {publisher} {McGraw-Hill, New York},\
  \bibinfo {year} {1971})\BibitemShut {NoStop}%
\bibitem [{\citenamefont {Coveney}\ and\ \citenamefont {Tew}()}]{Coveney25}%
  \BibitemOpen
  \bibfield  {author} {\bibinfo {author} {\bibfnamefont {C.~J.}\ \bibnamefont
  {Coveney}}\ and\ \bibinfo {author} {\bibfnamefont {D.~P.}\ \bibnamefont
  {Tew}},\ }\bibfield  {title} {\bibinfo {title} {{Non-Hermitian Green's
  function theory with $ N $-body interactions: the coupled-cluster similarity
  transformation}},\ }\href@noop {} {\bibinfo  {journal} {arXiv:2503.06586}\
  }\BibitemShut {NoStop}%
\end{thebibliography}%

\end{document}